\documentclass[american,british]{article}
\usepackage[T1]{fontenc}
\usepackage[latin9]{inputenc}
\usepackage{geometry}
\usepackage{float}
\usepackage{amsmath}
\usepackage{amssymb}
\usepackage{graphicx}
\usepackage{wasysym}

\makeatletter
\newcommand{\lyxaddress}[1]{
	\par {\raggedright #1
	\vspace{1.4em}
	\noindent\par}
}

\numberwithin{equation}{section}
\usepackage{hyperref}

\makeatother

\usepackage{babel}
\begin{document}
\title{Finite volume form factors in integrable theories}
\author{Zoltan Bajnok\textsuperscript{1}, Georgios Linardopoulos\textsuperscript{1}\thanks{Currently at the Asia Pacific Center for Theoretical Physics (APCTP);
Hogil Kim Memorial Building \#501 POSTECH, 77 Cheongam-Ro Nam-gu,
Pohang Gyeongsangbuk-do, 37673 Korea.}, Istvan M. Szecsenyi\textsuperscript{2}, and Istvan Vona\textsuperscript{1}}
\maketitle

\lyxaddress{\begin{center}
\textsuperscript{1}\emph{Wigner Research Centre for Physics}\\
\emph{Konkoly-Thege Miklós u. 29-33, 1121 Budapest , Hungary}\\
\emph{}\\
\textsuperscript{2}\emph{Nordita, KTH Royal Institute of Technology
and Stockholm University,}\\
\emph{Hannes Alfvéns väg 12, SE-106 91 Stockholm, Sweden}
\par\end{center}}
\begin{abstract}
\noindent We develop a new method to calculate finite size corrections
for form factors in two-dimensional integrable quantum field theories.
We extract these corrections from the excited state expectation value
of bilocal operators in the limit when the operators are far apart.
We elaborate the finite size effects explicitly up to the 3rd Lüscher
order and conjecture the structure of the general form. We also fully
recover the explicitly known massive fermion finite volume form factors.
\end{abstract}

\section{Introduction}

Recently, there has been a growing interest in the finite size corrections
of form factors that arise in integrable systems. The motivation comes
from several, not directly related places. Finite volume/tempe-rature
form factors are the building blocks of finite volume/temperature
correlation functions. In turn, these are the fundamental observables
and measurable quantities of two-dimensional integrable systems that
show up in statistical, condensed matter, quantum field, and high-energy
theoretical physics \cite{Mussardo:2010mgq,Bajnok:2013sa}. In finite
temperature statistical physics, the authors \cite{CortesCubero:2018mpj}
formulated form factor axioms and used them to calculate correlation
functions \cite{Cubero:2019rcy} which are relevant in generalised
hydrodynamics. In condensed matter physics, the quantities of interest
are finite temperature form factors on the lattice. Significant progress
has been made in developing the thermal form factor expansion \cite{Dugave:2013xya,Dugave:2016hep,Gohmann:2017mke}
for lattice observables in integrable models, as well as applying
it to the computation of real-time correlation functions \cite{Babenko:2020spo,Babenko:2020nzg}.
In quantum field theory, finite volume theories interpolate between
the infrared scattering description and the ultraviolet Lagrangian
formulation, where the volume serves as the renormalisation group
parameter. Thus the focus in these theories is mainly on finite volume
expectation values. There is an approach which exploits a fermionic
basis originating from a lattice discretisation \cite{Jimbo:2009ja,Jimbo:2011bc,Negro:2013wga,Bajnok:2019yik,Babenko:2019tvv},
while one can also derive expectation values directly from the lattice
by taking the continuum limit \cite{Hegedus:2017zkz,Hegedus:2019rju}.
Finite volume form factors are also related to AdS/CFT 3-point functions
\cite{Bajnok:2014sza,Bajnok:2016xxu,Jiang:2015bvm,Jiang:2016dsr,Bajnok:2017mdf}
which, together with 2-point functions (or the spectrum of scaling
dimensions) \cite{Beisert:2010jr,Gromov:2013pga}, characterise these
theories completely.

The integrable formulation of an integrable quantum field theory aims
to express all of its physical observables (e.g. the finite volume
spectrum and correlation functions) purely in terms of bootstrapable
quantities at infinite volume \cite{Luscher:1985dn,Luscher:1986pf}.
These quantities include masses, scattering matrices and infinite
volume form factors which are the matrix elements of local operators
between asymptotic multiparticle states. All these quantities can
be determined by completing the S-matrix and form factor bootstraps
\cite{Zamolodchikov:1978xm,Dorey:1996gd,Bajnok:2013sa,Mussardo:2010mgq,Smirnov:1992vz,Babujian:2003sc}.
The leading polynomial volume corrections originate from finite volume
momentum quantisation, which can be formulated in terms of pairwise
scatterings \cite{Luscher:1986pf}. The subleading exponentially suppressed
volume corrections are due to virtual particles \cite{Luscher:1985dn}
(which scatter among themselves) and physical particles. All such
contributions must be summed up for an exact formulation.

In the case of the \emph{energy spectrum}, the polynomial volume corrections
come from the quantization of momentum, which implies the Bethe-Yang
equations. The leading exponential finite size corrections of multiparticle
states include the modification of the Bethe-Yang equations and the
direct contribution of the sea of virtual particles \cite{Bajnok:2008bm}.
The subleading energy corrections also involve the scattering of virtual
particles among themselves \cite{Bombardelli:2013yka}. The total
contribution of virtual particles is summed up for the ground state
by the thermodynamic Bethe ansatz (TBA), which comes from evaluating
the torus partition function in two alternative ways, i.e. by choosing
two different time evolutions along the two orthogonal cycles \cite{Zamolodchikov:1989cf}.
Excited state energies can be obtained either by analytically continuing
the ground-state result \cite{Dorey:1996re}, or by calculating the
continuum limit of integrable lattice regularisations \cite{Fioravanti:1996rz,Teschner:2007ng,Feverati:1998dt,Destri:1997yz}.

Finite size corrections for \emph{diagonal} and \emph{non-diagonal}
matrix elements of local operators (form factors) are quite different.
The simplest \emph{diagonal} matrix element, namely the finite volume/temperatu-re
vacuum expectation value, can be obtained by evaluating the torus
one-point function in two alternative ways (i.e. just like the groundstate
energy). The result can be expressed in terms of infinite volume connected
form factors and the TBA pseudo-energy through the LeClair-Mussardo
(LM) formula \cite{Leclair:1999ys}. Analytic continuation of this
expression provides the expectation values of excited states \cite{Pozsgay:2013jua,Pozsgay:2014gza},
i.e. \emph{diagonal} finite volume form factors. The situation for
\emph{non-diagonal} form factors has yet to be understood at the same
level, and our paper aims to advance precisely this direction. In
other words, we would like to formulate an LM-type description for
\emph{non-diagonal} form factors at finite volume by going beyond
the results which are available in the literature.

In the case of non-diagonal finite volume form factors, the polynomial
corrections are due to changing the normalisation of the states \cite{Pozsgay:2007kn}.
The leading exponential corrections can be calculated by examining
the analytical structure of two-point functions at finite volume \cite{Bajnok:2018fzx,Bajnok:2019cdf}.
By identifying and evaluating the singularities of two-point functions
in momentum space, both the finite size spectrum and the finite volume
form factors can be systematically computed. Although this approach
works beyond the leading Lüscher correction, it is technically very
challenging to calculate higher-order Lüscher terms. A significant
simplification would involve the LM-type formula which was obtained
for two-point functions, i.e. for bilocal operators \cite{Pozsgay:2018tns}.
By analysing the large separation limit of the two operators and inserting
a complete system of finite volume states between them, it is possible
to focus on the contribution of each excited state. We would expect
this approach to allow us to extract finite volume form factors, however
the projection to a given excited state does not turn out to be very
straightforward. That is why we take a different route here.

Our approach introduces the LM-type formulation of the \emph{excited
state} expectation value for two-point functions and analyses their
large separation limit. This is in spirit similar to the approach
that was taken in \cite{Basso:2022nny} for calculating certain 3-point
functions in the AdS/CFT correspondence. The dominant contribution
in this limit comes from the vacuum, which is easy to separate and
elaborate. The resulting computational framework allows us to determine
systematically the finite size corrections of non-diagonal form factors.

Here's the outline of our paper. In Section 2, we introduce all the
fundamental quantities that are needed for our calculations. These
include the infinite volume scattering matrix and form factors (together
with their relevant properties), the definition of finite volume states
and form factors, the exact description of the finite volume spectrum
in terms of pseudo-energies (which satisfy the excited state TBA equations),
and the leading behaviour of the finite volume form factor. At the
beginning of Section 3, we recall the LM-type formula for bilocal
operators and its large separation limit, together with the relation
between the physical and the mirror channels. We then generalise this
formula for excited states. We show that its leading behaviour in
the large separation limit contains finite volume form factors and
a factor which grows exponentially in the exact finite volume energy
difference between the ground and the excited state. Section 4 deals
with the evaluation of finite volume form factors in the first three
Lüscher orders. We proceed order by order by gradually introducing
simplifications. We provide detailed evaluations for the first two
Lüscher orders, while presenting only the idea of the calculation
and the results for the third order. Finally, we conjecture the generic
structure of the all-order result and present our findings (up to
third order) in this language. Section 5 contains the definition of
all-order non-diagonal connected form factors and the graph rules
with which we evaluate them. Section 6 explains how our result can
be extended from a one-particle state to multiparticle states. In
Section 7 we focus on the finite size form factors of the non-local
$\sigma$-field in the free massive fermion theory. We demonstrate
that our approach indeed reproduces the non-trivial result of the
literature \cite{Fonseca:2001dc}. We conclude in Section 8 by also
providing an outlook. The various technical details are relegated
to five Appendices. In Appendix A we expand the energy difference
in Lüscher orders, while in Appendix B we do it in the excited state
filling fraction. Appendix C explains the singularity structure of
one of the simplest connected form factors. Appendix D contains the
calculation of the most involved third order diagram, while Appendix
E provides details on the calculation of free fermion form factors
at finite size.

\section{Preliminaries}

Our aim is to express the finite volume matrix element (or form factor)
of a local operator ${\cal O}$ in terms of the infinite volume form
factor $F^{{\cal O}}$ and the scattering matrix of the theory $S$.
We consider integrable relativistic theories with a single particle
type of mass $m$. We also neglect bound state formation.\footnote{Boundstate formation implies that a single particle is described by
more than one rapidity in the TBA formulation. This would make the
presentation more technical, however, our multiparticle result, with
appropriately placed rapidities can describe those theories, too.} The $2\to2$ scattering matrix $S(\theta_{1}-\theta_{2})$ is a single
function of the rapidity difference, which satisfies unitarity $S(\theta)S(-\theta)=1$
and crossing $S(i\pi-\theta)=S(\theta)$. We have the sinh-Gordon
theory in mind, but our considerations can be easily generalised to
any theory with diagonal scattering.

\begin{figure}[H]
\begin{centering}
\includegraphics[width=1.5cm]{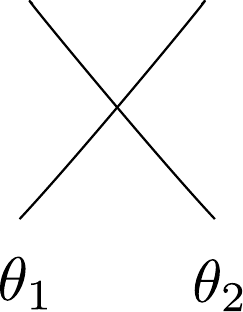}
\par\end{centering}
\caption{Graphical representation of the scattering matrix $S(\theta_{1}-\theta_{2})$.}
\end{figure}

A $N$-particle state at finite volume can be parametrised by the
momenta of the particles $\{\bar{p}_{i}\}$, or alternatively by their
rapidities $\{\bar{\theta}_{i}\},$ where $\bar{p}_{i}=m\sinh\bar{\theta}_{i}$.
The energy of the particle will be denoted by $\bar{e}_{i}=m\cosh\bar{\theta}_{i}$.
We distinguish the physical rapidities from the rapidities appearing
in the thermal formulation by a bar. The physical rapidities are determined
by the exact quantisation conditions 
\begin{equation}
\epsilon_{N}(\bar{\theta}_{i}+\frac{i\pi}{2})=i\pi(2k_{i}+1),
\end{equation}
where $\{k_{i}\}$ are integers and the pseudo-energy satisfies the
excited state TBA equation 
\begin{equation}
\epsilon_{N}(\theta)=mL\cosh\theta+\sum_{i=1}^{N}\log S(\theta-\bar{\theta}_{i}-\frac{i\pi}{2})-\int\frac{d\theta'}{2\pi}\varphi(\theta-\theta')\log(1+e^{-\epsilon_{N}(\theta')}).\label{eq:exTBA}
\end{equation}
Here $L$ is the volume, the integral kernel is related to the scattering
matrix by $\varphi(\theta)=-i\partial_{\theta}\log S(\theta)$, while
$P=\sum_{i}\frac{2\pi k_{i}}{L}$ is the total momentum. In all integrals,
if not explicitly stated otherwise, we integrate along the real line.
The energy of a $N$-particle state is given by 
\begin{equation}
E_{N}(\{\bar{\theta}_{i}\})=\sum_{i=1}^{N}m\cosh\bar{\theta}_{i}-m\int\frac{d\theta}{2\pi}\cosh\theta\,\log(1+e^{-\epsilon_{N}(\theta)}).\label{eq:exEnergy}
\end{equation}
For the vacuum state $N=0$ and the sums, as well as the quantisation
conditions, are absent. The finite volume spectrum is discrete and
the finite volume states 
\begin{equation}
\vert\bar{\theta}_{1},\dots,\bar{\theta}_{N}\rangle_{L},
\end{equation}
are symmetric and normalised to Kronecker delta functions\footnote{The phase of the state is not fixed and we have the freedom to choose
it in a convenient way.}
\begin{equation}
_{L}\langle\bar{\theta}_{1},\dots,\bar{\theta}_{N}\vert\bar{\theta}_{1}',\dots,\bar{\theta}_{N'}'\rangle_{L}=\delta_{NN'}\delta_{k_{1},k_{1}'}\dots\delta_{k_{N},k_{N}'}.
\end{equation}
We are interested in the finite volume form factors:
\begin{equation}
_{L}\langle0\vert{\cal O}\vert\bar{\theta}_{1},\dots,\bar{\theta}_{N}\rangle_{L},
\end{equation}
which we would like to express in terms of the pseudo energies $\{\epsilon_{0},\epsilon_{N}\}$
and the infinite volume form factors 
\begin{equation}
\langle0\vert{\cal O}\vert\theta_{1},\dots,\theta_{N}\rangle=F^{{\cal O}}(\theta_{1},\dots,\theta_{N}).\label{eq:infinite_volume_FF}
\end{equation}
The infinite volume form factors (\ref{eq:infinite_volume_FF}) are
the matrix elements of the local operator ${\cal O}$ between asymptotic
states. The initial states $\vert\theta_{1},\dots,\theta_{N}\rangle$
(with $\theta_{1}>\dots>\theta_{N}$) are connected to the final states
$\vert\theta_{N},\dots,\theta_{1}\rangle$ by the multiparticle S-matrix,
which factorises into two-particle scatterings. As a result (infinite
volume) form factors satisfy the permutation symmetry property 
\begin{equation}
F^{{\cal O}}(\theta_{1},\dots,\theta_{i},\theta_{i+1},\dots,\theta_{N})=S(\theta_{i}-\theta_{i+1})F^{{\cal O}}(\theta_{1},\dots,\theta_{i+1},\theta_{i},\dots,\theta_{N}).\label{eq:permutation}
\end{equation}
More complicated matrix elements can be obtained from the crossing
property which reads

\begin{figure}[H]
\begin{centering}
\includegraphics[width=1.8cm]{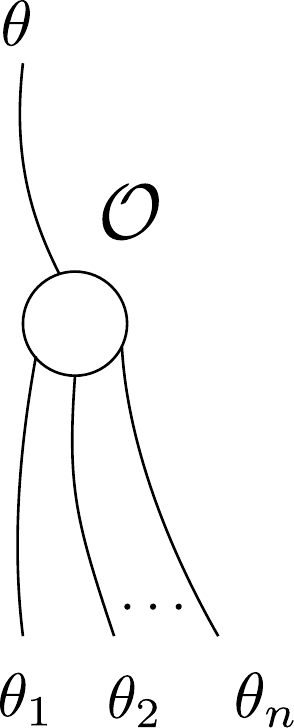}~~~~~~~~~~~~~~~~\includegraphics[width=3.1cm]{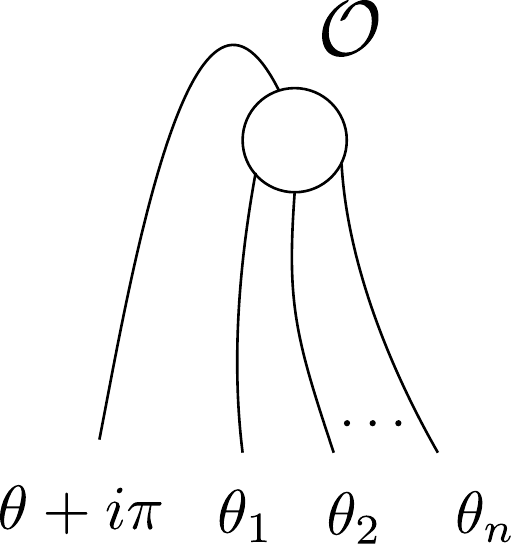}~~~~~~~~~~~~~~~~\includegraphics[width=3cm]{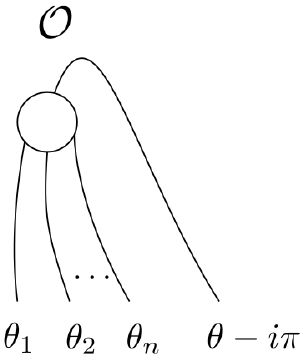}
\par\end{centering}
\caption{Graphical representation of the form factor crossing property.}
\end{figure}

\begin{equation}
\langle\theta\vert{\cal O}\vert\theta_{1},\dots,\theta_{N}\rangle=F^{{\cal O}}(\theta+i\pi,\theta_{1},\dots,\theta_{N})=F^{{\cal O}}(\theta_{1},\dots,\theta_{N},\theta-i\pi),\label{eq:periodicity}
\end{equation}
and we have assumed that $\theta\neq\theta_{i}$ ($i=1,...,N$). In
case of coinciding rapidities we have extra singular terms, which
manifest themselves as kinematical singularities of form factors 
\begin{equation}
F^{{\cal O}}(\theta+i\pi+i\varepsilon,\theta,\theta_{1},\dots,\theta_{N})=\frac{1}{\varepsilon}(1-\prod_{i=1}^{N}S(\theta-\theta_{i}))F^{{\cal O}}(\theta_{1},\dots,\theta_{N})+\dots\label{eq:kinsing}
\end{equation}

Infinite volume states are normalised to the Dirac delta function
as $\langle\theta\vert\theta'\rangle=2\pi\delta(\theta-\theta')$,
while finite volume states to the Kronecker delta function. In changing
between the two bases in the large volume limit we need the Jacobian
\begin{equation}
\rho_{N}=\det_{j,k}\partial_{\bar{\theta}_{k}}(-i\epsilon_{N}(\bar{\theta}_{j}+i\pi/2)).
\end{equation}
At large distances (when exponentially small corrections in the volume
are neglected) the integral terms are absent from both the TBA (\ref{eq:exTBA})
and the energy equations (\ref{eq:exEnergy}). At this polynomial
order the finite and infinite volume form factors are related by simply
changing the normalisation of states 
\begin{equation}
_{L}\langle0\vert{\cal O}\vert\bar{\theta}_{1},\dots,\bar{\theta}_{N}\rangle_{L}=\frac{F^{{\cal O}}(\bar{\theta}_{1},\dots,\bar{\theta}_{N})}{\sqrt{\rho_{N}\prod_{i<j}S(\bar{\theta}_{i}-\bar{\theta}_{j})}}+O(e^{-mL}).\label{eq:finite_volume_FF}
\end{equation}
The \foreignlanguage{american}{normalization} of the finite volume
state does not fix its phase, that is why we have included the phase
factor $\prod_{i<j}S(\bar{\theta}_{i}-\bar{\theta}_{j})$, which makes
the finite volume state symmetric.

The aim of our paper will be to systematically calculate all the exponentially
suppressed corrections in (\ref{eq:finite_volume_FF}). These corrections
appear in $\rho_{N}$, but also modify the form factor in the numerator
by the contribution of virtual particles (which circle around the
finite volume and are created/absorbed by the operator). We will extract
these terms by analysing the large separation behaviour of the excited
state two-point functions.

\section{Excited state expectation values of bilocal operators}

In order to extract finite volume form factors, we analyse the \emph{excited
state} expectation values of bilocal operators in the limit when the
two operators are taken far apart. As a warmup, we first go through
the analogous procedure for \emph{vacuum state} expectation values.

\subsection{Vacuum expectation values of bilocal operators}

\begin{figure}
\begin{centering}
\includegraphics[width=2.5cm]{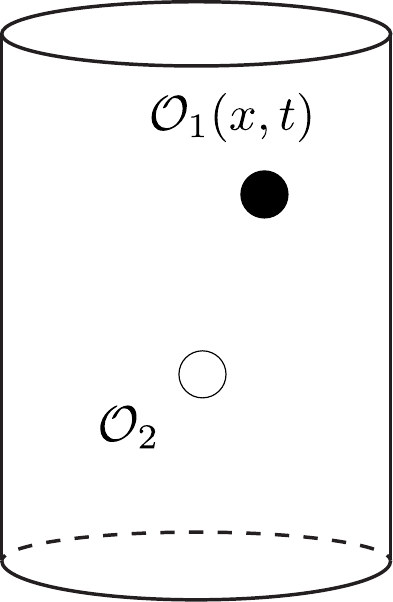}~~~~~~~~~~\includegraphics[width=4cm]{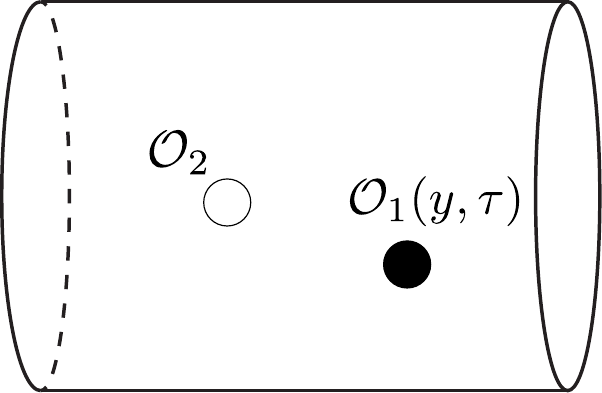}
\par\end{centering}
\caption{Graphical representation of the finite volume two-point function in
the physical (finite volume) and in the mirror (finite temperature)
channel.}

\label{O12vev}
\end{figure}

Let us analyse the following finite volume matrix element 
\begin{equation}
_{L}\langle0\vert\mathcal{O}_{1}(x,t)\mathcal{O}_{2}(0,0)\vert0\rangle_{L},
\end{equation}
 where we assume that $t>0$, i.e. the operators are time ordered
(see the left of Figure \ref{O12vev}). By inserting a complete system
of finite volume energy-momentum eigenstates we can write
\begin{align}
_{L}\langle0\vert\mathcal{O}_{1}(x,t)\mathcal{O}_{2}(0,0)\vert0\rangle_{L} & =\sum_{\vert\bar{\theta}_{1},\dots,\bar{\theta}_{N}\rangle_{L}}\,_{L}\langle0\vert\mathcal{O}_{1}\vert\bar{\theta}_{1},\dots,\bar{\theta}_{N}\rangle_{L}\,_{L}\langle\bar{\theta}_{1},\dots,\bar{\theta}_{N}\vert\mathcal{O}_{2}\vert0\rangle_{L}\,e^{-it(E_{N}-E_{0})+ixP_{N}},
\end{align}
where we used that $e^{iHt-iPx}\mathcal{O}(0,0)e^{iPx-iHt}=\mathcal{O}(x,t)$
and denoted ${\cal O}(0,0)$ by ${\cal O}$. We then put $x=0$ and
analytically continue to imaginary time $t=-iy$ with $y>0$. This
way we can suppress the contribution of excited states, so that in
the large separation limit ($y\to\infty$), only the ground state
survives:
\begin{equation}
_{L}\langle0\vert\mathcal{O}_{1}(0,-iy)\mathcal{O}_{2}\vert0\rangle_{L}\to\ _{L}\langle0\vert\mathcal{O}_{1}\vert0\rangle_{L}\ _{L}\langle0\vert\mathcal{O}_{2}\vert0\rangle_{L},\label{eq:LMbilocalclustering}
\end{equation}
and the results factorise. That is why this limit is often called
the clustering limit.

The finite volume expectation value of the bilocal operator can be
calculated in the thermal channel, when the operators are space-like
separated. In order to connect the finite volume formulation (\textquotedbl physical
channel\textquotedbl ) to the finite temperature one (\textquotedbl mirror
channel\textquotedbl ), we first need to continue the time $(x,t)\to(x,y=it)$
from Minkowskian to Euclidean signature. We then perform a rotation
$(x,y)\to(y,-x),$ and analytically continue back to Minkowskian signature
$(y,-x)\to(y,\tau=-ix)$. With this procedure we obtain (see Figure
\ref{O12vev}), 
\begin{equation}
_{L}\langle0\vert\mathcal{O}_{1}(x,t)\mathcal{O}_{2}(0,0)\vert0\rangle_{L}\to\langle\Omega\vert\mathcal{O}_{2}(0,0)\mathcal{O}_{1}(y,\tau)\vert\Omega\rangle,
\end{equation}
where we assumed that the operators do not have any spin(otherwise
they should also be rotated). The thermal state (corresponding to
the inverse temperature $L$) which minimizes the free energy is denoted
by $\Omega$. A LM-type of formula was derived for this expectation
value in \cite{Pozsgay:2018tns}: 
\begin{equation}
\langle\Omega\vert\mathcal{O}_{2}(0,0)\mathcal{O}_{1}(y,\tau)\vert\Omega\rangle=\sum_{n=0}^{\infty}\frac{1}{n!}\prod_{i=1}^{n}\int\frac{d\theta_{i}}{2\pi}\frac{1}{1+e^{\epsilon_{0}(\theta_{i})}}F_{c}^{12}(\theta_{1},\dots,\theta_{n}),\label{eq:LMbilocal}
\end{equation}
where $\epsilon_{0}$ is the pseudo energy of the ground state TBA,
and $F_{c}^{12}(\theta_{1},\dots,\theta_{n})$ is the connected diagonal
form factor of the bilocal operator $\mathcal{O}_{2}(0,0)\mathcal{O}_{1}(y,\tau)$,
which is given by the finite $\varepsilon$-independent part of the
almost diagonal matrix element
\begin{align}
F_{c}^{12}(\theta_{1},\dots,\theta_{n}) & =\mathrm{FP.}F^{12}(\theta_{1}+i\varepsilon_{1},\dots,\theta_{n}+i\varepsilon_{n}\vert\theta_{n},\dots,\theta_{1}).\label{eq:F12conndef}\\
 & =\mathrm{FP.}\langle\theta_{1}+i\epsilon_{1},\dots,\theta_{n}+i\epsilon_{n}\vert\mathcal{O}_{2}(0,0)\mathcal{O}_{1}(y,\tau)\vert\theta_{n},\dots,\theta_{1}\vert0\rangle.\nonumber 
\end{align}

The main result of the paper \cite{Pozsgay:2018tns} was to express
the form factors of bilocal operators in terms of the form factors
of their constituent operators:

\begin{align}
F^{12}(\{\vartheta\}_{I_{n}}\vert\{\theta\}_{I_{m}}) & =\sum_{N=0}^{\infty}\frac{1}{N!}\int_{\mathbb{R}-i\alpha}\prod_{i=1}^{N}\frac{d\mu_{i}}{2\pi}\sum_{A^{+}\cup A^{-}=I_{m}}\sum_{B^{+}\cup B^{-}=I_{n}}K_{y,\tau}(\{\mu\},\{\vartheta\}_{B^{-}}\vert,\{\theta\}_{A^{+}})\nonumber \\
 & F^{2}(\{\vartheta\}_{B^{+}}+i\pi,\{\theta\}_{A^{-}},\{\mu\}_{<})F^{1}(\{\vartheta\}_{B^{-}}+i\pi,\{\mu\}_{>}+i\pi,\{\theta\}_{A^{+}})\nonumber \\
 & S(\{\theta\}_{A^{-}},\{\theta\}_{A^{+}})S(\{\vartheta\}_{B^{-}},\{\vartheta\}_{B^{+}}),\label{eq:F12conn}
\end{align}
 where the sets $I_{m},A^{+},A^{-}$ are ordered increasingly (e.g.
$I_{m}=\{1,\dots,m\}$), while the sets $I_{n},B^{-},B^{+}$ are ordered
decreasingly. The corresponding ordering of the $\mu$-sets is indicated
by their subscripts ($><$ respectively). We also denote the rapidities
of the incoming particles by $\{\theta\}$, whereas the outgoing rapidities
are denoted by $\{\vartheta\}$. The kinematical factor is then given
by 
\begin{equation}
K_{y,\tau}(\{\alpha\}\vert\{\beta\})=e^{im\tau(\sum_{j}\cosh\alpha_{j}-\sum_{k}\cosh\beta_{k})}e^{-imy(\sum_{j}\sinh\alpha_{j}-\sum_{k}\sinh\beta_{k})},
\end{equation}
by assuming also that $y^{2}-\tau^{2}>0$, for $y>0$. For $y<0$
there is an analogous ordering with oppositely shifted $\mu$-integration.
A graphical representation is shown in Figure \ref{bilocalLM} below,
for $n=m=4$ and a specific choice of the sets $A^{\pm},B^{\pm}$.
To obtain the connected form factor we need to take $\vartheta_{j}=\theta_{j}+i\varepsilon_{j}$
and project onto the $\varepsilon$-independent term. The connected
form factor is symmetric in all its arguments and regular for coinciding
rapidities. Actually for coinciding rapidities, the form factor can
be expressed in terms of connected form factors with less particles
\cite{Pozsgay:2013jua}. This follows from the kinematical singularity
property of the form factors (\ref{eq:kinsing}). Note also that this
property extends to bilocal operators, as do other form factor properties
(e.g. \ref{eq:permutation}, \ref{eq:periodicity}).

In the large separation limit, we take $\tau=0$ and $y\to\infty$,
which implies that some exponents oscillate fast. The leading $y$-independent
behaviour comes from terms without $\mu$ integrals and for $B^{-}=A^{+}$.
In these terms, the $K$-factor is absent, the $S$-matrix factors
cancel out and the connected form factors can be calculated separately
for each of the two operators \cite{Pozsgay:2018tns}. Thus the whole
formula factorises into the product of two usual LM formulae (one
for each operator) and we recover the clustering behaviour (\ref{eq:LMbilocalclustering}).
In the following, we repeat the above analysis for the excited state
expectation value of bilocal operators.

\begin{figure}
\begin{centering}
\includegraphics[width=4cm]{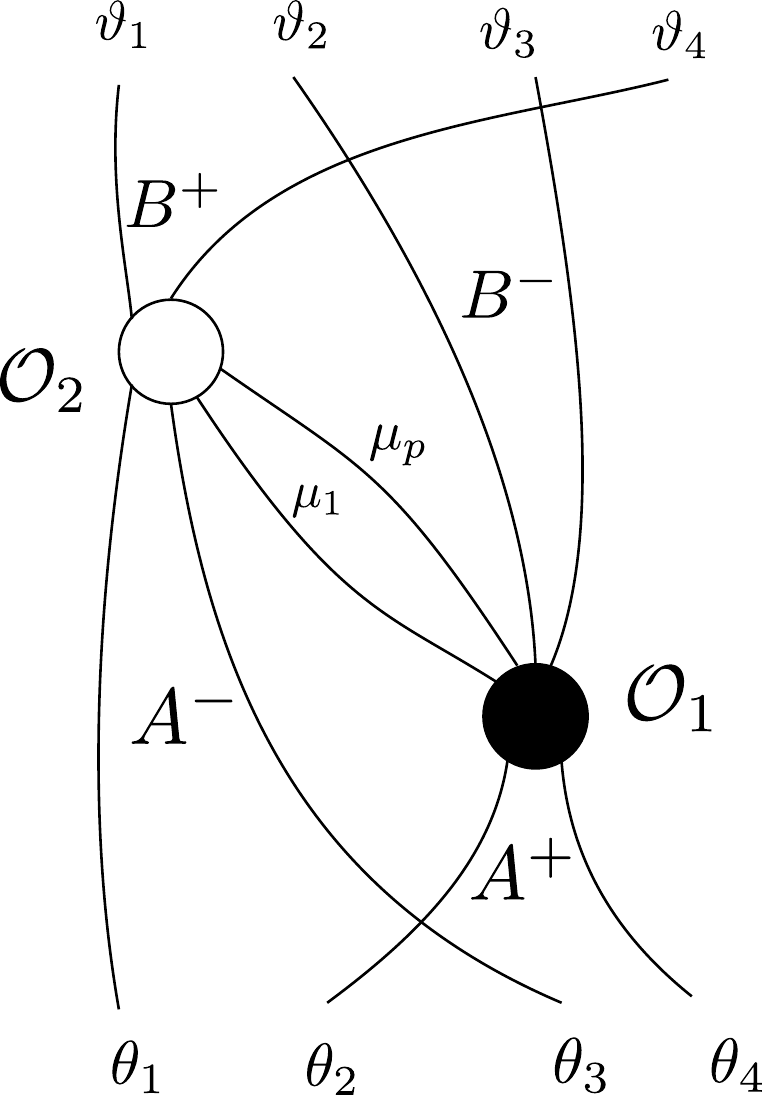}
\par\end{centering}
\caption{Graphical representation of the expansion of the bilocal form factor,
for $n=m=4$ and specific $A^{\pm}$ and $B^{\pm}$. Line crossings
indicate S-matrix factors. The upper and lower lines connecting the
two circles denote the outgoing/incoming particles of the form factors.
The black blob indicates that the space-time dependent factor $K$
is associated to the operator ${\cal O}_{1}$. To obtain the connected
form factor, we take $\vartheta_{j}=\theta_{j}+i\varepsilon_{j}$
and collect all the $\varepsilon$-independent terms.}

\label{bilocalLM}
\end{figure}

\subsection{Excited state expectation value of bilocal operators}

Let us now analyse the following excited state two point function
on the cylinder: 
\begin{equation}
_{L}\langle\bar{\theta}\vert\mathcal{O}_{1}(x,t)\mathcal{O}_{2}(0,0)\vert\bar{\theta}\rangle_{L},
\end{equation}
 where again $t>0$ and $\vert\bar{\theta}\rangle_{L}$ is a finite
volume one-particle state. We start with a one-particle state, but
later we explain how to generalise to multiparticle states. We can
again insert a complete finite volume basis to get
\begin{align}
_{L}\langle\bar{\theta}\vert\mathcal{O}_{1}(x,t)\mathcal{O}_{2}(0,0)\vert\bar{\theta}\rangle_{L} & =\sum_{\vert\bar{\theta}_{1},\dots,\bar{\theta}_{N}\rangle_{L}}\,_{L}\langle\bar{\theta}\vert\mathcal{O}_{1}\vert\bar{\theta}_{1},\dots,\bar{\theta}_{N}\rangle_{L}\,_{L}\langle\bar{\theta}_{1},\dots,\bar{\theta}_{N}\vert\mathcal{O}_{2}\vert\bar{\theta}\rangle_{L}e^{it(E_{1}-E_{N})-ix(P_{1}-P_{N})}.
\end{align}
By taking $x=0$ and continuing to imaginary time $t=-iy$ (where
$y>0$), we can suppress the excited states in the $y\to\infty$ limit,
so that the ground state's contribution will dominate and diverge
as 
\begin{equation}
_{L}\langle\bar{\theta}\vert\mathcal{O}_{1}(0,-iy)\mathcal{O}_{2}\vert\bar{\theta}\rangle_{L}\to\,_{L}\langle\bar{\theta}\vert\mathcal{O}_{1}\vert0\rangle_{L}\,_{L}\langle0\vert\mathcal{O}_{2}\vert\bar{\theta}\rangle_{L}e^{(E_{1}-E_{0})y}+O(1).\label{eq:clustering}
\end{equation}
Thus the leading exponentially growing behaviour factorises into three
terms: one depending only on the operator $\mathcal{O}_{1}$, another
that depends only on $\mathcal{O}_{2}$, and the third one which is
given by the space-time $y$-dependent exponential (where $y$ is
multiplied by the exact finite volume energy difference of the vacuum
and the one-particle state). We are after the two finite volume form
factors of $\mathcal{O}_{1}$ and $\mathcal{O}_{2}$.

\begin{figure}
\begin{centering}
\includegraphics[width=2cm]{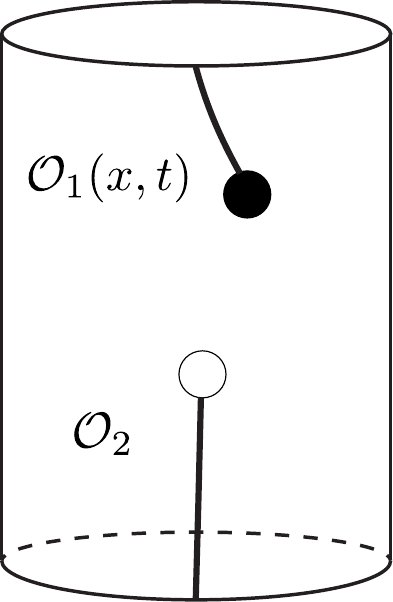}~~~~~~~~~~\includegraphics[width=3.2cm]{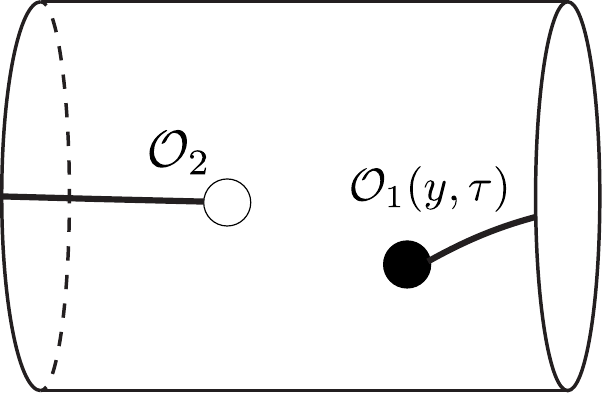}
\par\end{centering}
\caption{Graphical representation of the finite volume excited state two-point
function in the physical (finite volume) and in the mirror (finite
temperature) channel.}

\label{O12ex}
\end{figure}

Let us see how they can be calculated in the thermal channel (see
Figure \ref{O12ex}). After performing a double Wick rotation, we
arrive at the formula:
\begin{equation}
_{L}\langle\bar{\theta}\vert\mathcal{O}_{1}(x,t)\mathcal{O}_{2}(0,0)\vert\bar{\theta}\rangle_{L}\to\langle\Omega_{1}\vert\mathcal{O}_{2}(0,0)\mathcal{O}_{1}(y,\tau)\vert\Omega_{1}\rangle,
\end{equation}
where $\Omega_{1}$ refers to the excited state in the thermal formulation,
that is the state which minimizes the free energy in the presence
of a physical particle.

Analytical continuation can be used to extend the vacuum state LM
formula (\ref{eq:LMbilocal}) to the expectation values of local operators
in excited states \cite{Pozsgay:2013jua}. The derivation relies on
the special property of connected form factors that describes their
behavior for coinciding rapidities. As this property originates from
the kinematical singularity axiom (\ref{eq:kinsing}), it is also
shared by connected form factors of bilocal operators. The generalisation
of the LM formula turns out to be very similar. Let us spell out the
details for the case of an excited one-particle state. It consists
two pieces:
\begin{equation}
\langle\Omega_{1}\vert\mathcal{O}_{1}(x,t)\mathcal{O}_{2}(0,0)\vert\Omega_{1}\rangle=\frac{{\cal D}_{1}}{\rho_{1}(\bar{\theta})}+{\cal D}_{0}.\label{eq:LMbilocalex}
\end{equation}
The simpler piece ${\cal D}_{0}$ looks very much like the vacuum
formula \ref{eq:LMbilocal} 
\begin{equation}
\mathcal{D}_{0}=\sum_{n=0}^{\infty}\frac{1}{n!}\prod_{i=1}^{n}\int\frac{d\theta_{i}}{2\pi}\frac{1}{1+e^{\epsilon_{1}(\theta_{i})}}F_{c}^{12}(\theta_{1},\dots,\theta_{n}),
\end{equation}
but now $\epsilon_{1}$ is the pseudo-energy of the excited state
TBA (\ref{eq:exTBA}). The complicated part ${\cal D}_{1}$, involves
the particle's rapidity $\bar{\theta}$. It reads: 
\begin{equation}
\mathcal{D}_{1}=\sum_{n=0}^{\infty}\frac{1}{n!}\prod_{i=1}^{n}\int\frac{d\theta_{i}}{2\pi}\frac{1}{1+e^{\epsilon_{1}(\theta_{i})}}F_{c}^{12}(\theta_{1},\dots,\theta_{n},\bar{\theta}+\frac{i\pi}{2}).\label{eq:D1}
\end{equation}
The connected form factor of the bilocal operator $F_{c}^{12}$ is
given by the same infinite volume quantity that we defined for the
vacuum state expectation values in (\ref{eq:F12conndef}), when one
of its arguments is analytically continued to the physical channel
$\theta\to\bar{\theta}+\frac{i\pi}{2}$. Because connected form factors
are symmetric in all their arguments, it does not matter which argument
is analytically continued. Here, we found it slightly simpler to analytically
continue the last argument.

Let us now locate the exponentially growing term in the clustering
limit. In doing so we set $x=i\tau=0$ and analyse the limit $y\to\infty$.
Terms in ${\cal D}_{0}$ behave qualitatively as the vacuum state
expectation value and do not lead to exponential growth. The exponentially
growing term can only come from $K_{y,0}(\{\mu\},\{\vartheta\}_{B^{-}}\vert,\{\theta\}_{A^{+}})$
in ${\cal D}_{1}$:
\begin{equation}
K_{y,0}(\{\mu\},\{\vartheta\}_{B^{-}}\vert\{\theta\}_{A^{+}})=e^{-imy(\sum_{B^{-}}\sinh\vartheta_{j}+\sum_{k}\sinh\mu_{k}-\sum_{A^{+}}\sinh\theta_{j})}.
\end{equation}
In $F_{c}^{12}(\theta_{1},\dots,\theta_{n},\theta)$, the last argument
is analytically continued to $\theta\to\bar{\theta}+i\pi/2$. Since
$\sinh\theta\to i\cosh\bar{\theta}$, a diverging term of the form
$e^{ym\cosh\bar{\theta}}$ requires that $\vartheta=\theta+i\epsilon\in B^{-}$
and $\theta\notin A^{+}$.

In the following, we analyse systematically the clustering limit,
order by order in ${\cal D}_{1}$. This involves expanding (\ref{eq:D1})
in the number of thermal particles $(\theta_{i}$ integrals). The
expected behaviour is given by (\ref{eq:clustering}), 
\begin{equation}
{\cal D}_{1}\to\bar{F}^{1}(\bar{\theta})_{L}F^{2}(\bar{\theta})_{L}e^{y(E_{1}-E_{0})},
\end{equation}
where the sought for finite volume form factors appear as 
\begin{equation}
\,_{L}\langle\bar{\theta}\vert\mathcal{O}_{1}\vert0\rangle_{L}=\frac{\bar{F}^{1}(\bar{\theta})_{L}}{\sqrt{\rho_{1}(\bar{\theta})}},\qquad{}_{L}\langle0\vert\mathcal{O}_{2}\vert\bar{\theta}\rangle_{L}=\frac{F^{2}(\bar{\theta})_{L}}{\sqrt{\rho_{1}(\bar{\theta})}}.
\end{equation}
Observe that we are free to move an operator-independent (phase) factor
between the two expressions, as they cancel in the product. This is
related to the freedom we have in choosing the phase of the one-particle
state $\vert\bar{\theta}\rangle_{L}$.

The fact that the exact finite volume energy difference between the
one-particle and the vacuum state exponentiates is highly non-trivial.
We are going to calculate each quantity systematically by taking into
account higher and higher order exponential volume corrections. We
organize the results according to the order of the exponential volume
corrections:
\begin{align}
E_{1}-E_{0} & =m\cosh\bar{\theta}+\Delta_{1}E+\Delta_{2}E+\dots\label{eq:expansion}\\
F^{2}(\bar{\theta})_{L} & =F^{2}\left(1+\Delta_{1}F^{2}+\Delta_{2}F^{2}+\dots\right)\\
\bar{F}^{1}(\bar{\theta})_{L} & =F^{1}\left(1+\Delta_{1}\bar{F}^{1}+\Delta_{2}\bar{F}^{1}+\dots\right),
\end{align}
where we have also used the fact that for scalar operators the infinite
volume one-particle form factors $F^{1}$ and $F^{2}$ are constants.
By inspecting the large volume behaviour of the TBA pseudo-energies,
we see that their leading term is always $mL\cosh\theta$. As a result,
the expansion can be organised in the small parameter $e^{-mL\cosh\theta_{i}}$
with integrations for $\theta_{i}$. In this case $\Delta_{k}$ denotes
the product of $k$ such terms. Alternatively, we could choose either
$e^{-\epsilon_{0}}$ or $e^{-\epsilon_{1}}$ as small parameters and
expand the other one around it. It will turn out to be even more advantageous
to choose $n=(1+e^{\epsilon_{1}})^{-1}$ as the small parameter, in
which case $\Delta_{k}$ denotes the products of $k$ such terms.

\section{Evaluating the clustering limit order by order}

In this section we evaluate the clustering limit of the excited state
expectation value of the bilocal operator order by order in the exponentially
small finite volume corrections.

\subsection{Zero order: infinite volume result}

\begin{figure}
\begin{centering}
\includegraphics[width=2cm]{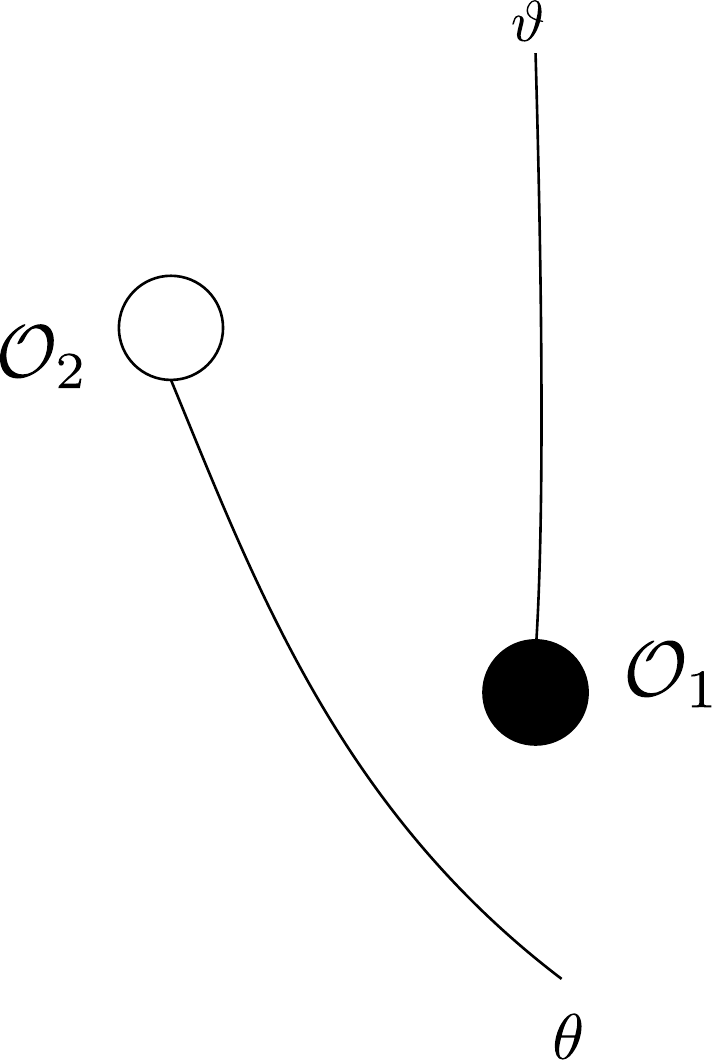}~~~~~~~~~~1.\includegraphics[width=2cm]{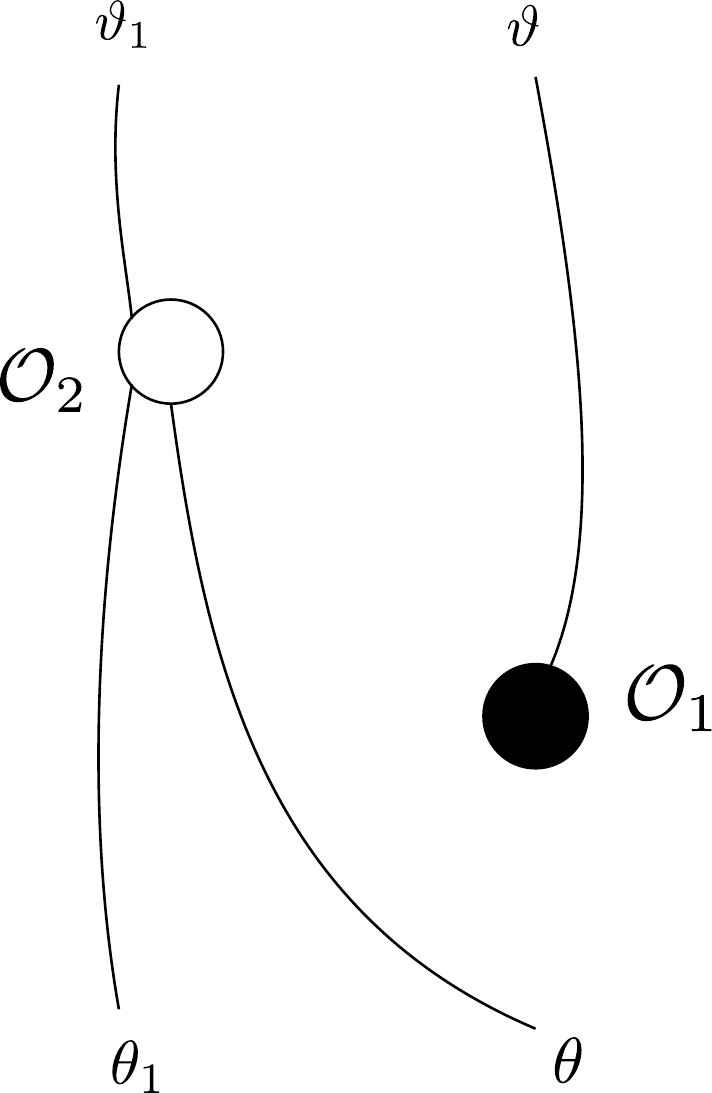}~~~~~~2.\includegraphics[width=2cm]{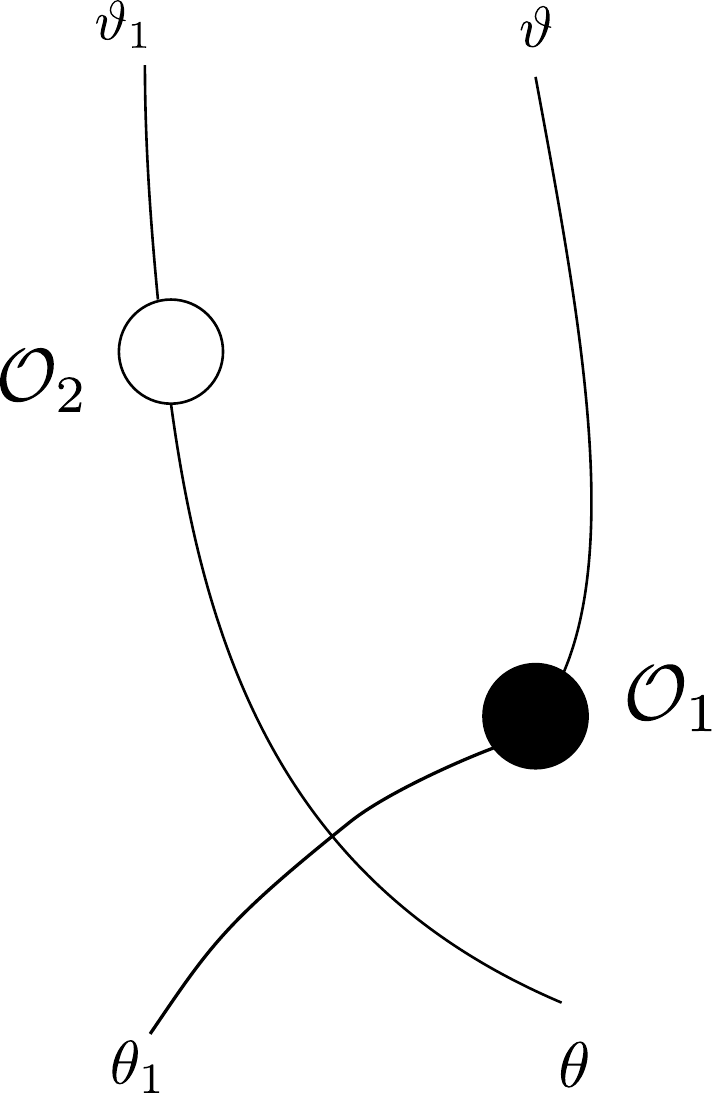}~~~~~~3.\includegraphics[width=2cm]{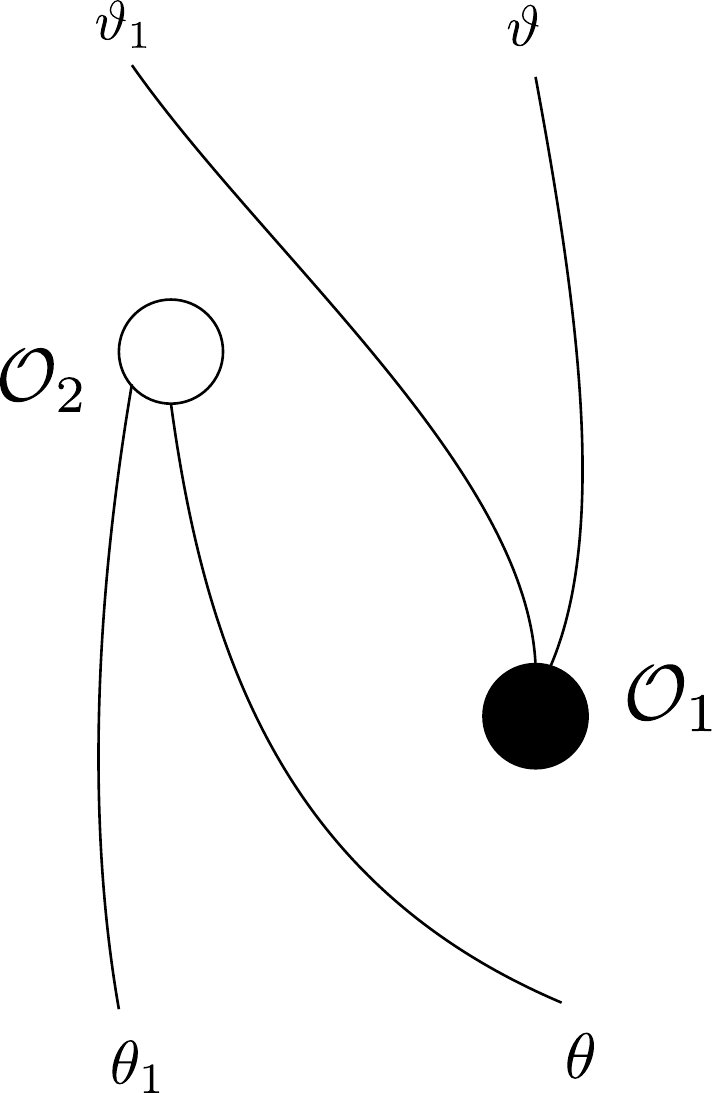}~~~~~~4.\includegraphics[width=2cm]{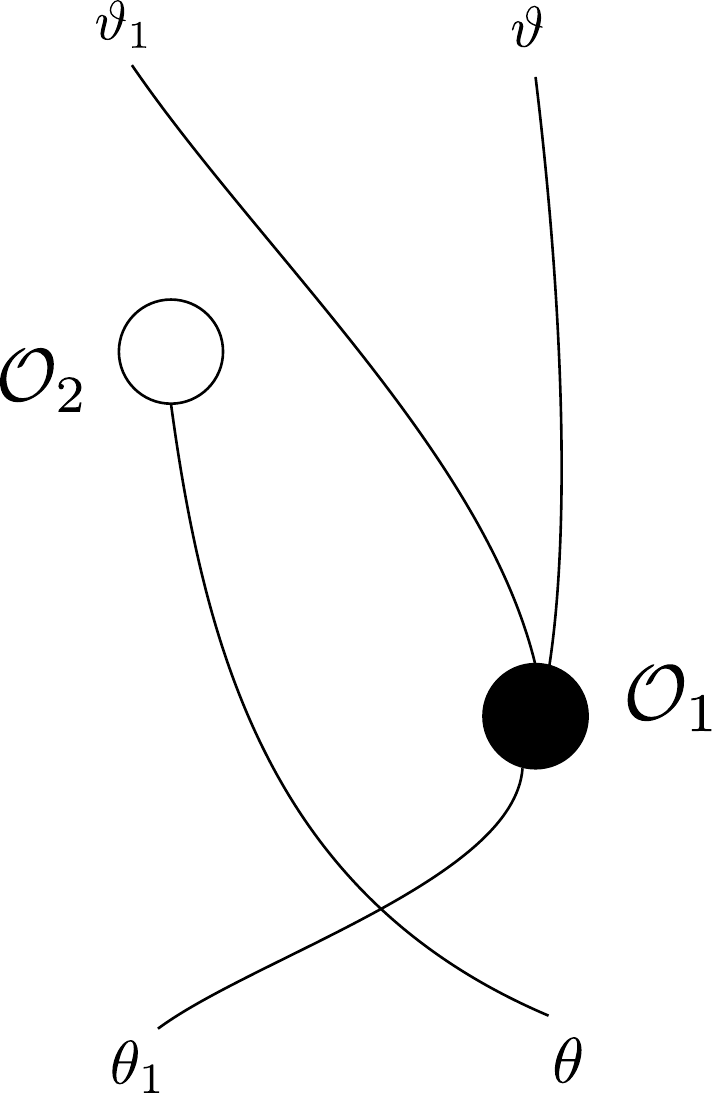}
\par\end{centering}
\caption{Zero and first order diagrams in the connected bilocal form factors.}

\label{1storder}
\end{figure}

By recalling the leading order behaviour of the finite volume form
factors as well as the energy differences in (\ref{eq:expansion}),
the clustering limit should take the form 
\begin{equation}
{\cal D}_{1}\to F^{1}F^{2}e^{ym\cosh\bar{\theta}}=F^{1}F^{2}e^{y\bar{e}}\,.
\end{equation}
To recover this result we have to take the $n=0$ term in ${\cal D}_{1}$.
This amounts to evaluate the connected form factor with only one rapidity
$\vartheta=\theta+i\varepsilon$ and $\theta$. Exponential growth
requires $B^{-}=\{\vartheta\},B^{+}=\emptyset$ and $A^{-}=\{\theta\},A^{+}=\emptyset$
. This term is indicated on the left diagram of Figure \ref{1storder}.
As the expression is regular, we can take $\varepsilon=0$. The leading
term after the $\theta\to\bar{\theta}+i\frac{\pi}{2}$ analytical
continuation takes the expected form

\begin{align}
F_{c}^{12}(\bar{\theta}+i\pi/2) & =F^{1}F^{2}e^{y\bar{e}}\,.
\end{align}
Let us note that for each $\mu$ integral there is a corresponding
exponential factor $e^{-iym\sinh\mu}$ which oscillates fast and together
with the shifts becomes suppressed in the $y\to\infty$ limit. Observe
also that in doing the analytical continuation we do not hit any (kinematical)
singularity of the form factor, thus all the $\mu$ integrals can
be neglected. This extends to any higher order terms, too.

\subsection{First order: Lüscher correction}

At the next order in the exponentially suppressed volume corrections
the terms which survive in the clustering limit in (\ref{eq:expansion})
take the form
\begin{align}
{\cal D}_{1} & \to F^{1}F^{2}e^{y\bar{e}}(1+\Delta_{1}\bar{F}^{1})(1+\Delta_{1}F^{2})e^{y\Delta_{1}E}\label{eq:LO}\\
 & \to F^{1}F^{2}e^{y\bar{e}}(1+\Delta_{1}\bar{F}^{1}+\Delta_{1}F^{2}+y\Delta_{1}E)+\dots\,.\nonumber 
\end{align}
At this order we have to take the $n=1$ term in (\ref{eq:D1}), i.e.
we have a single integral with a thermally suppressed factor 
\begin{equation}
\int\frac{d\theta_{1}}{2\pi}\frac{F_{c}^{12}(\theta_{1},\theta)}{1+e^{\epsilon_{1}(\theta_{1})}}\,.
\end{equation}
 The exponentially growing term requires that $\vartheta\in B^{-}$
and $\theta\in A^{-}$. As these extra particles always connect to
different operators, the corresponding form factor is never singular
in the $\varepsilon\to0$ limit, which limit can be taken from the
start. We thus have to analyse the $F^{12}(\theta,\vartheta_{1}\vert\theta_{1},\theta)$
form factor with $\vartheta_{1}=\theta_{1}+i\varepsilon_{1}$. By
definition the connected form factor is the O(1) term in $\varepsilon_{1}$.
Actually, being the connected form factor of a bilocal operator, it
should be regular for $\varepsilon_{1}\to0$. This is true for the
full expression, but it is not true term by term as we will see.

The rapidities $\vartheta_{1},\theta_{1}$ can be connected to each
operator in two different ways: $B^{+}=\{\vartheta_{1}\}$ or $B^{+}=\{\emptyset\}$
and independently $A^{+}=\{\theta_{1}\}$ or $A^{+}=\{\emptyset\}$.
We can combine them in all possible ways, see Figure \ref{1storder},
which we analyse one by one:
\begin{enumerate}
\item $B^{+}=\{\vartheta_{1}\}$, $A^{+}=\{\emptyset\}$ and the contribution
is 
\begin{equation}
F^{2}(\vartheta_{1}+i\pi,\theta_{1},\theta)F^{1}e^{-iym\sinh\theta}\,.
\end{equation}
The form factor of the first operator has a singular piece in $\varepsilon_{1}$
originating from the kinematical singularity axiom of the form 
\begin{equation}
F^{2}(\vartheta_{1}+i\pi,\theta_{1},\theta)=\frac{s_{1}}{\varepsilon_{1}}F^{2}+F_{c}^{2}(\theta_{1}+i\pi,\theta_{1},\theta)+O(\epsilon)\,,\label{eq:kinsin1}
\end{equation}
where $s_{1}=1-S_{1}$, with $S_{1}=S(\theta_{1}-\theta)=S(\theta_{1}-\bar{\theta}-i\frac{\pi}{2})$
and we defined the O(1), finite term to be the connected part of this
partially diagonal form factor. This is not the same, how the connected
form factor was defined in \cite{Bajnok:2018fzx,Bajnok:2019cdf} and
differs in an O(1) term. It is related to the freedom, how we normalize
the individual states and the freedom, that we can freely move terms
between the form factors of incoming and outgoing states. Of course
when we put together the two form factors in the two-point function
the result has to be invariant. We will come back to this freedom
when we formulate an ansatz for the all order finite volume form factor.
\item $B^{+}=\{\vartheta_{1}\}$ and $A^{+}=\{\theta_{1}\}$ with contribution
\begin{equation}
F^{2}(\vartheta_{1}+i\pi,\theta)F^{1}(\theta+i\pi,\theta_{1})S_{1}e^{-iym(\sinh\theta-\sinh\theta_{1})}\,.
\end{equation}
This term is regular for $\varepsilon_{1}\to0$. Clearly the remaining
contribution is not factorising due to the integration for $\theta_{1}$,
which connects the two operators. The exponent $e^{iym\sinh\theta_{1}}$
however, upon integration, will suppress the contribution in the $y\to\infty$
limit. To make this more precise we could shift the $\theta_{1}$
integration as $\theta_{1}\to\theta_{1}+i\delta_{1}$ with $\delta_{1}>0$
, but infinitesimally small. Then the exponent will vanish in the
$y\to\infty$ limit and this term will not contribute to the clustering
limit.
\item $B^{+}=\{\emptyset\}$ and $A^{+}=\{\emptyset\}$ contributing as
\begin{equation}
F^{2}(\theta_{1},\theta)F^{1}(\theta+i\pi,\vartheta_{1}+i\pi)e^{-iym(\sinh\theta+\sinh\vartheta_{1})}\,.
\end{equation}
Using similar argumentations to the previous case, we can see that
this term will not contribute either. In particular, the $\theta_{1}\to\theta_{1}+i\delta_{1}$
shift with $\delta_{1}>0$ can be analytically continued to $\delta_{1}<0$
without hitting any singularity of the integrand, which guaranties
a decaying exponent in the $y\to\infty$ limit.
\item Finally, $B^{+}=\{\emptyset\}$ and $A^{+}=\{\theta_{1}\}$ gives
\begin{equation}
F^{2}F^{1}(\theta+i\pi,\vartheta_{1}+i\pi,\theta_{1})S_{1}e^{-iym(\sinh\theta+\sinh\vartheta_{1}-\sinh\theta_{1})}\,.\label{eq:diffexp}
\end{equation}
The singular piece takes the form 
\begin{equation}
F^{1}(\vartheta_{1}+i\pi,\theta_{1},\theta-i\pi)=-\frac{s_{1}S_{1}^{-1}}{\varepsilon_{1}}F^{1}+F_{c}^{1}(\theta_{1}+i\pi,\theta_{1},\theta-i\pi)+O(\epsilon)\,.
\end{equation}
The singular piece has two effects. First, it cancels the similar
singular term coming from 1, such that the total expression is finite
in the $\varepsilon_{1}\to0$ limit. Second, in the limit we also
have to take into account the $\varepsilon_{1}$-dependence in the
exponent in (\ref{eq:diffexp}) coming from $\vartheta_{1}=\theta_{1}+i\epsilon_{1}$,
thus it gives an extra term by differentiating the exponent:
\begin{equation}
-s_{1}F^{2}F^{1}ym\cosh\theta_{1}e^{-iym\sinh\theta}\,.
\end{equation}
\end{enumerate}
The total contribution after the analytical continuation is then 
\begin{align}
e^{y\bar{e}}\left(F_{c}^{2}(\theta_{1}+i\pi,\theta_{1},\bar{\theta}+i\frac{\pi}{2})F^{1}+F^{2}F_{c}^{1}(\theta_{1}+i\pi,\theta_{1},\bar{\theta}-i\frac{\pi}{2})S_{1}-s_{1}F^{1}F^{2}ym\cosh\theta_{1}\right)\,.
\end{align}
These terms should agree with the terms in (\ref{eq:LO}) one by one.
Let us see how they match.

When we organise the expansion in powers of the symbol $L_{i}=e^{-mL\cosh\theta_{i}}$
and keep the leading order (denoted by $\Delta_{1})$ we have to expand
the integration measure as
\begin{equation}
\frac{1}{1+e^{\epsilon_{1}(\theta_{1})}}=e^{-\epsilon_{1}(\theta_{1})}+\dots=S_{1}^{-1}L_{1}+\dots\quad;\qquad L_{1}=e^{-mL\cosh\theta_{1}}\,.
\end{equation}
This gives the following $y$-dependent piece 
\begin{equation}
\Delta_{1}E=-y\int\frac{d\theta_{1}}{2\pi}(S_{1}^{-1}-1)e_{1}L_{1}\quad;\quad e_{i}=m\cosh\theta_{i}\,.
\end{equation}
This is indeed the leading exponentially small term in the energy
difference, see Appendix \ref{app:Ediff} for the expansion of the
energy difference.

The analogous correction for the form factors are 
\begin{equation}
F^{2}\Delta_{1}F^{2}=\int\frac{d\theta_{1}}{2\pi}F_{c}^{2}(\theta_{1}+i\pi,\theta_{1},\bar{\theta}+i\frac{\pi}{2})S_{1}^{-1}L_{1}\,,
\end{equation}
and 
\begin{equation}
F^{1}\Delta_{1}\bar{F}^{1}=\int\frac{d\theta_{1}}{2\pi}F_{c}^{1}(\theta_{1}+i\pi,\theta_{1},\bar{\theta}-i\frac{\pi}{2})L_{1}\,.
\end{equation}
Let us note that these expressions agree with \cite{Bajnok:2018fzx}
up to an operator independent phase factor, which are related to different
normalizations of the one-particle states. This is also related how
we defined the connected form factors. The alternative definitions
in \cite{Bajnok:2018fzx} add a term to $F^{2}\Delta_{1}F^{2}$ and
subtract the same term from $F^{1}\Delta_{1}\bar{F}^{1}$, such that
the sum is the same.

In summarising, we have seen that the singular terms in $\varepsilon$
completely cancelled each other. This is indeed expected from the
connected form factor and it must happen also at higher orders. We
have also seen that the $\mu$ integrals are decaying in the clustering
limit, so we can completely neglect them. By shifting the $\theta_{1}$
integral we could also get rid off other terms with unbalanced exponential
factors. This will be also true at higher orders.

\subsection{Second Lüscher correction}

At the second Lüscher order when we take the clustering limit the
correction terms (\ref{eq:expansion}) have the form 
\begin{equation}
\frac{1}{2}y^{2}(\Delta_{1}E)^{2}+y\Delta_{2}E+y\Delta_{1}E(\Delta_{1}F^{1}+\Delta_{1}F^{2})+\Delta_{1}F^{1}\Delta_{1}F^{2}+\Delta_{2}F^{1}+\Delta_{2}F^{2}\,.
\end{equation}
If we were interested only in $\Delta_{2}F$ then we could just locate
the contributing diagrams and evaluate them. For consistency, however
we decided to evaluate all terms as we also would like to confirm
that our method is consistent. Indeed, we will see that this approach
pays off, since there are terms whose contributions are easy to miss,
but they are relevant for the correctness of the results.

We start by pointing out that the already calculated first order terms
also contribute at the second and higher Lüscher orders. Indeed, by
expanding the measure 
\begin{align}
\int\frac{d\theta_{1}}{2\pi}\frac{1}{1+e^{\epsilon_{1}(\theta_{1})}} & =\int\frac{d\theta_{1}}{2\pi}(e^{-\epsilon_{1}(\theta_{1})}-e^{-2\epsilon_{1}(\theta_{1})}+\dots)\\
 & =\int\frac{d\theta_{1}}{2\pi}\left\{ S_{1}^{-1}L_{1}\left(1+\int\frac{d\theta_{2}}{2\pi}\varphi_{12}S_{2}^{-1}L_{2}+\dots\right)-S_{1}^{-2}L_{1}^{2}+\dots\right\} \,,\nonumber 
\end{align}
 we get corrections which contribute to both form factors and energy
differences. Here we just displayed the $\Delta_{1}$ and $\Delta_{2}$
terms, but they appear at any $\Delta_{k}$. It is thus technically
simpler to perform the expansion directly in the excited state filling
fraction
\begin{equation}
n_{i}=\frac{1}{1+e^{\epsilon_{1}(\theta_{i})}}\,,
\end{equation}
and express the energy difference at every order in terms of polynomials
of $n_{i}$ and S-matrix factors, since this term will not contribute
at any higher $n_{i}$ orders. This is completely analogous to the
usual LM formula, which uses the filling fraction and connected form
factors to organize the result. At the leading $\Delta_{1}$ order
expansion in $n_{i}$ or $L_{i}$ give the same result. Since whenever
the symbol $n_{i}$ appears we also have an integration $\int\frac{d\theta_{i}}{2\pi}$,
we do not write this integration out explicitly. With this convention
the first order results read as 
\begin{equation}
\Delta_{1}E=-ye_{1}s_{1}n_{1}\quad;\quad\Delta_{1}F^{2}=F^{2}(1)n_{1}\quad;\quad\Delta_{1}\bar{F}^{1}=\bar{F}^{1}(1)n_{1}\,,
\end{equation}
where we also streamlined the notation by introducing 
\begin{equation}
F^{2}(1)=F_{c}^{2}(\theta_{1}+i\pi,\theta_{1},\bar{\theta}+i\frac{\pi}{2})/F^{2}\quad;\quad\bar{F}^{1}(1)=F_{c}^{1}(\theta_{1}+i\pi,\theta_{1},\bar{\theta}-i\frac{\pi}{2})S_{1}/F^{1}\,.
\end{equation}
Clearly these terms will not contribute to higher orders in the expansion
in the $n_{i}$-s. The k-th order term in the $n_{i}$ expansion contains
exactly $k$ number of $n_{i}$ factors. In the following $\Delta_{k}$
in (\ref{eq:expansion}) collects the contribution of those terms.
We are now ready to calculate the second order.

\begin{figure}[H]
\begin{centering}
1.\includegraphics[width=1.6cm]{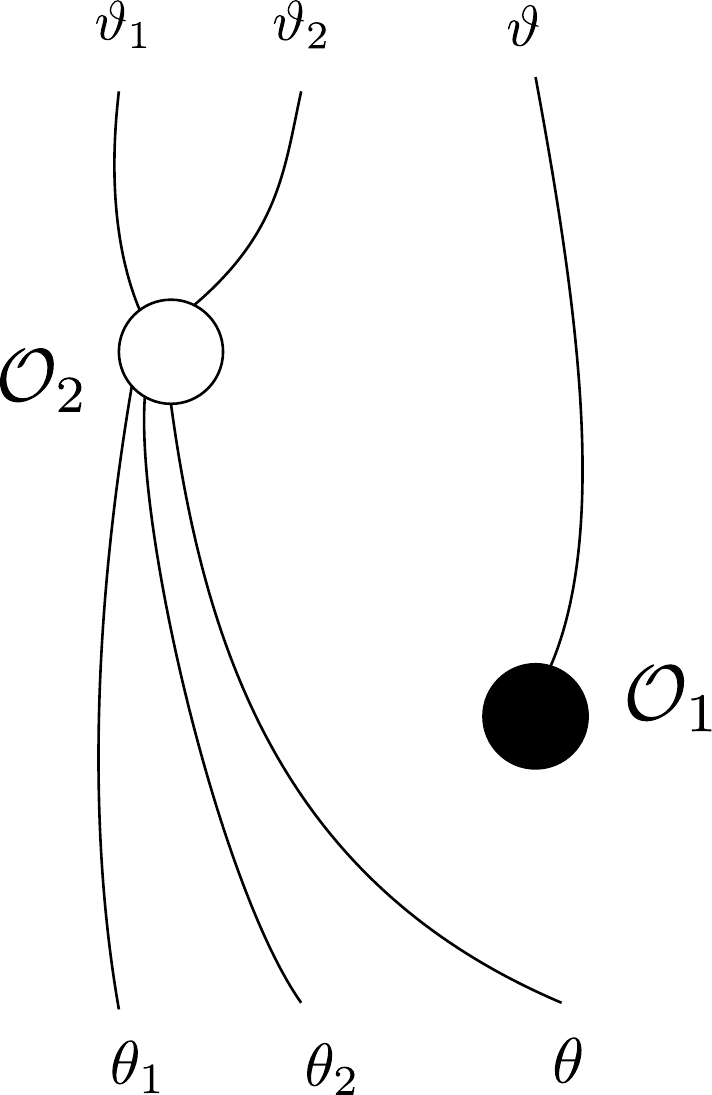}~~~~~~2.\includegraphics[width=1.6cm]{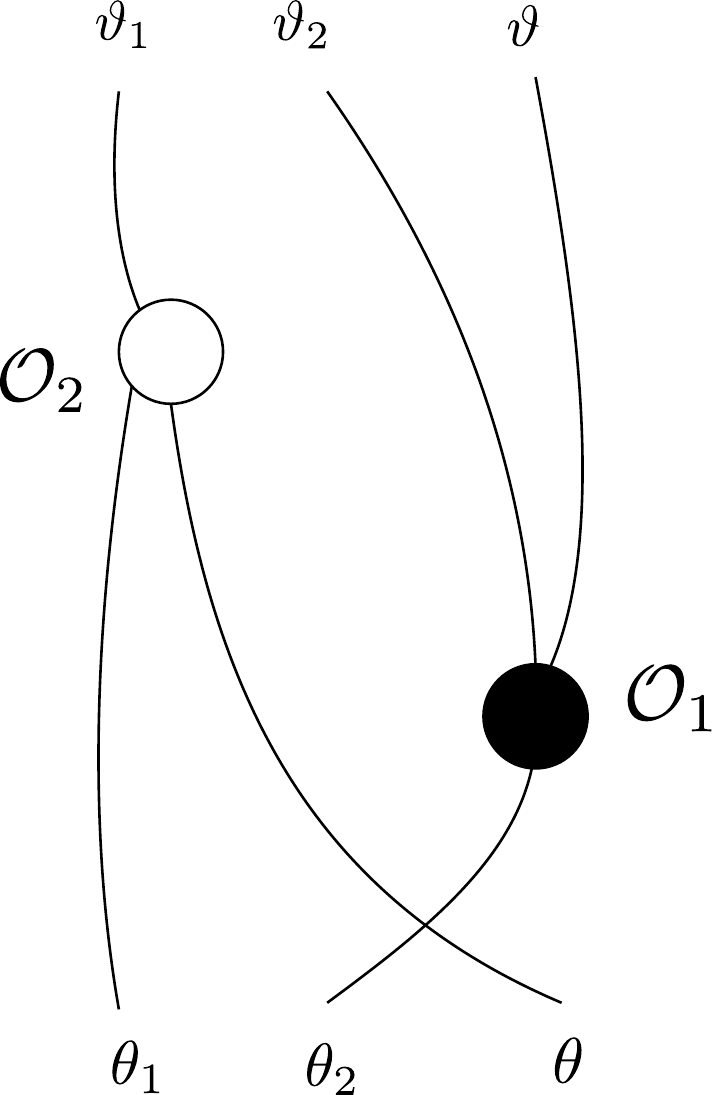}~~~~~~3.\includegraphics[width=1.6cm]{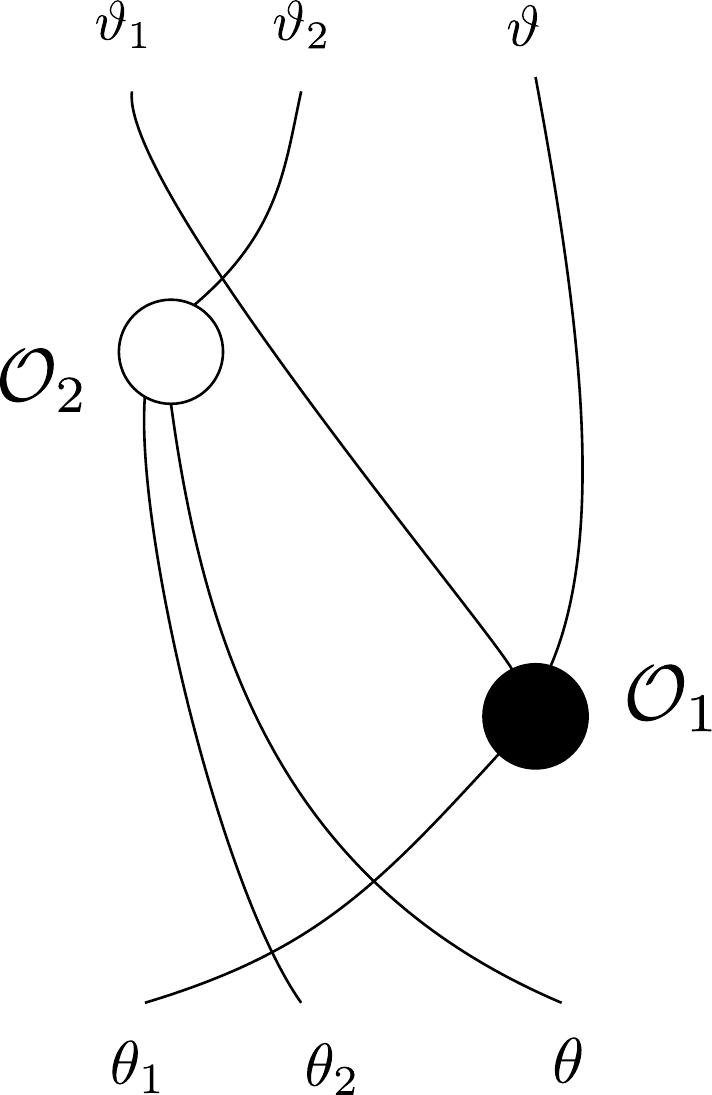}~~~~~~4.\includegraphics[width=1.6cm]{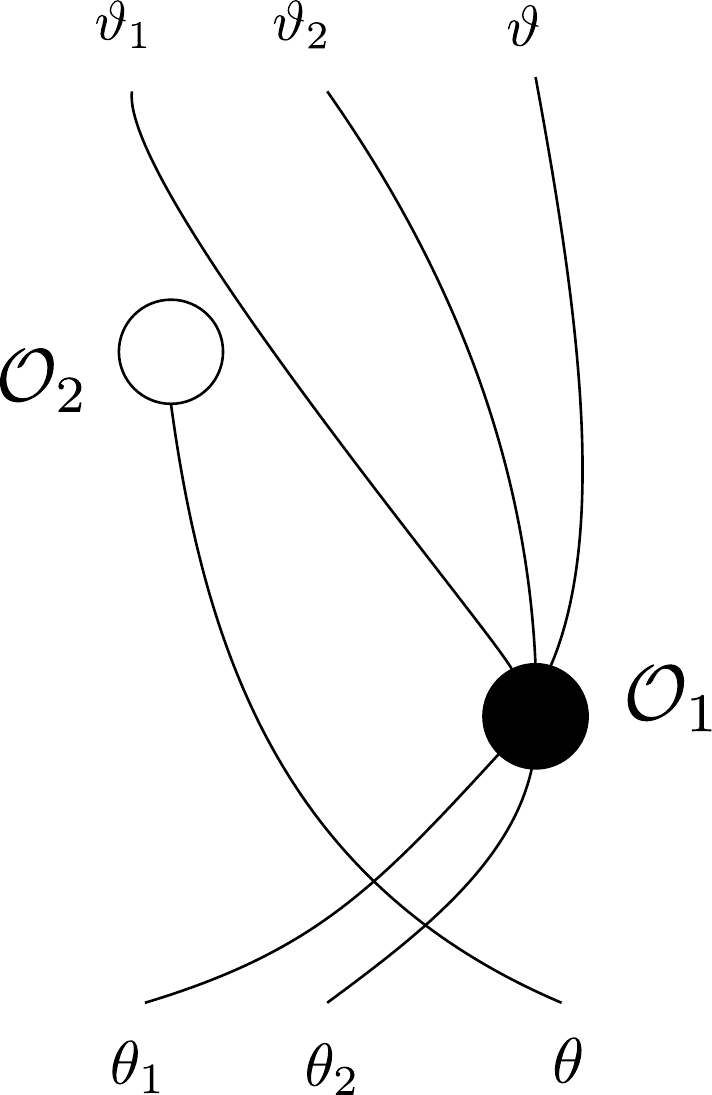}~~~~~~5.\includegraphics[width=1.6cm]{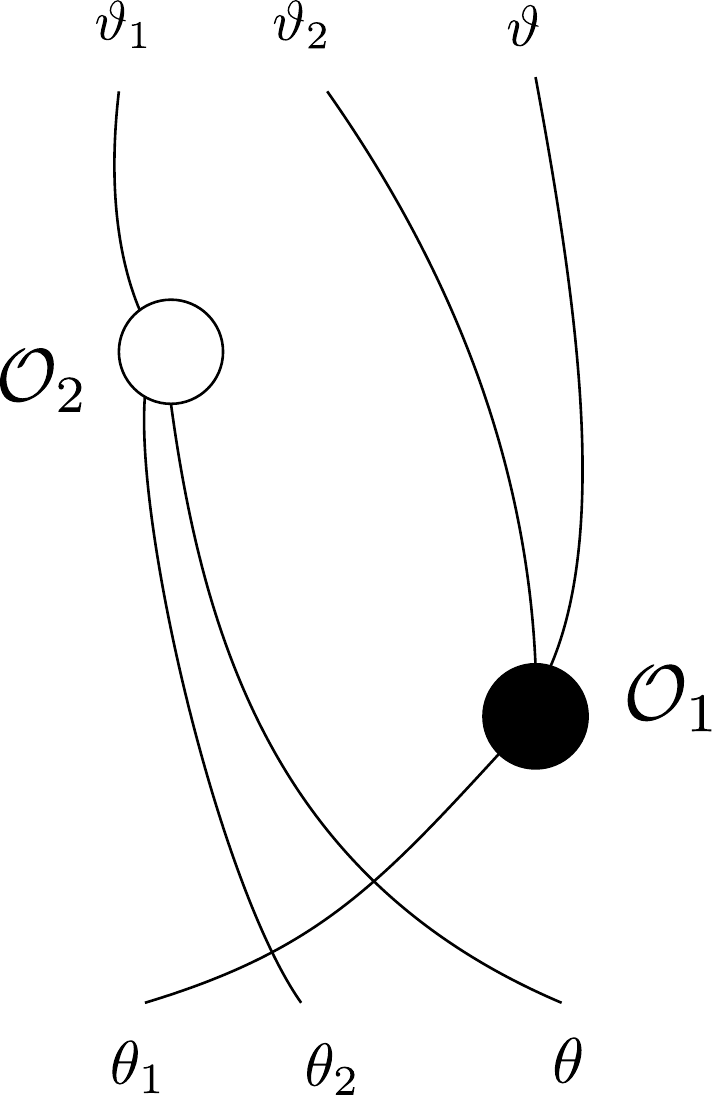}~~~~~~6.\includegraphics[width=1.6cm]{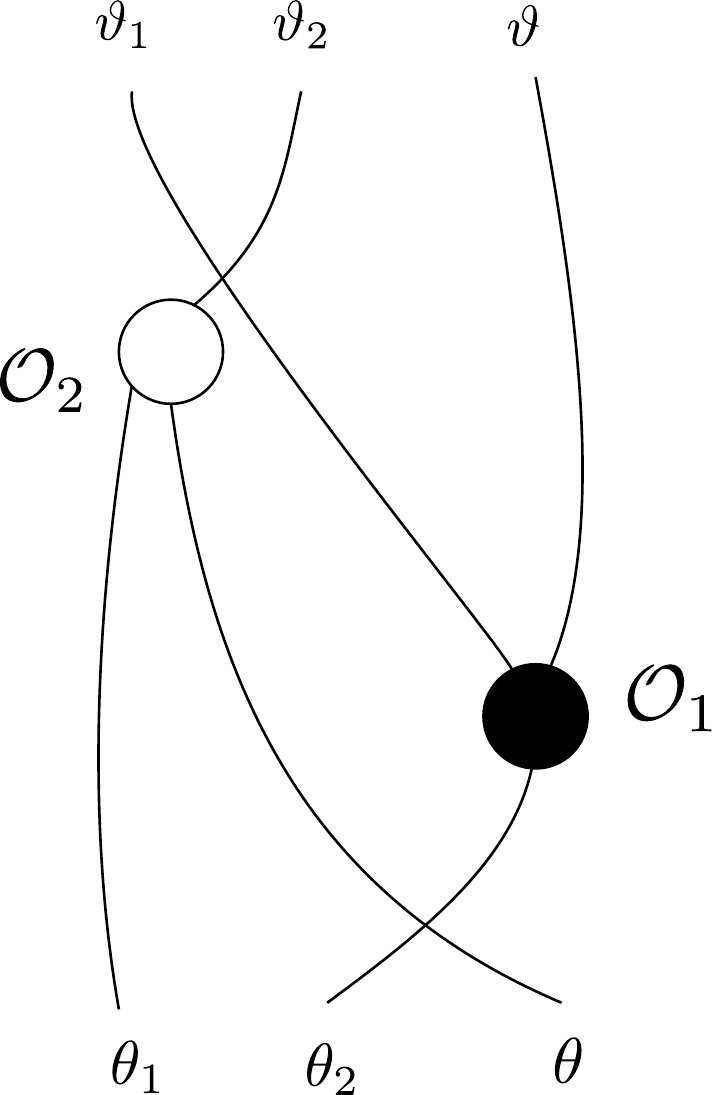}
\par\end{centering}
\caption{Contributing diagrams at second order in the clustering limit.}

\label{NLOcontrib}
\end{figure}

At the second order we take the $n=2$ term in ${\cal D}_{1},$ which
has two integrations (not written out explicitly)
\begin{equation}
\frac{1}{2}n_{1}n_{2}F_{c}^{12}(\theta_{1},\theta_{2},\theta)\,.
\end{equation}
We thus need to evaluate $F_{c}^{12}(\theta_{1},\theta_{2},\theta)$
and continue in $\theta$ to $\theta\to\bar{\theta}+i\pi/2$. There
are $16$ diagrams which contribute to the exponential growth $e^{my\cosh\bar{\theta}}$
(after the analytical continuation). Let us premise that those diagrams
in which $\vert B_{-}\vert\neq\vert A_{+}\vert$ will not survive
in the clustering limit. These diagrams have different number of incoming
$\theta_{i}$ and outgoing $\vartheta_{j}$ rapidities. Consequently,
in the exponent some unbalanced, oscillating $\sinh\theta_{i}$ or
$\sinh\theta_{j}$ terms remain (after putting all $\varepsilon$-s
to zero) which suppress the contribution. This is similar what happened
at the previous order. Thus the contributing diagrams are those, which
are displayed in Figure \ref{NLOcontrib}, which we analyse one by
one. In order to focus on the corrections we factor out the leading
order result $F^{1}F^{2}e^{iym\sinh\theta}$ from each term. We start
with the first four diagrams, which individually are singular in the
$\varepsilon$-s, but regular when summed up.

\begin{enumerate}
\item Let us see the contribution of the first diagram: 
\begin{equation}
F^{2}(\vartheta_{2}+i\pi,\vartheta_{1}+i\pi,\theta_{1},\theta_{2},\theta)/F^{2}\,.
\end{equation}
Since $\vartheta_{j}=\theta_{j}+i\varepsilon_{j}$ we have singular
terms in the $\varepsilon s$, originating from the kinematical singularities
of the form factor. Applying successively the kinematical singularity
property we arrive at 
\begin{equation}
F^{2}(\theta_{2}+i\pi+i\varepsilon_{2},\theta_{1}+i\pi+i\varepsilon_{1},\theta_{1},\theta_{2},\theta)/F^{2}=\frac{A_{12}}{\varepsilon_{1}\varepsilon_{2}}+\frac{A_{1}}{\varepsilon_{1}}+\frac{A_{2}}{\varepsilon_{2}}+F^{2}(1,2)+\dots\,,
\end{equation}
where the ellipses denote terms with ratios or higher order terms
in $\varepsilon$s, which do not contribute to the connected evaluation.
We abbreviated the connected form factor after removing the zeroth
order term as 
\begin{equation}
F_{c}^{2}(\theta_{2}+i\pi,\theta_{1}+i\pi,\theta_{1},\theta_{2},\theta)=F^{2}F^{2}(1,2)\,.
\end{equation}
The $A$-coefficients turn out to be (see section \ref{sec:Graph_rules})
\begin{equation}
A_{12}=s_{1}s_{2}\quad;\quad A_{1}=s_{1}F^{2}(2)+S_{1}s_{2}\varphi_{12}\quad;\quad A_{2}=s_{2}F^{2}(1)+s_{1}\varphi_{21}\,,
\end{equation}
where, as before, $s_{i}=1-S_{i}$. Clearly, the whole expression
is not symmetric in $\theta_{1}$ and $\theta_{2}$. Since similar
objects appear at every order in the calculation we develop a diagrammatic
technique in Section \ref{sec:Graph_rules} to evaluate these expressions
and define the basis of connected form factors, which appears at the
various orders in the finite volume expansion. Singular terms in $\varepsilon$-s
must be cancelled by other terms in the expansion. Let us see how
this happens.
\item The contribution of the second diagram is 
\begin{align}
F^{2}(\vartheta_{1}+i\pi,\theta_{1},\theta)/F^{2}F^{1}(\vartheta_{2}+i\pi,\theta_{2},\vartheta-i\pi)/F^{1}S_{2}e^{-imy(\sinh\vartheta_{2}-\sinh\theta_{2})} & =\nonumber \\
\text{\ensuremath{\left(\frac{s_{1}}{\varepsilon_{1}}+F^{2}(1)\right)}}\left(-\frac{s_{2}}{\varepsilon_{2}}+\bar{F}^{1}(2)\right)\left(1+y\varepsilon_{2}e_{2}\right)+\dots & \,.
\end{align}
\item The similar contribution of the diagram in which we crossed the particles
is
\begin{align}
F^{2}(\vartheta_{2}+i\pi,\theta_{2},\theta)/F^{2}F^{1}(\vartheta_{1}+i\pi,\theta_{1},\vartheta-i\pi)/F^{1}S(\vartheta_{2}-\vartheta_{1})S_{12}S_{1}e^{-imy(\sinh\vartheta_{1}-\sinh\theta_{1})} & =\nonumber \\
\left(\frac{s_{2}}{\varepsilon_{2}}+F^{2}(2)\right)\left(-\frac{s_{1}}{\varepsilon_{1}}+\bar{F}^{1}(1)\right)\left(1+(\varepsilon_{1}-\varepsilon_{2})\varphi_{12}-\varepsilon_{1}\varepsilon_{2}\left(i\varphi'_{12}+\varphi_{12}^{2}\right)\right)\;\left(1+y\varepsilon_{1}e_{1}\right)+\dots & \,.
\end{align}
where $S_{12}=S(\theta_{1}-\theta_{2})$. Due to this crossing we
also had to expand the S-matrix factor $S(\vartheta_{2}-\vartheta_{1}$)
in the $\varepsilon$-s which introduced further asymmetry in $1$
and $2$.
\item The contribution of the fourth diagram is 
\begin{align}
F^{1}(\vartheta_{2}+i\pi,\vartheta_{1}+i\pi,\theta_{1},\theta_{2},\vartheta-i\pi)/F^{2}S_{1}S_{2}e^{-im(\sinh\vartheta_{1}+\sinh\vartheta_{2}-\sinh\theta_{1}-\sinh\theta_{2})} & =\nonumber \\
\left(\frac{\bar{A}_{12}}{\varepsilon_{1}\varepsilon_{2}}+\frac{\bar{A}_{1}}{\varepsilon_{1}}+\frac{\bar{A}_{2}}{\varepsilon_{2}}+\bar{F}^{1}(1,2)\right)\left(1+y\varepsilon_{1}e_{1}+y\varepsilon_{2}e_{2}+y^{2}\varepsilon_{1}e_{1}\varepsilon_{2}e_{2}\right)+\dots & \,,
\end{align}
where $\bar{A}$ can be obtained from $A$ by replacing $S_{i}$ with
$S_{i}^{-1}$ and multiplying with $S_{1}S_{2}$: 
\begin{equation}
\bar{A}_{12}=s_{1}s_{2}\quad;\quad\bar{A}_{1}=-s_{1}\bar{F}^{1}(2)-s_{2}\varphi_{12}\quad;\quad\bar{A}_{2}=-s_{2}\bar{F}^{1}(1)-S_{2}s_{1}\varphi_{21}\,.
\end{equation}
\item The contribution of the fifth diagram is 
\begin{align}
F^{2}(\vartheta_{1}+i\pi,\vartheta_{2},\theta)/F^{2}F^{1}(\vartheta_{2}+i\pi,\theta_{1},\vartheta-i\pi)/F^{1}S_{12}S_{1}e^{-iym(\sinh\vartheta_{2}-\sinh\theta_{1})} & =\nonumber \\
F^{2}(\theta_{1}+i\pi,\theta_{2},\theta)/F^{2}F^{1}(\theta_{2}+i\pi,\theta_{1},\theta-i\pi)/F^{1}S_{12}S_{1}e^{-iy(p_{2}-p_{1})}+\dots & \,,
\end{align}
where we could safely put the $\varepsilon$-s to zero and denoted
the momentum by $p_{i}=m\sinh\theta_{i}$.
\item The similar contribution of the last diagram is 
\begin{align}
F^{2}(\vartheta_{2}+i\pi,\theta_{1},\theta)/F^{2}F^{1}(\vartheta_{1}+i\pi,\theta_{2},\vartheta-i\pi)/F^{1}S(\vartheta_{2}-\vartheta_{1})S_{2}e^{-iym(\sinh\vartheta_{1}-\sinh\theta_{2})} & =\nonumber \\
F^{2}(\theta_{2}+i\pi,\theta_{1},\theta)/F^{2}F^{1}(\theta_{1}+i\pi,\theta_{2},\theta-i\pi)/F^{1}S_{21}S_{2}e^{-iy(p_{1}-p_{2})}+\dots & \,,
\end{align}
where we could again safely put the $\varepsilon$-s to zero.
\end{enumerate}
The connected form factor $F_{c}^{12}(\theta_{1},\theta_{2},\theta)$
is the finite, $\varepsilon$-independent part of the sum of the six
diagrams above. One can easily check that all singular terms in the
$\varepsilon$-s cancel. Being a connected form factor the result
is a symmetric function in the rapidities, and regular whenever they
coincide. In particular, the result is symmetric in $1$ and $2$
and regular for $\theta_{1}=\theta_{2}$. This is true for the sum
of the diagrams, but not for the individual diagrams.

We are interested in the clustering limit of the result. The contribution
of the first four terms does not have any exponentially oscillatory
part and survive in the $y\to\infty$ limit. The last two terms are
more tricky. Naively we would drop these terms due to the oscillations
in the exponent, however this is not correct as the integrands develop
singularities for $\theta_{1}=\theta_{2}$, whose residues do not
oscillate. In oder to calculate their contributions carefully we shift
the integration contours as $\theta_{1}\to\theta_{1}+i\delta_{1}$
and $\theta_{2}\to\theta_{2}+i\delta_{2}$ with $\delta_{1}>\delta_{2}>0$.
Since the connected form factor is regular in $\theta_{1}$ and $\theta_{2}$
in the vicinity of the real line, this is a safe operation which does
not change the result. With this regularization each individual diagram
gives a finite contribution and we are ready to investigate the $y\to\infty$
limit. In order to see how the exponents behave we note that $e^{-iyp_{j}}\sim e^{ye_{j}\sin\delta_{j}}$.
This implies that $e^{-iy(p_{2}-p_{1})}$ grows as $e^{ye_{2}\sin\delta_{2}}$
and we need to analytically continue in $\delta_{2}$ to negative,
i.e. we need to shift the $\theta_{2}$ integration below the real
line. As $\delta_{1}>\delta_{2}$ there is no singularity in $\theta_{2}$
and after the shift we can safely drop the contribution of the fifth
diagram. In the sixth diagram, however, we have a growing factor $e^{-iyp_{1}}\sim e^{ye_{1}\sin\delta_{1}}$
and we have to shift the $\theta_{1}$ integration below the real
line. In doing so we have to pick up the residue of the integrand
at $\theta_{2}$, which follows from the kinematical singularity property
\begin{equation}
-2\pi i\text{Res}_{\theta_{1}=\theta_{2}}\frac{n_{1}}{2\pi}\frac{n_{2}}{2\pi}\biggl(\frac{is_{1}}{\theta_{2}-\theta_{1}}+F^{2}(1)+\dots\biggr)\biggl(-\frac{is_{2}}{\theta_{1}-\theta_{2}}+\bar{F}^{1}(2)+\dots\biggr)S_{21}e^{-iy(p_{1}-p_{2})}\,.
\end{equation}
After evaluating the residues and integrating by parts in the term
$in'_{2}n_{2}s_{2}^{2}$ we arrive at 
\begin{equation}
n_{2}^{2}s_{2}\left\{ -s_{2}\left(ye_{2}+\varphi(0)\right)+\bar{F}^{1}(2)+F^{2}(2)\right\} \,.
\end{equation}
These residues are the terms where higher powers of the filling fractions
appear, hence they are instrumental to reproduce the exact energies.

By putting all the contributions together we successfully reproduce
the lower order terms and can extract the sought correction for the
finite volume form factors 
\begin{align}
\Delta_{2}F^{1}+\Delta_{2}F^{2} & =\frac{1}{2}n_{2}^{2}s_{2}\left\{ -s_{2}\varphi_{22}+\bar{F}^{1}(2)+F^{2}(2)\right\} +\label{eq:res2nd}\\
 & \qquad\frac{1}{2}n_{1}n_{2}\Bigg\lbrace\bar{F}^{1}(1,2)+F^{2}(1,2)-\left[s_{2}\bar{F}^{1}(1)+s_{1}F^{2}(2)\right]\varphi_{12}+s_{1}s_{2}\varphi_{12}^{2}\Bigg\rbrace\,,\nonumber 
\end{align}
where we have dropped the antisymmetric term $\frac{1}{2}n_{1}n_{2}s_{1}s_{2}i\varphi'_{12}$,
which vanishes under integration.

Let us make a remark here: in integrating the connected form factors
\begin{equation}
\frac{1}{2}\int\frac{d\theta_{1}}{2\pi}\int\frac{d\theta_{2}}{2\pi}n_{1}n_{2}\left(\bar{F}^{1}(1,2)+F^{2}(1,2)\right),
\end{equation}
we have to be careful, as these objects are singular for $\theta_{1}=\theta_{2}$
(see Appendix \ref{sec:Pole-structure-of}) and contain a second order
pole:\footnote{As mentioned before, $F_{c}^{12}(\theta_{1},\theta_{2},\theta)$ is
regular for $\theta_{1}=\theta_{2}$, since the singularities cancel
between diagrams 1, 4, 5 and 6, which are individually regulated by
the contour shifts.}
\begin{equation}
F(1,2)=-\frac{s_{1}s_{2}}{(\theta_{1}-\theta_{2})^{2}}+\mathcal{O}(1).\label{eq:connsing}
\end{equation}
Thus keeping the prescription $\theta_{1}\to\theta_{1}+i\delta_{1}$
and $\theta_{2}\to\theta_{2}+i\delta_{2}$ for the connected form
factor integral is still necessary. The double pole, however, will
not contribute. This is because it gets multiplied with a measure
factor $\frac{1}{2}n_{1}n_{2}$ leading to the integrand of the form
$-\frac{1}{2}f(\theta_{1})f(\theta_{2})/(\theta_{1}-\theta_{2})^{2}$
with $f(\theta_{1})=n(\theta_{1})s(\theta_{1})$. For an arbitrary
function $f(\theta)$ the residue of such a term is a total derivative:
\begin{equation}
\text{Res}_{\theta_{1}=\theta_{2}}\frac{f(\theta_{1})f(\theta_{2})}{(\theta_{1}-\theta_{2})^{2}}=\frac{1}{2}\frac{d}{d\theta_{2}}f^{2}(\theta_{2}),
\end{equation}
which vanishes under $\theta_{2}$-integration when $f(\theta)$ decays
at the infinities, as in the case of the filling fraction.

\subsection{Third order correction}

At the second order there were 5 graphs contributing in the clustering
limit out of all the 16 graphs. At the third order we have 64 graphs
out of which only 14 will have a non-zero contribution in the clustering
limit (see Figure \ref{NNLOcontrib}). As the calculation is quite
cumbersome we merely summarize the result here. In Appendix \ref{sec:Contour-deformation-for}
we demonstrate the most technically involved calculation on diagram
11, when we had to deform two contours and after picking up the residues
we arrived at a single-integral term with measure factor $n_{1}^{3}$.

\begin{figure}[H]
\begin{centering}
1.\includegraphics[width=1.4cm]{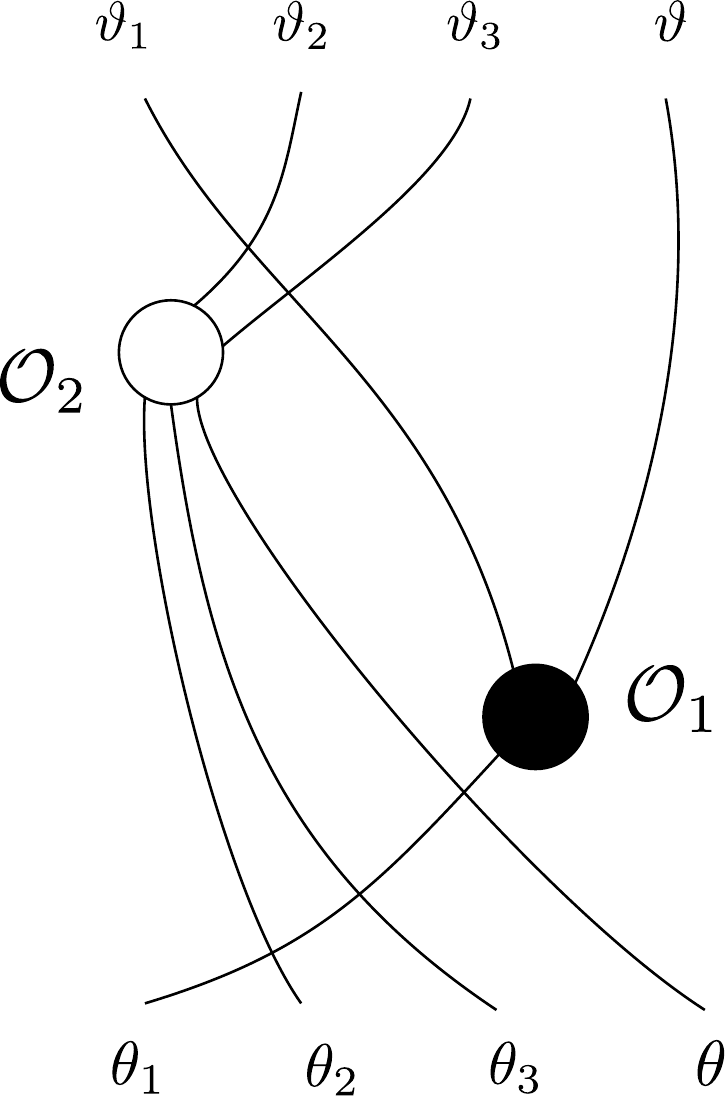}~~~~~~2.\includegraphics[width=1.4cm]{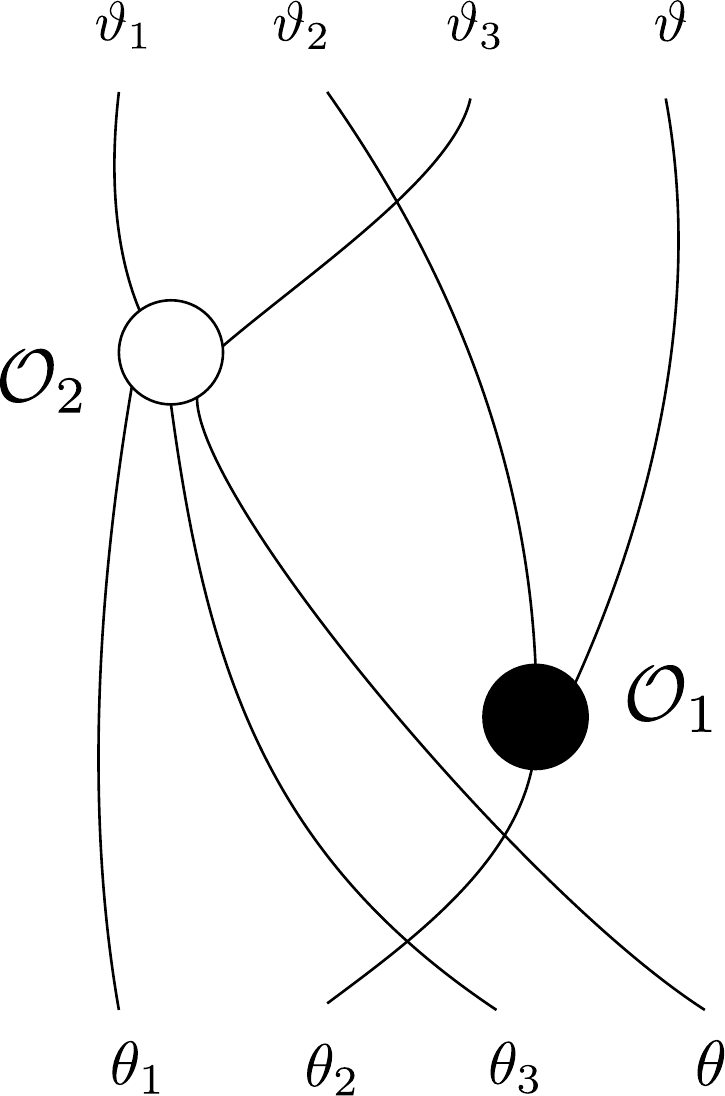}~~~~~~3.\includegraphics[width=1.4cm]{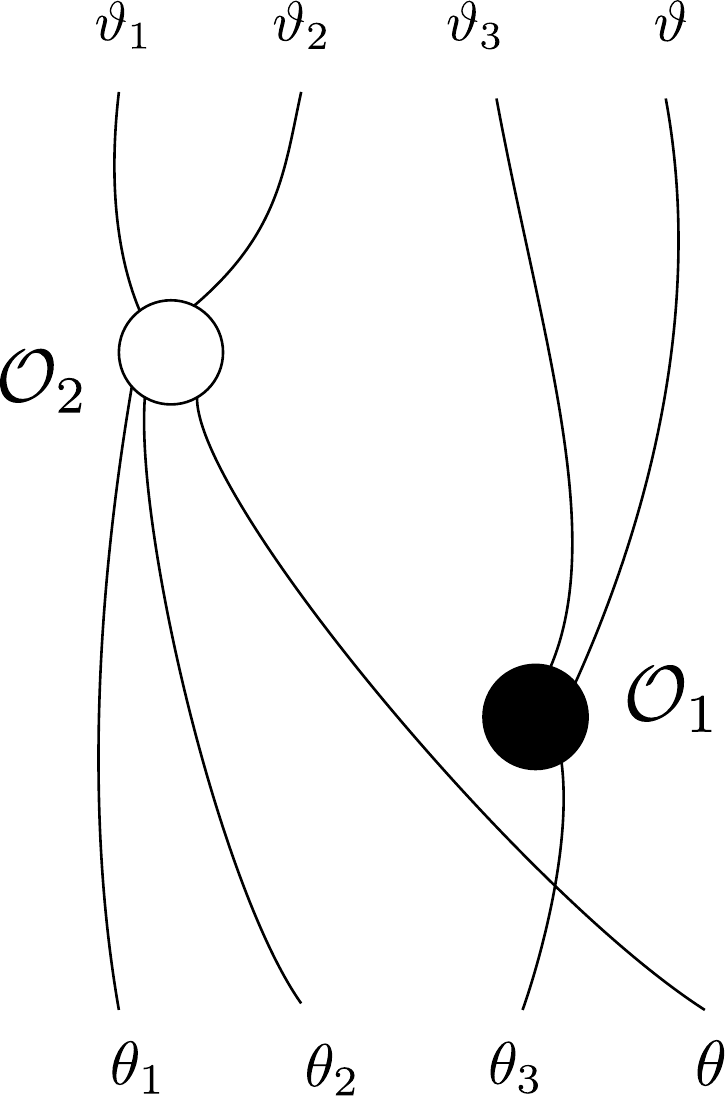}~~~~~~4.\includegraphics[width=1.4cm]{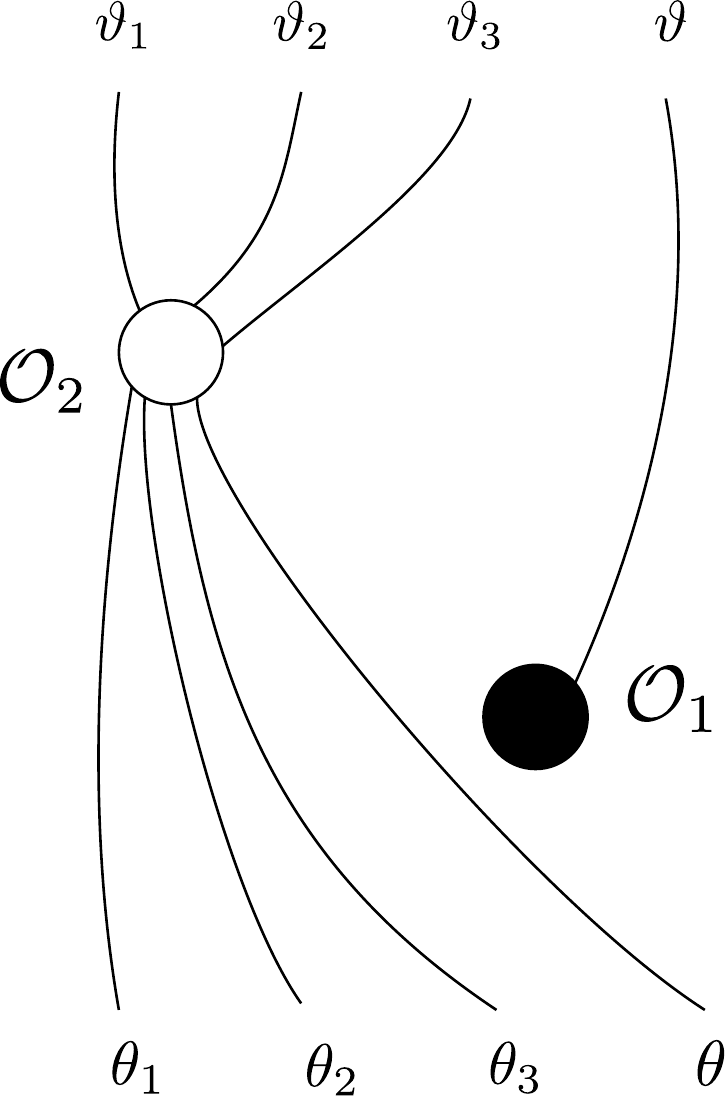}~~~~~~5.\includegraphics[width=1.4cm]{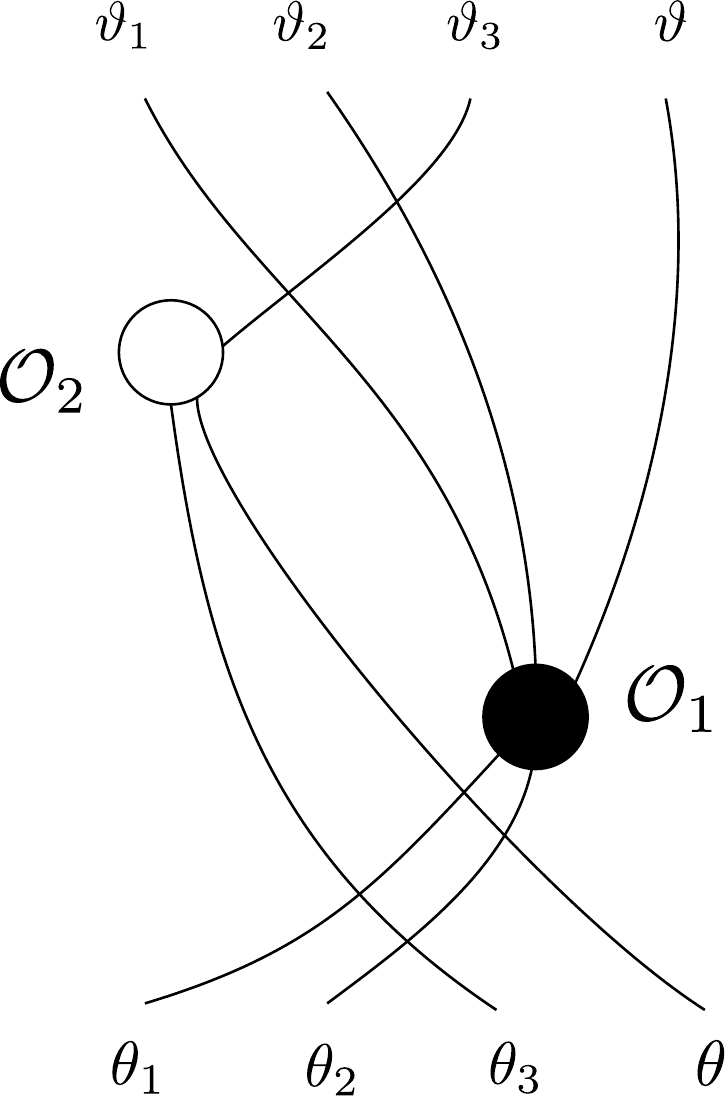}~~~~~~6.\includegraphics[width=1.4cm]{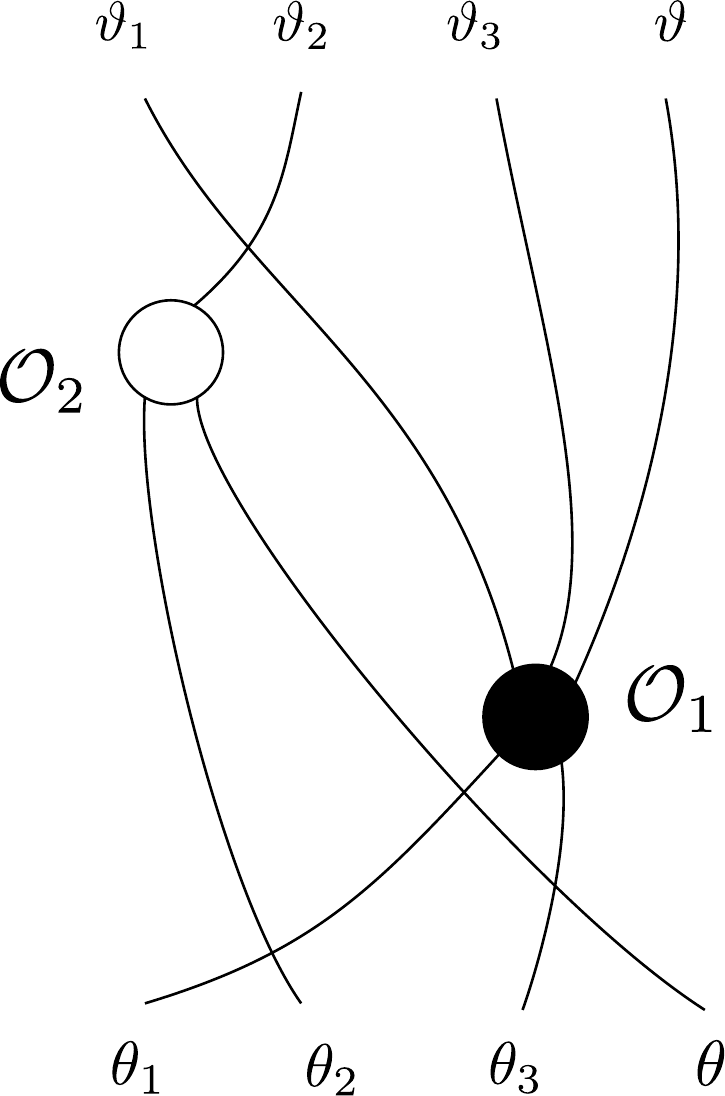}~~~~~~7.\includegraphics[width=1.4cm]{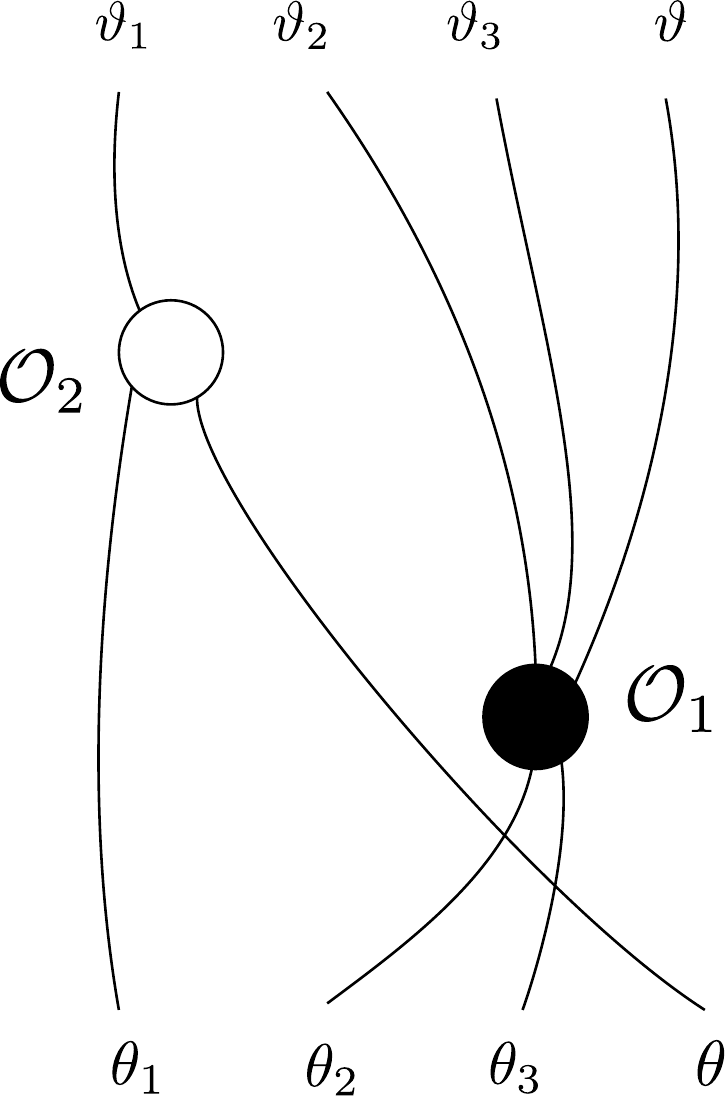}
\par\end{centering}
\medskip{}

\begin{centering}
8.\includegraphics[width=1.4cm]{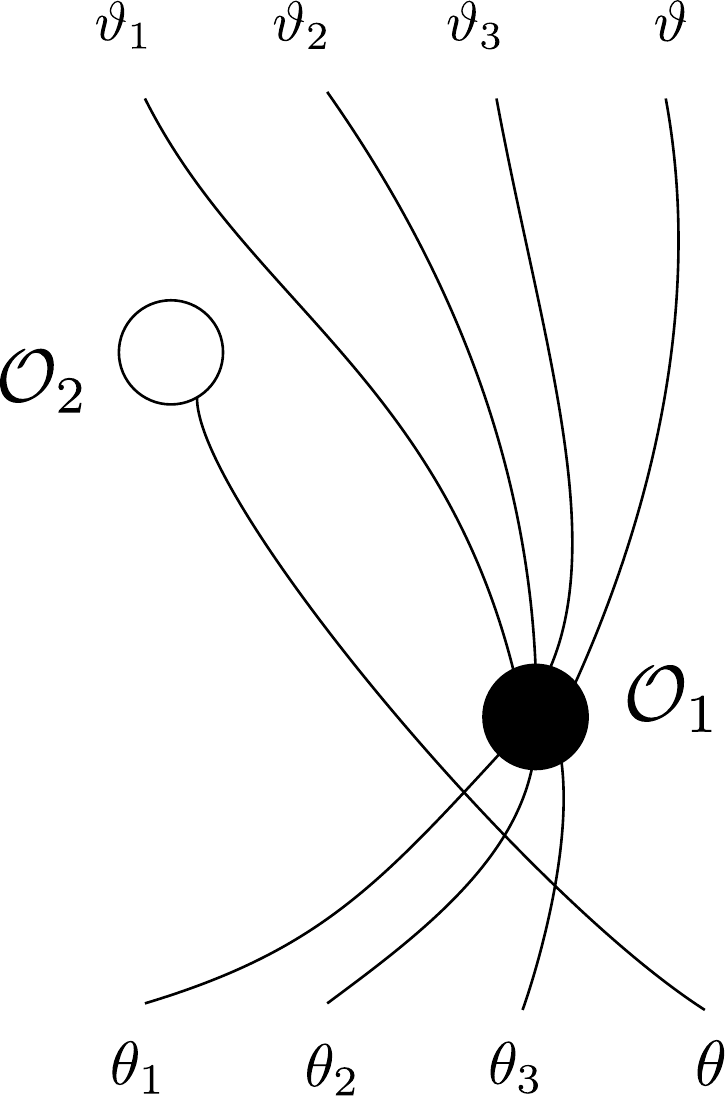}~~~~~~9.\includegraphics[width=1.4cm]{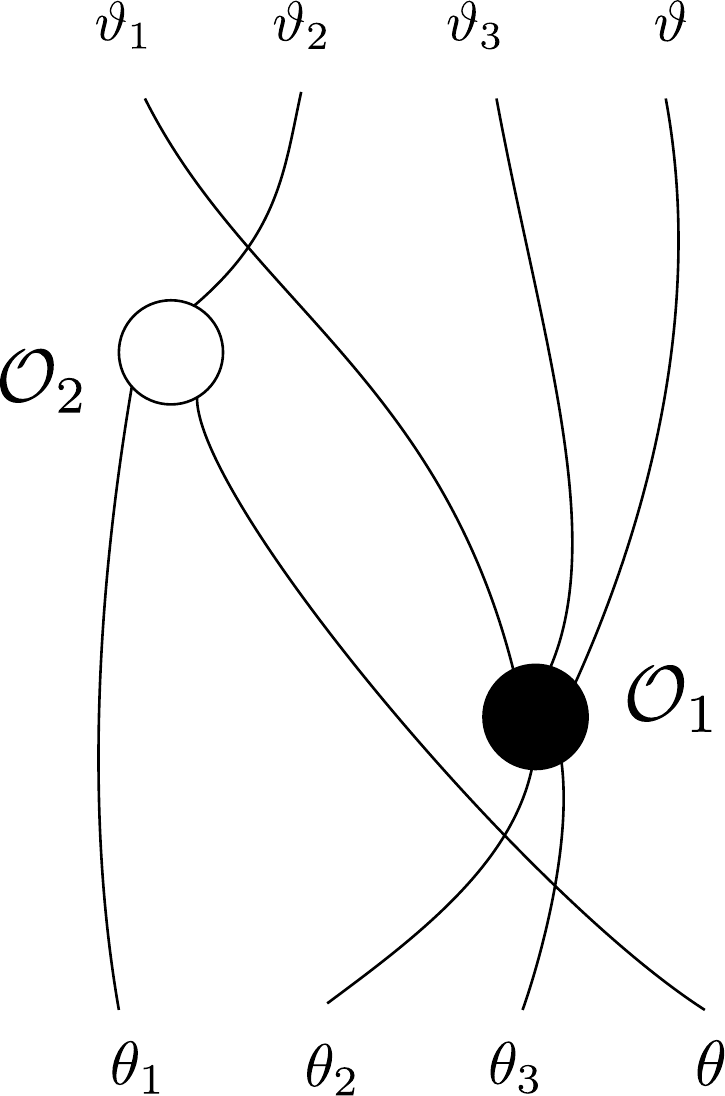}~~~~~~10.\includegraphics[width=1.4cm]{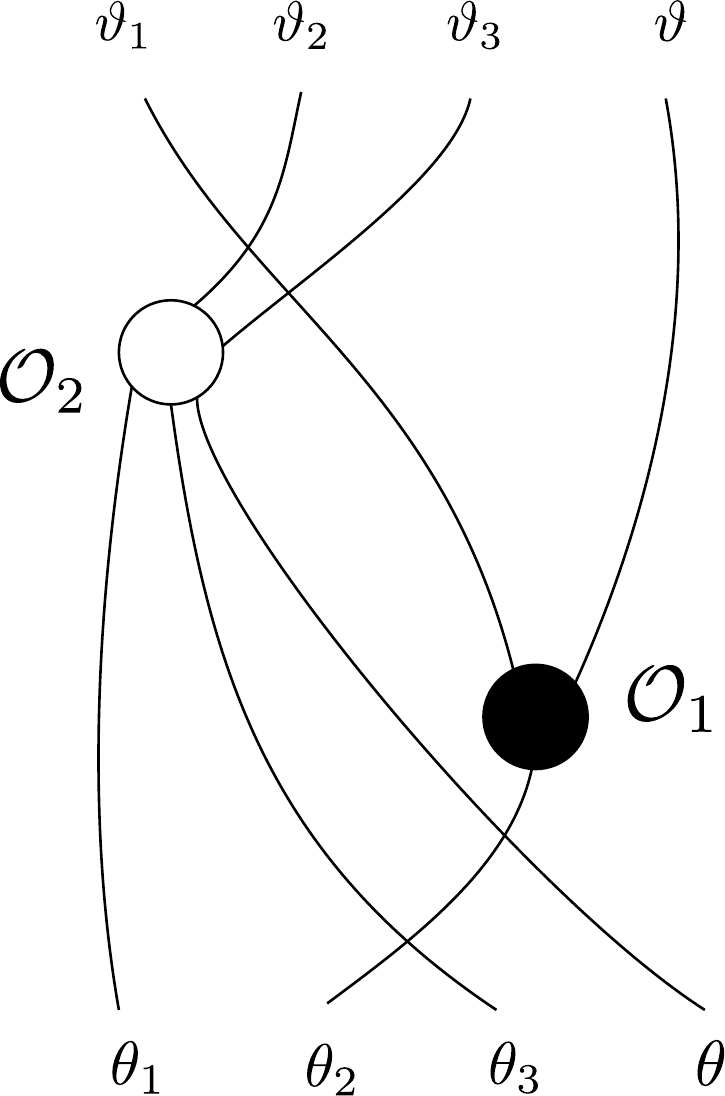}~~~~~~11.\includegraphics[width=1.4cm]{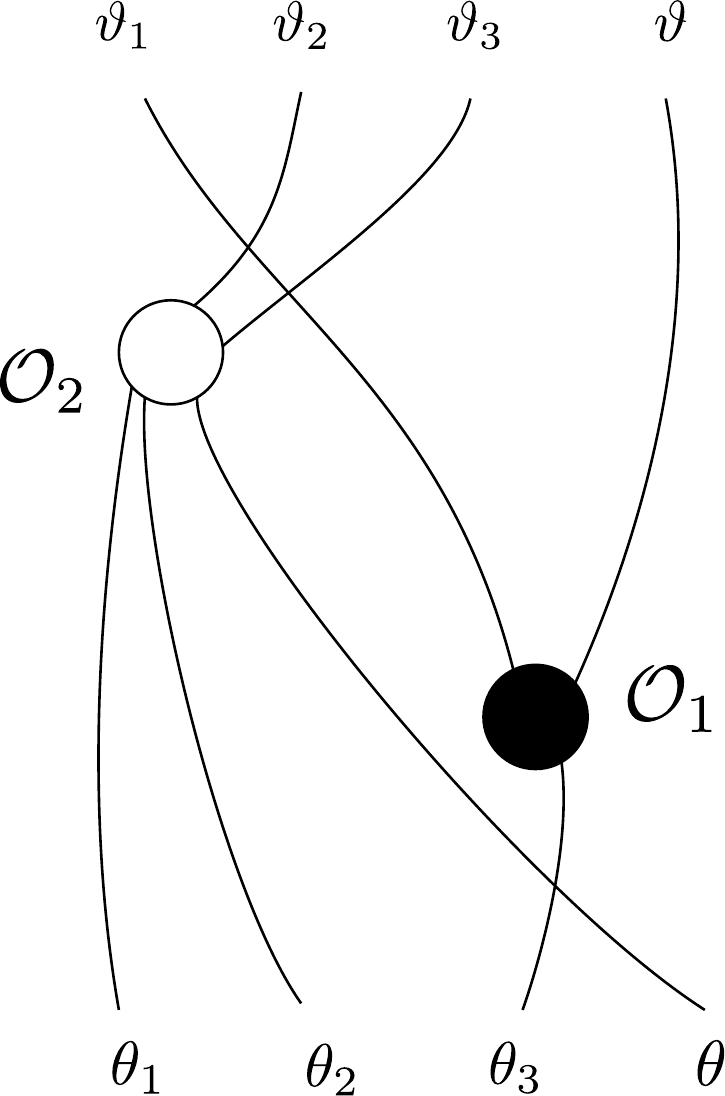}~~~~~~12.\includegraphics[width=1.4cm]{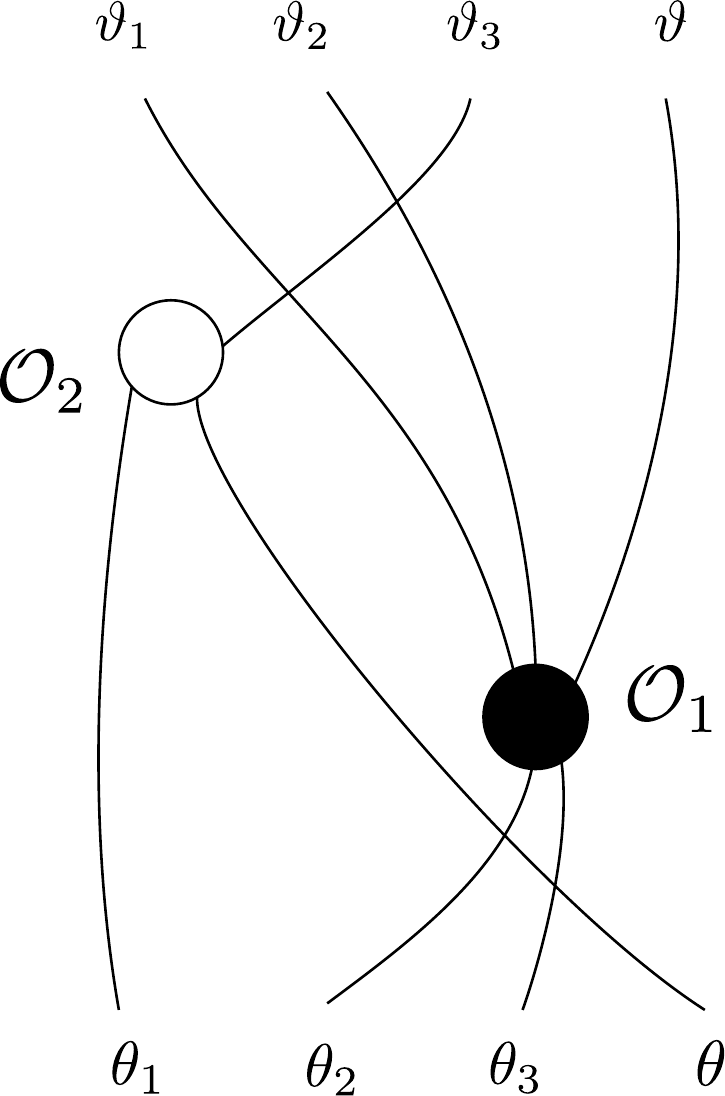}~~~~~~13.\includegraphics[width=1.4cm]{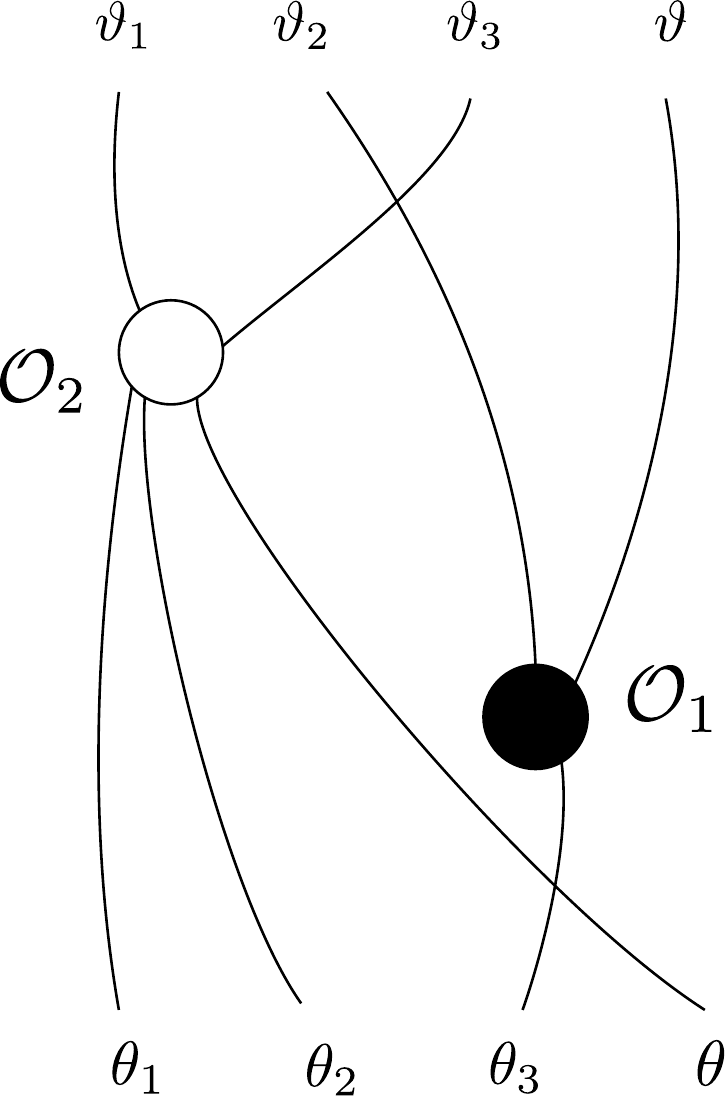}~~~~~~14.\includegraphics[width=1.4cm]{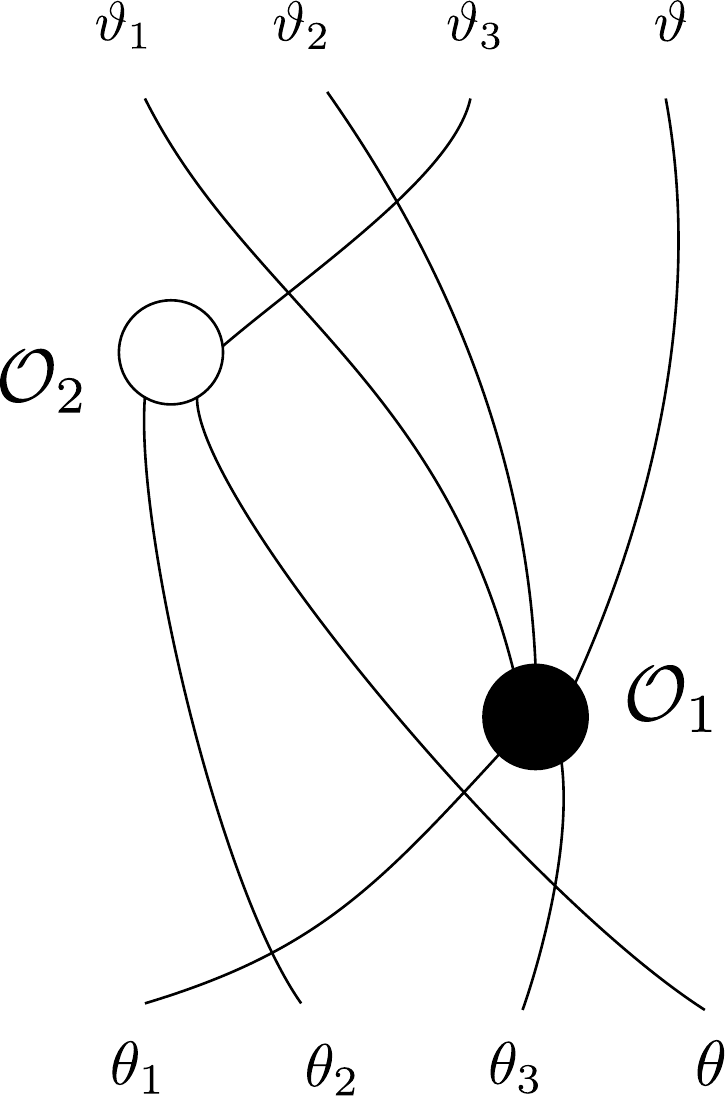}
\par\end{centering}
\caption{Contributing diagrams at third order in the clustering limit.}

\label{NNLOcontrib}
\end{figure}

After evaluating the diagrams, one needs to compare the result to
the product of finite volume form factor corrections and the exponentiated
energy-difference (\ref{eq:expansion}) at third order in the excited
state filling fraction $n_{i}$. The $y$-dependent terms are 
\begin{align}
\frac{1}{3!} & y^{3}(\Delta_{1}E)^{3}+\frac{1}{2}y^{2}\left((\Delta_{1}E)^{2}(\Delta_{1}\bar{F}^{1}+\Delta_{1}F^{2})+2\Delta_{2}E\right)\nonumber \\
+ & y\left(\Delta_{3}E+\Delta_{2}E(\Delta_{1}\bar{F}^{1}+\Delta_{1}F^{2})+\Delta_{1}E(\Delta_{1}\bar{F}^{1}\Delta_{1}F^{2}+\Delta_{2}\bar{F}^{1}+\Delta_{2}F^{2})\right).
\end{align}
which can be verified using the second order result for the form factor
corrections and the direct expansion (\ref{eq:Lexpansion}) of the
energy difference $\Delta E$ in terms of $n_{i}$.

We found it useful to reorganize the $y$-independent part, i.e. the
product of the finite volume form factor corrections. This is because
already at the second order (\ref{eq:res2nd}) we encountered terms
such as $-\frac{1}{2}n_{2}^{2}s_{2}^{2}\varphi_{22}+\frac{1}{2}n_{1}n_{2}s_{1}s_{2}\varphi_{12}^{2}$,
which cannot be associated naturally to any of the operators. If we
rather recollect these terms into a normalizing factor $\mathcal{N}^{2}=1+\Delta\mathcal{N}^{2}$
we can write 
\begin{equation}
\left(1+\Delta\bar{F}^{1}\right)\left(1+\Delta F^{2}\right)=\left(1+\Delta\mathcal{N}^{2}\right)\left(1+\Delta\bar{\mathsf{F}}^{1}\right)\left(1+\Delta\mathcal{\mathsf{F}}^{2}\right),
\end{equation}
where $\Delta$ means the full correction, i.e. the sum of all the
$\Delta_{k}$ orders. The newly defined $\Delta_{k}\bar{\mathsf{F}}^{1},\Delta_{k}\mathsf{F}^{2}$
constitute only from those terms at the $k^{\text{th}}$ order in
$n_{i}$, which contain connected form factors of the respective operator.
Then these corrections up to the second order take the form:
\begin{align}
\Delta_{1}\mathcal{N}^{2}=0\quad;\qquad\Delta_{2}\mathcal{N}^{2} & =-\frac{1}{2}n_{2}^{2}s_{2}^{2}\varphi_{22}+\frac{1}{2}n_{1}n_{2}s_{1}s_{2}\varphi_{12}^{2}\\
\Delta_{1}\bar{\mathcal{\mathsf{F}}}^{1}=\bar{F}^{1}(1)n_{1}\quad;\qquad\Delta_{2}\bar{\mathsf{F}}^{1} & =\frac{1}{2}n_{2}^{2}s_{2}\bar{F}^{1}(2)+\frac{1}{2}n_{1}n_{2}\Bigg\lbrace\bar{F}^{1}(1,2)-s_{2}\bar{F}^{1}(1)\varphi_{12}\Bigg\rbrace,\nonumber 
\end{align}
and one gets $\Delta\mathcal{\mathsf{F}}^{2}$ from $\Delta\mathcal{\bar{\mathsf{F}}}^{1}$
by replacing the connected form factors $\bar{F}^{1}(1,2,\ldots,N)$
with $F^{2}(1,2,\ldots,N)$.

The redefined expansion at third order looks as:

\begin{equation}
\Delta_{1}\bar{\mathsf{F}}^{1}\Delta_{2}\mathsf{F}^{2}+\Delta_{2}\bar{\mathsf{F}}^{1}\Delta_{1}\mathsf{F}^{2}+\Delta_{2}\mathcal{N}^{2}\left(\Delta_{1}\bar{\mathsf{F}}^{1}+\Delta_{1}\mathsf{F}^{2}\right)+\Delta_{3}\mathcal{N}^{2}+\Delta_{3}\bar{\mathsf{F}}^{1}+\Delta_{3}\mathsf{F}^{2},
\end{equation}
and after subtracting the first few terms, which are already known
from the previous orders, we arrive at our new result (after permuting
the integration variables many times to simplify it): 
\begin{align}
\Delta_{3}\mathcal{N}^{2}+\Delta_{3}\bar{\mathsf{F}}^{1}+\Delta_{3}\mathsf{F}^{2}= & \;\varphi_{11}\left(\frac{1}{3}n_{1}^{3}s_{1}^{2}(1-2s_{1})-n_{1}n_{2}^{2}\varphi_{12}s_{1}s_{2}S_{2}\right)+\frac{1}{3}n_{1}n_{2}^{2}\varphi_{12}^{2}s_{2}(3s_{1}s_{2}-s_{1}-2s_{2})\nonumber \\
 & +n_{1}n_{2}n_{3}\left\{ \varphi_{12}^{2}\varphi_{13}s_{2}s_{3}S_{1}+\frac{1}{3}\varphi_{12}\varphi_{23}\varphi_{13}(s_{1}s_{2}+s_{1}s_{3}-s_{1}s_{2}s_{3})\right\} +\nonumber \\
 & +\Bigg\lbrace\frac{1}{3}n_{1}^{3}s_{1}^{2}+n_{1}n_{2}\varphi_{12}s_{2}\left(\frac{1}{2}n_{1}-\frac{1}{6}n_{2}s_{2}-\frac{5}{6}n_{1}s_{1}\right)\\
 & \quad\quad+n_{1}n_{2}n_{3}\left(\frac{1}{6}s_{2}s_{3}\varphi_{12}\varphi_{13}-\frac{1}{2}S_{2}s_{3}\varphi_{12}\varphi_{2,3}\right)\Bigg\rbrace\left(\bar{F}^{1}(1)+F^{2}(1)\right)\nonumber \\
 & +\frac{1}{3!}\left(\nu_{12}\bar{F}^{1}(1,2)+\nu_{21}F^{2}(1,2)\right)+\frac{1}{3!}n_{1}n_{2}n_{3}\left(\bar{F}^{1}(1,2,3)+F^{2}(1,2,3)\right)\nonumber 
\end{align}
where we introduced the measure
\begin{equation}
\nu_{12}=n_{1}n_{2}\left(\left(2n_{1}s_{1}+n_{2}s_{2}\right)-\left(\varphi_{23}+2\varphi_{13}\right)s_{3}n_{3}\right).
\end{equation}
We may rewrite the terms in which this non-symmetric measure $\nu_{12}$
appears in a nicer way (anti-symmetric terms disappear under integration)
:
\begin{equation}
\nu_{12}\bar{F}^{1}(1,2)+\nu_{21}F^{2}(1,2)=\nu_{12}^{S}\left(\bar{F}^{1}(1,2)+F^{2}(1,2)\right)+\nu_{12}^{A}\left(\bar{F}_{A}^{1}(1,2)-F_{A}^{2}(1,2)\right),\label{eq:symmetrization}
\end{equation}
where $\nu_{12}^{S}=(\nu_{12}+\nu_{21})/2,\;\nu_{12}^{A}=(\nu_{12}-\nu_{21})/2$
and the antisymmetric part of the connected form factors $F_{A}(1,2)=(F(1,2)-F(2,1))/2$
can be deduced from (\ref{eq:antisymmpart}).

The last term on the r.h.s. of (\ref{eq:symmetrization}) takes the
form
\begin{align}
 & \frac{1}{3!}\Bigg\lbrace n_{1}n_{2}\varphi_{12}^{2}(s_{1}-s_{2})(n_{1}s_{1}-\varphi_{13}s_{3}n_{3})\\
+ & \frac{1}{2}n_{1}n_{2}\left(\left(n_{1}s_{1}-n_{2}s_{2}\right)-\left(\varphi_{13}-\varphi_{23}\right)s_{3}n_{3}\right)\varphi_{12}s_{2}\left(\bar{F}^{1}(1)+F^{2}(1)\right)\Bigg\rbrace,\nonumber 
\end{align}
and gives a correction to $\Delta_{3}\mathcal{N}^{2}$, and also to
the measure which multiplies single-argument connected form factors
$\bar{F}^{1}(1)$ and $F^{2}(1)$. These corrections appear in the
formulae presented in Subsection \ref{subsec:Organisation-of-the}.
The symmetric part of the measure which multiplies the two-argument
connected form factor gives a term in $\Delta_{3}\bar{\mathsf{F}}^{1}$:
\begin{equation}
\frac{1}{3!}\nu_{12}^{S}\bar{F}^{1}(1,2)=\frac{1}{4}n_{1}n_{2}\left(\left(n_{1}s_{1}+n_{2}s_{2}\right)-\left(\varphi_{13}+\varphi_{23}\right)s_{3}n_{3}\right)\bar{F}^{1}(1,2).
\end{equation}
If we combine this term with that part of the second order correction
$\Delta_{2}\bar{\mathsf{F}}^{1}$ in which $\bar{F}^{1}(1,2)$ appears,
we get
\begin{equation}
\left(\frac{1}{2}n_{1}n_{2}+\frac{1}{3!}\nu_{12}^{S}\right)\bar{F}^{1}(1,2)=\frac{1}{2}n_{1}n_{2}\left\{ 1+\frac{1}{2}\left(\left(n_{1}s_{1}+n_{2}s_{2}\right)-\left(\varphi_{13}+\varphi_{23}\right)s_{3}n_{3}\right)\right\} \bar{F}^{1}(1,2).\label{eq:measurefactorization}
\end{equation}
We now collect what multiplies the single-argument connected form
factor up to second order, i.e. in $\Delta_{1}\bar{\mathcal{\mathsf{F}}}^{1}+\Delta_{2}\bar{\mathcal{\mathsf{F}}}^{1}$:
\begin{equation}
\mu_{1}\bar{F}^{1}(1)=n_{1}\left\{ 1+\frac{1}{2}\left(n_{1}s_{1}-n_{2}s_{2}\varphi_{12}\right)+\ldots\right\} \bar{F}^{1}(1),
\end{equation}
where we denoted this measure object as $\mu_{1}$, and by the ellipses
we mean, that it will get higher order corrections from $\Delta_{k\geq3}\bar{\mathsf{F}}^{1}$
as well. Notice that the measure appearing before the two-argument
connected form factor (\ref{eq:measurefactorization}) is nothing
but the product $\frac{1}{2}\mu_{1}\mu_{2}$ truncated at the third
order. As the LO of $\mu_{i}$ is nothing but $n_{i}$ itself, the
term $\frac{1}{3!}n_{1}n_{2}n_{3}\bar{F}(1,2,3)$ is also trivially
consistent with the idea that the integration measure is factorizing.
Note that as the multivariate connected form factors $F(1,2,\ldots,N)$
are not symmetric in their arguments, we could also use their symmetrized
version (\ref{eq:symmFc}) as a basis, because the product $\mu_{1}\ldots\mu_{N}$
of the measures projects out the non-symmetric part under integration.

\subsection{Organisation of the result}

\label{subsec:Organisation-of-the}

Now that we have calculated the finite volume corrections to the product
of form factors $\bar{F}^{1}(\bar{\theta})_{L}F^{2}(\bar{\theta})_{L}$
up to the third order, we would like to understand the structure of
the result. There are many ways to factorise it, but in each case
we expect a structure, which is similar to the usual LM formula:
\begin{equation}
F(\bar{\theta})_{L}=\mathcal{N}\left\{ \sum_{N=0}^{\infty}\frac{1}{N!}\mu_{1}\mu_{2}\ldots\mu_{N}\mathcal{F}(1,2,\ldots,N)\right\} ,\label{eq:structure}
\end{equation}
where we integrate for $\theta_{1},\dots,\theta_{N}$ with the factorising
integration measure $\mu(\theta_{i})$, which should be expressed
in terms of the ground state and excited state pseudo energies. The
$\mathcal{F}(1,2,\ldots,N)$ objects are related to the connected
form factors; while the factor $\mathcal{N}$ is some normalisation
factor not containing connected form factors. Choosing a different
basis for the form factor building blocks ${\cal F}(1,\dots,N)$ redefines
the measure and the normalisation factor. In order to demonstrate
our result, we choose the symmetrised versions of the connected form
factors 
\begin{equation}
{\cal F}(1,\dots N)=\frac{1}{N!}\sum_{\sigma\in P}F(\sigma_{1},\dots,\sigma_{N})\label{eq:symmFc}
\end{equation}
where we sum over all permutations. In the next section we provide
the all order definition of finite connected form factors in the generic
case.

We have checked that our result is consistent with the factorizing
structure. The perturbative expansion of the normalisation factor,
which does not contain any operator-dependent terms turns out to be
\begin{align}
{\cal N}^{2} & =1-\frac{1}{2}n_{1}^{2}s_{1}^{2}\varphi_{11}-\frac{1}{3}n_{1}^{3}s_{1}^{2}(1-2s_{1})\varphi_{11}+n_{1}n_{2}^{2}\varphi_{12}\varphi_{11}s_{1}s_{2}(1-s_{2})+\dots\label{eq:N2}\\
 & \quad+\frac{1}{2}n_{1}n_{2}s_{1}s_{2}\varphi_{12}^{2}+\frac{1}{2}n_{1}n_{2}^{2}\varphi_{12}^{2}s_{2}(2s_{1}s_{2}-s_{1}-s_{2})+\dots\nonumber \\
 & \quad+n_{1}n_{2}n_{3}(\varphi_{12}^{2}\varphi_{13}(s_{2}s_{3}(1-s_{1})-\frac{1}{6}(s_{1}-s_{2})s_{3}+\frac{1}{3}\varphi_{12}\varphi_{23}\varphi_{31}(s_{1}s_{2}+s_{1}s_{3}-s_{1}s_{2}s_{3}))+\dots\nonumber 
\end{align}
while the measure is
\begin{align}
\mu_{1} & =n_{1}+\frac{1}{2}n_{1}^{2}s_{1}+\frac{1}{3}n_{1}^{3}s_{1}^{2}-\frac{1}{2}n_{1}n_{2}s_{2}\varphi_{12}+\frac{1}{4}n_{1}n_{2}s_{2}\varphi_{12}\left(2n_{1}-n_{2}s_{2}-3n_{1}s_{1}\right)+\dots\label{eq:mu}\\
 & \quad+\frac{1}{12}n_{1}n_{2}n_{3}s_{3}\varphi_{12}\left(s_{2}(\varphi_{13}+\varphi_{23})-6(1-s_{2})\varphi_{23}\right)+\dots\nonumber 
\end{align}

In summarizing, up to the third explicilty calculated order, the finite
volume form factor takes the LM-type form (\ref{eq:structure}) in
the basis (\ref{eq:symmFc}) with (\ref{eq:N2},(\ref{eq:mu})). Our
framework provides a way to systematically calculate both the normalization
factor and the measure, but at higher orders they are getting more
and more involved. Unfortunately, we could not recognise any nice
structure in these terms, which could give a hint how higher order
terms should look like. Most probably a better definition of the connected
form factors could simplify these expressions. Later we analyse the
free fermion theory, where we can go to all orders and sum up the
appearing terms.

\section{Definition of connected form factors}

\label{sec:Graph_rules}

In this section we investigate the singular $\varepsilon$-dependence
of the form factor

\begin{equation}
F(\theta_{n}+i\pi+i\varepsilon_{n},\dots,\theta_{1}+i\pi+i\varepsilon_{1},\theta_{1},\dots,\theta_{n},\theta)/F(\theta)\,,
\end{equation}
where for scalar operators the form factor is a constant $F(\theta)=F$.
This singular behaviour is in stark contrast to the diagonal form
factor, which is regular in the $\varepsilon\to0$ limit, but the
result depends on the direction how we approach it. Here, due to the
extra particle, the expression is singular and we work out all the
singular terms. This calculation is the extension of the one in \cite{Bajnok:2017bfg}
by keeping all the terms. Our method is to use the kinematical singularity
axiom successively to eliminate all $\varepsilon$s and define the
connected form factors iteratively. From the repeated application
of the kinematical singularity axiom it follows that the singular
terms in $\varepsilon$ take the form:
\begin{equation}
\frac{A_{12\dots n}}{\varepsilon_{1}\dots\varepsilon_{n}}+\sum_{k=1}^{n}\varepsilon_{k}\frac{A_{1\dots k-1k+1\dots n}}{\varepsilon_{1}\dots\epsilon_{n}}+\dots+\sum_{k=1}^{n}\frac{A_{k}}{\varepsilon_{k}}\,.
\end{equation}
 Where all terms can be evaluated by using the following graphical
rules:
\begin{enumerate}
\item Draw $n$ labeled points (from $1$ to $n$) and colour them each
black or white all possible ways
\item Connect the points with arrows all possible ways respecting the rules:
each point has at most one incoming arrow, arrows can leave from white
points, such that at each point arrows can go either all to the left
or all to the right and there are no loops.
\item Calculate the contribution of each graph with the following rules
and drop those in which after cancelations $\varepsilon$ remains
in the numerator
\begin{enumerate}
\item black dot contributes as 
\[
\CIRCLE_{k}\quad=\frac{s_{k}}{\varepsilon_{k}}\,,
\]
\item incoming left/right arrow (independently whether it is black or white)
\[
\LEFTcircle_{k}\leftarrow\quad=\varepsilon_{k}\qquad;\qquad\rightarrow\LEFTcircle_{k}\quad=-\varepsilon_{k}\,,
\]
\item outgoing left/right arrow (could be more then one, but the contribution
does not depend on their number) 
\[
\leftarrow\Circle_{k}\quad=\frac{1}{\varepsilon_{k}}\qquad;\qquad\Circle_{k}\rightarrow\quad=-\frac{S_{k}}{\varepsilon_{k}}\,,
\]
\item each arrow (independently if it goes left or right or between different
colours) carries a factor 
\[
\LEFTcircle_{k}\to\LEFTcircle_{l}\quad=\varphi_{kl}=\varphi_{lk}\quad=\LEFTcircle_{k}\leftarrow\LEFTcircle_{l}\quad\,,
\]
\item white dots without arrows give the connected form factor
\end{enumerate}
\end{enumerate}
We can proof these rules recursively.

For $n=1$ we can draw only one point which can be either black or
white with contributions
\[
\CIRCLE_{1}=\frac{1-S_{1}}{\varepsilon_{1}}\quad;\qquad\Circle_{1}=F(1)=F_{c}(\theta_{1}+i\pi,\theta_{1},\theta)/F\,.
\]
This is simply the kinematical singularity axioms for $F^{2}(\theta_{1}+i\varepsilon_{1}+i\pi,\theta_{1},\theta)$
as we already used in (\ref{eq:kinsin1}).

In the generic case we check the singular term of the form $\varepsilon_{k}^{-1}$.
Such term can either come from a black dot $\CIRCLE_{k}$ or from
a white dot with outgoing arrows. In the kinematical singularity axioms
the singular term in $\varepsilon_{k}$ takes the form 
\begin{align}
F(\theta_{n}+i\pi+i\varepsilon_{n},\dots,\theta_{k}+i\pi+i\varepsilon_{k},\dots,\theta_{1}+i\pi+i\varepsilon_{1},\theta_{1},\dots,\theta_{k},\dots,\theta_{n},\theta)=\\
\frac{1}{\varepsilon_{k}}\bigl(\prod_{j<k}S(\theta_{j}-\theta_{k})S(\theta_{j}+i\pi+i\varepsilon_{j}-\theta_{k})-S(\theta_{k}-\theta)\prod_{j>k}S(\theta_{k}-\theta_{j})S(\theta_{k}-\theta_{j}+i\pi-i\varepsilon_{j})\bigr)\times\nonumber \\
F(\theta_{n}+i\pi+i\varepsilon_{n},\dots,\theta_{1}+i\pi+i\varepsilon_{1},\theta_{1},\dots,\theta_{n},\theta)_{\text{\ensuremath{k-\mathrm{removed}}}}\nonumber 
\end{align}
where we also used the permutation axiom. In the third line we have
a form factor similar to what we started with, but the $k$th particle
is missing, thus we can use induction. Clearly that form factor can
have at most single poles in the remaining $\varepsilon$s. This suggests
to expand the S-matrix factors as 
\begin{equation}
S(\theta_{j}-\theta_{k})S(\theta_{j}+i\pi+i\varepsilon_{j}-\theta_{k})=1+\varepsilon_{j}\varphi(\theta_{j}-\theta_{k})+\dots
\end{equation}
in the terms for $j<k$ and a similar expression but with $-\varepsilon_{j}$
for $j>k$. We are now ready to read off the graph rules for the terms
containing $\varepsilon_{k}^{-1}$. Keeping the ones in the product
we get a term proportional to $1-S_{k}=s_{k}$. This contribution
is denoted by the black dot. Terms coming from the $j<k$ product
are represented by arrows going to the left with no extra factors,
while terms from the $j>k$ product has an extra $-S(\theta_{k}-\theta)=-S_{k}$
factor as well as an extra minus sign in $-\varepsilon_{j}$. We attribute
this extra minus sign to the incoming arrow as more than one $\varepsilon$
can give contributions due to multiple $\varepsilon$s in the remaining
form factor. Clearly, we have either the $j<k$ or the $j>k$ products,
so arrows can be drawn either all to the left or all to the right.
Using these rules inductively, proves the correctness of our graph
rules.

Let us now see the example of the two particle term. For $n=2$ we
have the following contributions 
\begin{align*}
\CIRCLE_{1}\,\,\,\quad\CIRCLE_{2}\quad & =\frac{s_{1}}{\varepsilon_{1}}\frac{s_{2}}{\varepsilon_{2}}\,,\\
\CIRCLE_{1}\,\,\,\quad\Circle_{2}\quad & =\frac{s_{1}}{\varepsilon_{1}}F(2)\,,\\
\Circle_{1}\,\,\,\quad\CIRCLE_{2}\quad & =F(1)\frac{s_{2}}{\varepsilon_{2}}\,,\\
\Circle_{1}\,\,\,\quad\Circle_{2}\quad & =F(1,2)=F_{c}(\theta_{2}+i\pi,\theta_{1}+i\pi,\theta_{1},\theta_{2},\theta)/F\,,\\
\CIRCLE_{1}\leftarrow\Circle_{2}\quad & =\frac{s_{1}\varepsilon_{1}}{\varepsilon_{1}}\frac{1}{\varepsilon_{2}}\varphi_{21}\,,\\
\Circle_{1}\rightarrow\CIRCLE_{2}\quad & =\text{\ensuremath{\left(-\frac{S_{1}}{\varepsilon_{1}}\right)}}\frac{s_{2}(-\varepsilon_{2})}{\varepsilon_{2}}\varphi_{12}\,,
\end{align*}
we would also have terms with two white dots and an arrow, but there
some epsilon remains in the numerator, so we dropped them. By summing
all terms up we have the following form 
\begin{equation}
F(\theta_{2}+i\pi+i\epsilon_{2},\theta_{1}+i\pi+i\epsilon_{1},\theta_{1},\theta_{2},\theta)/F=\frac{A_{12}}{\varepsilon_{1}\varepsilon_{2}}+\frac{A_{1}}{\varepsilon_{1}}+\frac{A_{2}}{\varepsilon_{2}}+F(1,2)+O(\varepsilon/\varepsilon)\,,\label{eq:Fcexpansion}
\end{equation}
where 
\begin{equation}
A_{12}=s_{1}s_{2}\quad;\quad A_{1}=s_{1}F(2)+S_{1}s_{2}\varphi_{12}\quad;\quad A_{2}=F(1)s_{2}+s_{1}\varphi_{21}\,.
\end{equation}
We note that the connected form factor $F(1,2)$ is not symmetric.
We can relate $F(2,1)$ to $F(1,2)$ by using the form factor axioms
\begin{align}
F(\theta_{1}+i\pi+i\varepsilon_{1},\theta_{2}+i\pi+i\varepsilon_{2},\theta_{2},\theta_{1},\theta) & =S(\theta_{1}-\theta_{2}+i(\varepsilon_{1}-\varepsilon_{2}))S(\theta_{2}-\theta_{1})\times\nonumber \\
 & \qquad F(\theta_{2}+i\pi+i\varepsilon_{2},\theta_{1}+i\pi+i\varepsilon_{1},\theta_{1},\theta_{2},\theta)\,.
\end{align}
We need to expand the scattering matrix 
\begin{equation}
\frac{S(\theta+i\mathcal{\epsilon})}{S(\theta)}=1+i\mathcal{\varepsilon}\frac{S'(\theta)}{S(\theta)}-\frac{1}{2}\varepsilon^{2}\frac{S''(\theta)}{S(\theta)}+\dots=1-\varepsilon\varphi(\theta)+\frac{1}{2}\varepsilon^{2}\left(\varphi(\theta)^{2}-i\varphi'(\theta)\right)+\dots\,.
\end{equation}
where we used that 
\begin{equation}
i\varphi(\theta)=\frac{S'(\theta)}{S(\theta)}\quad;\qquad i\varphi'(\theta)=\frac{S''(\theta)}{S(\theta)}-\frac{S'(\theta)^{2}}{S(\theta)^{2}}=\frac{S''(\theta)}{S(\theta)}+\varphi(\theta)^{2}\,.
\end{equation}
Thus 
\begin{equation}
F(2,1)=F(1,2)+\varphi_{12}(A_{2}-A_{1})-A_{12}(\varphi_{12}^{2}-i\varphi'_{12})\,.
\end{equation}
A bit simplified form can be obtained as 
\begin{align}
F(2,1)-F(1,2) & =\varphi_{12}(F(1)s_{2}-s_{1}F(2))+i\varphi_{12}'s_{1}s_{2}+\varphi_{12}^{2}(S_{2}-S_{1})\,.\label{eq:antisymmpart}
\end{align}
which is clearly anti-symmetric for the exchange $1\leftrightarrow2$.
Actually this difference under symmetric integration vanishes.

Finally we note that the rules for the $\varepsilon$-dependence of
the form factor 
\begin{equation}
F(\theta_{n}+i\pi+i\varepsilon_{n},\dots,\theta_{1}+i\pi+i\varepsilon_{1},\theta_{1},\dots,\theta_{n},\theta+i\pi)/F(\theta)\,,
\end{equation}
 is analogous, we merely have to make the $S_{i}\to S_{i}^{-1}$ replacement.
This form factor always appears with a prefactor $S_{1}\dots S_{n}$
so it is natural to include this factor in the definition of the connected
form factor.

\section{Extension for multiparticle states}

In this section we explain how the results can be extended from the
simplest one-sided finite volume form factor to the generic case
\begin{equation}
_{L}\langle0\vert{\cal O}\vert\bar{\theta}\rangle_{L}\to{}_{L}\langle0\vert{\cal O}\vert\bar{\theta}_{1},\dots,\bar{\theta}_{N}\rangle_{L}\equiv{}_{L}\langle0\vert{\cal O}\vert\{\bar{\theta}\}\rangle_{L}
\end{equation}
We have to start by investigating the clustering behaviour of the
generic excited state expectation value of the bilocal operator
\begin{equation}
_{L}\langle\{\bar{\theta}\}\vert\mathcal{O}_{1}(x,t)\mathcal{O}_{2}(0,0)\vert\{\bar{\theta}\}\rangle_{L}\,,
\end{equation}
In the $y=it\to\infty$ limit the expression factorizes into the product
of the needed form factors and the exponentialized excited state energy
difference:
\begin{equation}
_{L}\langle\{\bar{\theta}\}\vert\mathcal{O}_{1}(0,-iy)\mathcal{O}_{2}\vert\{\bar{\theta}\}\rangle_{L}\to\,_{L}\langle\{\bar{\theta}\}\vert\mathcal{O}_{1}\vert0\rangle_{L}\,_{L}\langle0\vert\mathcal{O}_{2}\vert\{\bar{\theta}\}\rangle_{L}e^{(E_{N}-E_{0})y}+O(1)\,.
\end{equation}

We have to calculate the same limit in the crossed channel for the
excited state expectation value, which has the form \cite{Pozsgay:2013jua}
\begin{equation}
\langle\Omega_{N}\vert\mathcal{O}_{1}(x,t)\mathcal{O}_{2}(0,0)\vert\Omega_{N}\rangle=\sum_{\alpha\cup\bar{\alpha}}\frac{{\cal D}_{\alpha}\bar{\rho}_{\bar{\alpha}}}{\rho_{N}(\bar{\theta})}\,
\end{equation}
Here $\vert\Omega_{N}\rangle$ denotes the thermal state related to
the solution of the excited state TBA. We have to sum up for all partitions
$\alpha=\{i_{1},\dots,i_{\vert\alpha\vert}\}$ of the set $\{1,\dots,N\}=\alpha\cup\bar{\alpha}$
and 
\begin{equation}
\mathcal{D}_{\alpha}=\sum_{n=0}^{\infty}\frac{1}{n!}\prod_{i=1}^{n}\int\frac{d\theta_{i}}{2\pi}\frac{1}{1+e^{\epsilon_{N}(\theta_{i})}}F_{c}^{12}(\theta_{1},\dots,\theta_{n},\{\bar{\theta}+i\frac{\pi}{2}\}_{\alpha})
\end{equation}
where $\{\bar{\theta}\}_{\alpha}=\{\bar{\theta}_{i_{1}},\dots,\bar{\theta}_{i_{\vert\alpha\vert}}\}$
and $\bar{\rho}_{\bar{\alpha}}$ denotes the corresponding subdeterminant
for the $\bar{\alpha}$ rapidity set. By investigating the exponential
growth of the various ${\cal D}_{\alpha}$ contributions one can see
that the expected $e^{(E_{N}-E_{0})y}$ behaviour comes only from
the ${\cal D}_{N}$ term. Even more, it can come only from diagrams
when all the incoming particles are connected to operator ${\cal O}_{2}$,
while all the outgoing particles to operator ${\cal O}_{1}$, just
as it happened for the one particle case. By inspecting the details
of the order by order calculations one can show that all steps generalizes
naturally. The filling fraction has to be replaced with the excited
state filling fraction: 
\begin{equation}
n_{i}=\frac{1}{1+e^{\epsilon_{N}(\theta_{i})}}
\end{equation}
In drawing the various diagrams one can realize that the only thing
one has to replace is our spectator particle of rapidity $\theta$
with the group of such particles leading to the modification of only
the S-matrix factor 
\begin{equation}
s_{i}=1-S_{i}=1-\prod_{k=1}^{N}S(\theta_{i}-\bar{\theta}_{k}-\frac{i\pi}{2})
\end{equation}
which now contains the contributions of all physical particles. Similarly
the connected form factor also includes all the physical particles
as spectators.
\begin{equation}
F(1,\dots,k)F(\bar{\theta}_{1}+i\frac{\pi}{2},\dots\bar{\theta}_{N}+i\frac{\pi}{2})={\rm FP}.F(\theta_{1}+i\pi+i\varepsilon_{1},\dots\theta_{k}+i\pi+i\varepsilon_{k},\theta_{k},\dots,\theta_{1},\bar{\theta}_{1}+i\frac{\pi}{2},\dots\bar{\theta}_{N}+i\frac{\pi}{2})
\end{equation}
The graph rules apply also in this case with these replacements and
our final formulas (\ref{eq:structure}) describe the generic one-sided
excited state finite volume form factors.

\section{Free fermion finite volume form factors}

In the work \cite{Fonseca:2001dc}, Fonseca and Zamolodchikov derived
the exact finite volume form factor of the spin field in the thermally
perturbed Ising model, which is nothing but the field theory of a
free massive fermion. The $\sigma$ field is a non-local operator,
which interpolates between the Ramond and the Neveu-Schwarz sectors.
Its simplest excited state matrix element takes the form
\begin{equation}
_{\mathrm{NS}}\langle0\vert\sigma\vert\{\bar{\theta}\}\rangle_{\mathrm{R}}=S(L)g(\bar{\theta}_{1})\dots g(\bar{\theta}_{N})F_{N}\,,
\end{equation}
where $F_{N}=F_{N}(\{\bar{\theta}\})$ is the infinite volume form
factor, 
\begin{equation}
S(L)=_{\mathrm{NS}}\langle0\vert\sigma\vert0\rangle_{\mathrm{R}}=\exp\left\{ \frac{(mL)^{2}}{2}\int\hspace{-2.5mm}\int_{-\infty}^{\infty}\frac{d\theta_{1}d\theta_{2}}{(2\pi)^{2}}\frac{\sinh\theta_{1}\sinh\theta_{2}\,\log\coth\vert\frac{\theta_{1}-\theta_{2}}{2}\vert}{\sinh(mL\cosh\theta_{1})\sinh(mL\cosh\theta_{2})}\right\} \,,
\end{equation}
is the finite volume form factor of the $\sigma$ operator, which
creates the Neveu-Schwarz vacuum from the Ramond. The excited state-dependent
factor contains the norm of the state $\rho_{1}$ and takes also an
exponentiated form
\begin{equation}
g(\bar{\theta})=\frac{e^{\kappa(\bar{\theta})}}{\sqrt{mL\cosh\bar{\theta}}}\quad;\quad\kappa(\bar{\theta})=\int_{-\infty}^{\infty}\frac{d\theta}{2\pi}\frac{1}{\cosh(\bar{\theta}-\theta)}\log\frac{1-e^{-mL\cosh\theta}}{1+e^{-mL\cosh\theta}}\,.
\end{equation}

Let us manipulate these expressions by observing that 
\begin{equation}
\partial_{\theta}\mathcal{L}\equiv\partial_{\theta}\log\frac{1-e^{-mL\cosh\theta}}{1+e^{-mL\cosh\theta}}=\frac{mL\sinh\theta}{\sinh(mL\cosh\theta)}\,.
\end{equation}
Integration by parts twice leads to the expression
\begin{equation}
S(L)=\exp\left\{ \frac{1}{2}\int\hspace{-2.5mm}\int_{-\infty}^{\infty}\frac{d\theta_{1}d\theta_{2}}{(2\pi)^{2}}{\cal L}(\theta_{1}){\cal L}(\theta_{2})f(\theta_{1}-\theta_{2})\right\} =e^{\frac{1}{2}{\cal L}_{1}{\cal L}_{2}f_{12}}\,,
\end{equation}
where 
\begin{equation}
f_{ij}=f(\theta_{i}-\theta_{j})=-\frac{\cosh(\theta_{i}-\theta_{j})}{\sinh(\theta_{i}-\theta_{j})^{2}}\,.
\end{equation}
In the following we recover these results from our approach. We start
with the vacuum amplitude $S(L)$, we then turn to deriving the $g(\bar{\theta})$
factor.

\subsection{Calculation of the vacuum amplitude}

We start by recovering the $S(L)$ factor, which can be interpreted
as the vacuum amplitude of the non-local operator. As this operator
changes the NS vacuum to the R one it connects the true ground state
to an excited state and can be recovered by analysing the clustering,
$y\to\infty$, limit of the excited state expectation value of the
two-point function
\begin{equation}
_{R}\langle0\vert\sigma\sigma(y)\vert0\rangle_{\mathrm{R}}={}_{\mathrm{R}}\langle0\vert\sigma\vert0\rangle_{\mathrm{NS}\,\mathrm{NS}}\langle0\vert\sigma\vert0\rangle_{\mathrm{R}}\,e^{y\Delta E}+\dots=S(L)^{2}e^{y\Delta E}+\dots\,,
\end{equation}
where the energy difference is 
\begin{equation}
\Delta E=E_{\mathrm{R}}-E_{\mathrm{NS}}=-m\int\frac{d\theta_{1}}{2\pi}\cosh\theta_{1}\,{\cal L}(\theta_{1})=-e_{1}{\cal L}_{1}\,,
\end{equation}
and we integrate for $\theta_{i}$ with a $\frac{1}{2\pi}$ factor,
whenever the symbol ${\cal L}_{i}$ appears. Since 
\begin{equation}
{\cal L}_{1}=\log\frac{1-L_{1}}{1+L_{1}}=-2\left\{ L_{1}+\frac{L_{1}^{3}}{3}+\frac{L_{1}^{5}}{5}+\dots\right\} \quad;\quad L_{i}=e^{-mL\cosh\theta_{i}}\,,
\end{equation}
is negative, $\Delta E$ is positive, and we are indeed focusing on
the leading exponentially growing term. In recovering the exact result
we observe that it exponentiates $_{R}\langle0\vert\sigma\sigma(y)\vert0\rangle_{\mathrm{R}}\to e^{a+b}+\dots$
with $a={\cal L}_{1}{\cal L}_{2}f_{12}$ and $b=-e_{1}{\cal L}_{1}$.
In our approach we calculate directly the expansion of these exponential
terms $e^{a+b}=\sum_{n,k=0}^{\infty}\frac{a^{n}}{n!}\frac{b^{k}}{k!}$,
thus the various contributions should be factorised and each term
should be divided by its symmetry factor. Having checked this property
it is enough to compare the exponent $a+b$ to the connected terms.

In our approach we use the excited state LM type formula for the bilocal
operator
\begin{equation}
_{R}\langle0\vert\sigma\sigma(y)\vert0\rangle_{\mathrm{R}}=\sum_{N=0}^{\infty}\frac{1}{N!}\prod_{i=1}^{N}\int\frac{d\theta_{i}}{2\pi}n(\theta_{i})F_{c}^{\sigma\sigma(y)}(\theta_{1},\dots,\theta_{N})\,,\label{eq:sigmavev}
\end{equation}
with the measure factor, which corresponds to the Ramond groundstate
\begin{equation}
n(\theta_{i})=n_{i}=\frac{1}{1-L_{i}^{-1}}\,.
\end{equation}
This expansion factor is related to the one appearing in the exact
result ${\cal L}$ as:
\begin{align}
{\cal {\cal L}}_{i} & =\log\frac{1-L_{i}}{1+L_{i}}=-\log(1-2n_{i})=2n_{i}+\frac{(2n_{i})^{2}}{2}+\frac{(2n_{i})^{3}}{3}+\dots=\sum_{k=1}^{\infty}\frac{(2n_{i})^{k}}{k}\,.\label{eq:calL}
\end{align}

We now specify the expression (\ref{eq:F12conn}) by noting that the
scattering matrix is simply $S=-1$, implying that $\varphi_{ij}=0$.
We also drop the $\mu$ integrals and keep terms only when $\vert A_{+}\vert=\vert B_{-}\vert$
in order to have terms, which survive in the clustering limit:

\begin{align}
F^{\sigma\sigma(y)}(\{\vartheta\}_{I},\{\theta\}_{I}) & =\sum_{A^{+}\cup A^{-}=I}\sum_{B^{+}\cup B^{-}=I}K_{y}(\{\vartheta\}_{B^{-}}\vert\{\theta\}_{A^{+}})F^{2}(\{\vartheta\}_{B^{+}}+i\pi,\{\theta\})F^{1}(\{\vartheta\}_{B^{-}}+i\pi,\{\theta\}_{A^{+}})\,.\label{eq:Fsigmasigma}
\end{align}
The key simplification is the explicit use of the form factors of
the theory. The even infinite volume form factors of the $\sigma$
field are simply
\begin{equation}
F(\theta_{1},\dots,\theta_{2n})=i^{n}\prod_{j<k}\tanh((\theta_{j}-\theta_{k})/2)\,.
\end{equation}
As we are in a free theory there is an alternative form based on Wick-theorem:
\begin{equation}
F(\theta_{1},\dots,\theta_{2n})=\sum_{\mathrm{all\ pairings}}\quad\prod_{\mathrm{pairs}}F(\mathrm{pairs)}(-1)^{\#}\,,
\end{equation}
where the sign can be calculated as follows. We draw the $2n$ points
on a circle and connect them pairwise. $\#$ counts how many crossings
we have. Since the S-matrix is also $-1$ we just need to pair them
in all possible way with the usual conventions, that whenever we have
a crossing we associate an S-matrix for it. Actually there are $(2n-1)!!=\frac{(2n)!}{2^{n}n!}$
ways to form pairs and connect the points and each contribution is
factorised into two-particle terms. We should also keep in mind that
the $\sigma$ field is non-local and that $_{\mathrm{NS}}\langle0\vert\sigma\vert0\rangle_{\mathrm{R}}$
and $_{\mathrm{R}}\langle0\vert\sigma\vert0\rangle_{\mathrm{NS}}$
have opposite non-locality. As a consequence
\begin{equation}
F^{2}(\theta_{1},\dots,\theta_{2n})=(-1)^{n}F(\theta_{1},\dots,\theta_{2n})\quad;\qquad F^{1}(\theta_{1},\dots,\theta_{2n})=F(\theta_{1},\dots,\theta_{2n})\,,
\end{equation}
where it is assumed that $n$ particles are incoming and $n$ are
outgoing.

Let us see now why our formula (\ref{eq:sigmavev},\ref{eq:Fsigmasigma})
gives a factorising and exponentiating result. Clearly the measure
factor and the $K_{y}$ factor factorise into one-particle terms.
Moreover, each form factor is a sum of terms factorising into two
particle terms, thus the total contribution is a sum of factorised
terms. The only thing we have to check is that contributions appearing
multiple times are divided by the corresponding symmetry factors.
We argue in Appendix \ref{sec:Free-fermion} why this actually happens.
As a consequence it is enough to compare the connected part of our
formula to the exponent of the exact result. We start with the energy
difference.

In the case of the energy difference we would like to recover the
\begin{equation}
y\Delta E=-ye_{1}{\cal L}_{1}=-ye_{1}\left(2n_{1}+\frac{(2n_{1})^{2}}{2}+\frac{(2n_{1})^{3}}{3}+\dots\right)\,,
\end{equation}
term order by order. We need to show that at $N^{th}$ order the singly
differentiated $K_{y}$ factor comes with a $2^{N}/N$ factor. This
term originates from deforming $N-1$ contours and picking up $N-1$
times the residue. The main problem is to classify these diagrams
and evaluate them all. This is performed in Appendix \ref{sec:Free-fermion}
and we completely recovered the measure factor ${\cal L}_{1}$ multiplying
the energy difference.

In the form factor part we use again factorisation and compare the
exponent 
\begin{equation}
\int\hspace{-2.5mm}\int_{-\infty}^{\infty}\frac{d\theta_{1}d\theta_{2}}{(2\pi)^{2}}{\cal L}(\theta_{1}){\cal L}(\theta_{2})f(\theta_{1}-\theta_{2})\,,
\end{equation}
to the connected part of our result. In particular we compare the
expansion of only one of the ${\cal L}$s as the expression must be
symmetric. At the $k+1$ particle level it should give
\begin{equation}
\frac{(2n_{1})^{k}}{k}(2n_{2})f_{12}\,,
\end{equation}
which we test order by order. Since we cannot distinguish between
$n_{1}$ and $n_{2}$ in the calculation, for $k>1$ there is an extra
factor of $2$. The calculation is similar to the energy difference,
which we detail in Appendix \ref{sec:Free-fermion}. The outcome is
that we also recover completely the measure as well as the form factor
part.

\subsection{Excited state}

In order to extract the excited state form factor, we investigate
the clustering, $y\to\infty$, limit of the expectation value of the
excited state two-point function
\begin{equation}
_{R}\langle\bar{\theta}\vert\sigma\sigma(y)\vert\bar{\theta}\rangle_{\mathrm{R}}={}_{\mathrm{R}}\langle\bar{\theta}\vert\sigma\vert0\rangle_{\mathrm{NS}\,\mathrm{NS}}\langle0\vert\sigma\vert\bar{\theta}\rangle_{\mathrm{R}}\,e^{y\Delta\bar{E}}+\dots=S(L)^{2}g(\bar{\theta})^{2}e^{y\Delta\bar{E}}+\dots\,,
\end{equation}
where the energy difference contains also the contribution of the
moving particle 
\begin{equation}
\Delta\bar{E}=E_{\mathrm{R}}+m\cosh\bar{\theta}-E_{\mathrm{NS}}=m\cosh\bar{\theta}-e_{1}{\cal L}_{1}\,,
\end{equation}

We would like to recover this result from the expression (\ref{eq:F12conn})
by simplifying it with $S=-1$, $\varphi_{ij}=0$ and by dropping
the $\mu$ integrals and keeping only terms when $\vert A_{+}\vert=\vert B_{-}\vert$
as in (\ref{eq:Fsigmasigma}). The filling fraction is the same as
before $n(\theta_{i})=n_{i}=\frac{1}{1-L_{i}^{-1}}\,.$

At $k^{th}$ order we take rapidities $\vartheta_{j}=\theta_{j}+i\varepsilon_{j}$
for $j=1,\dots,k$, while the last argument $\vartheta=\theta$ will
be analytically continued to the physical rapidity $\theta\to\bar{\theta}+i\frac{\pi}{2}$.
Due to this extra physical particle we need the odd form factors of
the sigma field
\begin{equation}
F(\theta_{1},\dots,\theta_{2n+1})=i^{n}\prod_{j<k}\tanh((\theta_{j}-\theta_{k})/2)\,,
\end{equation}
which again can be written as 
\begin{equation}
F(\theta_{1},\dots,\theta_{2n+1})=\sum_{j=1}^{2n+1}(-1)^{j-1}F(\theta_{j})\sum_{\mathrm{all\ pairings}}\quad\prod_{\mathrm{pairs}}F(\mathrm{pairs)}(-1)^{\#}\,,
\end{equation}
where a pairing is understood for the even set missing $j$ and the
one-particle form factor in the above normalization is $F(\theta)=1$.

In checking the excited state formula we focus only on the $g(\bar{\theta})^{2}$
factor, in particular, only its exponent as we have a factorizing
result
\begin{equation}
\kappa(\bar{\theta})=\int_{-\infty}^{\infty}\frac{d\theta_{1}}{2\pi}\frac{1}{\cosh(\bar{\theta}-\theta_{1})}{\cal L}(\theta_{1})\,,
\end{equation}
which we expand in $n_{1}$ as in (\ref{eq:calL}). Thus at the $k^{th}$
order we need to check the contribution $\frac{1}{\cosh(\theta-\theta_{1})}\frac{(2n_{1})^{k}}{k}$.
The calculation is similar to the energy difference and the form factor,
which we detail in Appendix \ref{sec:Free-fermion}. The result is
that we completely recover this expression in our framework.

\section{Conclusion}

We set out to understand non-diagonal finite volume form factors in
integrable field theories beyond the first exponential Lüscher correction.
As the first step in our calculation, we introduced a LeClair-Mussardo
type formulation for the two-point function evaluated in excited states.
In the clustering limit, when the separation of the operators is significant,
an exponentially growing term dominates the expression, which is proportional
to the square of the finite volume form factor. The exponent is proportional
to the separation of the operators and the exact energy difference
between the excited and ground states.

In Section 4, we showed how to systematically use the bilocal form
factor formulation of the two-point function in the mirror channel
to extract the exponentially growing terms in the clustering limit.
Understanding the kinematical singularity structure of the form factors
in our expression was instrumental for this step. We developed a graphical
representation for the singularity structure, by generalising the
result for the connected expansion of diagonal form factors \cite{Pozsgay:2007gx}.

Two kinds of terms contribute to the exponentially growing part of
the two-point function. The first kind shows apparent exponential
growth in the clustering limit. Moreover, the integration measure
for the rapidities is linear in the filling fraction $n$. The second
kind is a collection of seemingly exponentially suppressed terms;
however, the integration region contains kinematical singularities
that modify the outcome and contribute to the clustering limit. To
calculate such terms, we shifted the integration contour of the rapidities
with infinitesimal imaginary parts with a specific ordering. By contour
manipulation, it was straightforward to calculate the contributing
residues. Consequently, the integration measure for such terms contains
higher powers of the filling fraction, crucial in reproducing the
exact energy differences in the expression.

As proof of the viability of our approach, we calculated the finite
volume form factor in the field theory of free massive fermion, which
is the integrable model describing the thermal perturbation of the
Ising conformal point. Due to the lack of interaction, the calculation
vastly simplifies, and we managed to derive the exact finite volume
form factor formulae presented in \cite{Fonseca:2001dc}.

For a general massive integrable field theory with a single particle
type that lacks bound state formation, we explicitly calculated the
clustering limit of the two-point function up to the third Lüscher
order. We can separate the terms contributing towards the energy factor
in the calculation. Up to the third order, they reproduce the energy
difference between the excited and ground states described by the
TBA equations and show clear signs of exponentiation. Hence we expect
our method to reproduce the expected exponential growth of the excited
state two-point function for all orders.

From the remaining terms, we conjectured the general structure of
the finite volume form factor.

\begin{equation}
\,_{L}\langle0\vert\mathcal{O}\vert\bar{\theta}_{1},\dots,\bar{\theta}_{N}\rangle_{L}=\frac{1}{\sqrt{\rho_{N}}}\mathcal{N}\left\{ \sum_{K=0}^{\infty}\frac{1}{K!}\prod_{j=1}^{K}\int d\theta_{j}\mu(\theta_{j})\mathcal{F}(\theta_{1},\theta_{2},\ldots,\theta_{K})\right\} 
\end{equation}

It has three building blocks: the exact density factor of the excited
states, an operator-independent normalisation factor (\ref{eq:N2}),
and a \textquotedbl dressed\textquotedbl{} version of the non-diagonal
form factor.

The density factor appears in the denominator of the two-point function
and propagates to the finite volume form factor. This form is consistent
with the exact diagonal finite volume form factor formula and the
polynomial correction to the form factors in large volumes.

The normalisation factor is independent of the properties of the operators,
assuming that they are spinless, as we did from the start of our calculation.
Similarly to the density factor, we symmetrically distribute the normalisation
factor from the two-point function between the two operators. The
origin of the normalisation factor roots in the different finite volume
states on the two sides of the operators, namely the finite volume
excited and ground states. We can think of it as the ratio of the
self-energies of the states.

The \textquotedbl dressing\textquotedbl{} of the non-diagonal form
factor comes from summing up virtual particles winding around the
finite volume cylinder with a certain measure. The form factor term
under the integral is the connected non-diagonal form factor defined
by the singular expansion.. We saw that the integration measure factorises
into single particle contributions, and we calculated its value up
until the third order (\ref{eq:mu}) in the filling fraction $n$.

With the conjectured structure for the finite volume form factor,
the following open question is to understand the introduced quantities
in all orders of the filling fraction and express them with other
physically meaningful quantities. For the former direction, we can
pursue the calculation of specific terms from the bilocal expansion
that contribute to only one specific quantity. However, our definitions
still have some freedom, e.g. the rapidity ordering inside the connected
form factors or the imaginary shift of the integration contours. Investigating
the dependence on these properties might give insight into a natural
choice that leads to simplification. For the latter, results obtained
by the fermionic base approach \cite{Jimbo:2009ja,Jimbo:2011bc,Negro:2013wga,Bajnok:2019yik}
and expression for three-point functions in N=4 SYM via integrable
techniques can give essential insight \cite{Basso:2015eqa,Basso:2015zoa,Eden:2015ija,Basso:2017muf,Basso:2022nny}.

We plan to return to these questions in a subsequent paper.

\subsection*{Acknowledgments}

We thank Benjamin Basso the useful discussions. This work was supported
by the NKFIH research Grant K134946. The work of IMSZ was supported
by the grant \textquotedblright Exact Results in Gauge and String
Theories\textquotedblright{} from the Knut and Alice Wallenberg foundation
and by RFBR grant 18-01-00460A and by Nordita that is supported in
part by NordForsk.

\appendix

\section{Large volume expansion of the energy difference\label{app:Ediff}}

In this Appendix we perform the large volume expansion of the TBA
energies. We do it in the usual way in terms of the small quantity
$e^{-mL\cosh\theta}$. The expansion of the ground state TBA equation
takes the form
\begin{equation}
\epsilon_{0}(\theta)=mL\cosh\theta-\int\frac{d\theta'}{2\pi}\varphi(\theta-\theta')e^{-mL\cosh\theta'}+\dots\,,
\end{equation}
which leads to the ground state energy up to the second Lüscher order
as 
\begin{align}
E_{0} & =-m\int\frac{d\theta}{2\pi}\cosh\theta\,\log(1+e^{-\epsilon_{0}(\theta)})\\
 & =-m\int\frac{d\theta}{2\pi}\cosh\theta\left[e^{-\epsilon_{0}(\theta)}-\frac{1}{2}e^{-2\epsilon_{0}(\theta)}+\dots\right]\nonumber \\
 & =-m\int\frac{d\theta}{2\pi}\cosh\theta\left[e^{-mL\cosh\theta}-\frac{1}{2}e^{-2mL\cosh\theta}+e^{-mL\cosh\theta}\int\frac{d\theta'}{2\pi}\varphi(\theta-\theta')e^{-mL\cosh\theta'}+\dots\right]\,.\nonumber 
\end{align}
Similar calculation for the excited state gives 
\begin{equation}
\epsilon_{1}(\theta)=mL\cosh\theta+\log S(\theta-\bar{\theta}-\frac{i\pi}{2})-\int\frac{d\theta'}{2\pi}\varphi(\theta-\theta')S(\frac{i\pi}{2}+\theta'-\bar{\theta})e^{-mL\cosh\theta'}+\dots,,
\end{equation}
\begin{align}
E_{\bar{\theta}} & =m\cosh\bar{\theta}-m\int\frac{d\theta}{2\pi}\cosh\theta\,\log(1+e^{-\epsilon_{1}(\theta)})\\
 & =m\cosh\bar{\theta}-m\int\frac{d\theta}{2\pi}\cosh\theta\,(e^{-\epsilon_{1}(\theta)}-\frac{1}{2}e^{-2\epsilon_{1}(\theta)}+\dots)\nonumber \\
 & =m\cosh\bar{\theta}-m\int\frac{d\theta}{2\pi}\cosh\theta\,\left[S(\frac{i\pi}{2}+\theta-\bar{\theta})e^{-mL\cosh\theta}-\frac{1}{2}S(\frac{i\pi}{2}+\theta-\bar{\theta})^{2}e^{-2mL\cosh\theta}\right.\nonumber \\
 & \qquad\qquad\qquad\hspace{1em}\qquad+\left.S(\frac{i\pi}{2}+\theta-\bar{\theta})e^{-mL\cosh\theta}\int\frac{d\theta'}{2\pi}\varphi(\theta-\theta')S(\frac{i\pi}{2}+\theta'-\bar{\theta})e^{-mL\cosh\theta'}\right]\,.\nonumber 
\end{align}
Clearly the difference between the ground state and the excited state
energy is the change $e^{-mL\cosh\theta}\to S(i\frac{\pi}{2}+\theta-\bar{\theta})e^{-mL\cosh\theta}$,
which can be substituted directly in any term of the large volume
expansion.

In order to streamline the notation we could introduce the symbol
$L_{i}=e^{-mL\cosh\theta_{i}}$ and understood an integration for
$\theta_{i}$, whenever $L_{i}$ appears. Thus the energy difference
takes a compact form 
\begin{align}
E_{\bar{\theta}}-E_{0} & =\bar{e}-e_{1}\,(S_{1}^{-1}-1)L_{1}+\frac{1}{2}e_{1}\,(S_{1}^{-2}-1)L_{1}^{2}-e_{1}\,\varphi_{12}\left(S_{1}^{-1}S_{2}^{-1}-1\right)L_{1}L_{2}+\dots\nonumber \\
 & =m\cosh\bar{\theta}+\Delta_{1}E+\Delta_{2}E\,,
\end{align}
where we used the previously introduced streamlined notations $S_{1}^{-1}=S(\frac{i\pi}{2}+\theta_{1}-\bar{\theta})$,
together with $\varphi_{ij}=\varphi(\theta_{i}-\theta_{j})$ and $e_{i}=m\cosh\theta_{i}$.
There is a nice graphical representation for the whole expansion in
\cite{Kostov:2018dmi}. It turns out that the expansion for the form
factor is more natural in terms of the filling fraction, so in the
next Appendix we perform that expansion.

\section{The energy difference in terms of the filling fraction}

As one can see from Appendix \ref{app:Ediff}, the energy difference
\begin{equation}
E_{\bar{\theta}}-E_{0}=m\cosh\bar{\theta}-\int\frac{\mathrm{d}\theta}{2\pi}\,m\cosh(\theta)\ln\left(\frac{1+e^{-\epsilon_{1}(\theta)}}{1+e^{-\epsilon_{0}(\theta)}}\right)\,,
\end{equation}
is expressible by integrating the difference of the logarithmic factors
corresponding to the pseudo energies of the excited and vacuum states:
\begin{equation}
\mathcal{L}_{i}=\ln\left(\frac{1+e^{-\epsilon_{1}(\theta_{i})}}{1+e^{-\epsilon_{0}(\theta_{i})}}\right).
\end{equation}
Let us note here that a similar quantity plays the role of the measure
in case of the Ising model form factors.

We would like to compare the above energy difference appearing in
the exponential factor of the large separation limit to the result
of the cluster expansion. For the ease of comparison, we will expand
the above quantity directly in terms of the $n_{i}$ excited state
filling fraction. That is, we would like to express $\mathcal{L}_{i}$
as a sum of contributions
\begin{equation}
\mathcal{L}_{i}=\Delta_{1}\mathcal{L}_{i}+\Delta_{2}\mathcal{L}_{i}+\Delta_{3}\mathcal{L}_{i}+\ldots+\Delta_{k}\mathcal{L}_{i}+\ldots\,,
\end{equation}
where $\Delta_{k}\mathcal{L}_{i}$ contains only terms exactly of
order $k$ in the $n_{i}$ filling fraction. Note that this expansion
at each order will regroup infinitely many terms of the Lüscher expansion.
Actually, at first we need to determine the Lüscher expansion of $\mathcal{L}_{1}$
and the filling fraction themselves (we follow the notations established
previously - except integration is not understood for the $\theta_{1}$
variable now, if we write $L_{1}$):
\begin{align}
\mathcal{L}_{1} & =L_{1}\left(S_{1}^{-1}-1\right) & \quad &  & n_{1} & =L_{1}S_{1}^{-1} & \text{LO}\nonumber \\
 & -\frac{1}{2}L_{1}^{2}\left(S_{1}^{-2}-1\right)+L_{1}L_{2}\varphi_{12}\left(S_{1}^{-1}S_{2}^{-1}-1\right) & \quad &  &  & -L_{1}^{2}S_{1}^{-2}+L_{1}L_{2}\varphi_{12}S_{1}^{-1}S_{2}^{-1} & \text{NLO}\\
 & +\mathcal{O}(e^{-3mL}) & \quad &  &  & +\mathcal{O}(e^{-3mL})\,.\nonumber 
\end{align}

At first we compare the leading orders and redefine the $\text{LO}$
of $\mathcal{L}(\theta_{1})$ such that it contains the exact filling
fraction, instead of the $\text{LO}$ of $n_{1}$ only:
\begin{equation}
\mathcal{L}_{1}^{(\text{LO})}=L_{1}\left(S_{1}^{-1}-1\right)=\left(1-S_{1}\right)n_{1}^{(\text{LO})}\quad\Rightarrow\quad\Delta_{1}\mathcal{L}_{1}=\left(1-S_{1}\right)n_{1}\,.
\end{equation}
Now we proceed order-by-order, and our next step is to compare (neglecting
$e^{-3mL}$ terms) the two sides of:
\begin{equation}
\mathcal{L}_{1}=\Delta_{1}\mathcal{L}_{1}+\Delta_{2}\mathcal{L}_{1}+\mathcal{O}(e^{-3mL})=\mathcal{L}_{1}^{(\text{LO})}+\mathcal{L}_{1}^{(\text{NLO})}+\mathcal{O}(e^{-3mL})\,.
\end{equation}
Thus we can determine $\Delta_{2}\mathcal{L}_{1}$ up to second Lüscher
order, and we get:
\begin{equation}
\Delta_{2}\mathcal{L}_{1}=\frac{1}{2}L_{1}^{2}\left(S_{1}^{-1}-1\right)^{2}+L_{1}L_{2}\varphi_{12}\left(S_{2}^{-1}-1\right)+\mathcal{O}(e^{-3mL})\,,
\end{equation}
which we may rewrite by using the relation $L_{i}=n_{i}S_{i}+\mathcal{O}(e^{-2mL})$
as this modifies only $\mathcal{O}(e^{-3mL})$ terms. In the end we
arrive at a definition
\begin{equation}
\Delta_{2}\mathcal{L}_{1}=\frac{1}{2}n_{1}^{2}\left(1-S_{1}\right)^{2}+n_{1}n_{2}\varphi_{12}S_{1}\left(1-S_{2}\right)\,,
\end{equation}
where it is understood that we integrate over the argument $\theta_{2}$
of $n_{2}$.

Here we also present the whole formula together with the third order
in a compact notation:
\begin{align}
\mathcal{L}_{1} & =n_{1}s_{1}+\frac{1}{2}n_{1}^{2}s_{1}^{2}+n_{1}n_{2}\varphi_{12}S_{1}s_{2}+\frac{1}{3}n_{1}^{3}s_{1}^{3}\nonumber \\
 & +n_{1}^{2}n_{2}\varphi_{12}S_{1}s_{1}s_{2}+\frac{1}{2}n_{1}n_{2}^{2}\varphi_{12}S_{1}s_{2}^{2}+n_{1}n_{2}n_{3}\left[\varphi_{12}\varphi_{23}S_{1}S_{2}s_{3}-\frac{1}{2}\varphi_{12}\varphi_{13}S_{1}s_{2}s_{3}\right]+\mathcal{O}(e^{-4mL})\,,\label{eq:Lexpansion}
\end{align}

where $s_{i}=1-S_{i}$.

\section{Pole structure of the connected form factor\label{sec:Pole-structure-of}}

In this appendix we would like to understand the behaviour of $F(1,2)$
when the two arguments approach each other $\theta_{1}\sim\theta_{2}$.
The antisymmetric part (\ref{eq:antisymmpart}) of the connected form
factor has a zero at $\theta_{1}=\theta_{2}$, thus we may write:
\begin{equation}
F(1,2)=\frac{R(\theta_{1},\theta_{2})}{(\theta_{1}-\theta_{2})^{2}}+\mathcal{O}(1),
\end{equation}
where $R(\theta_{1},\theta_{2})$ is a symmetric function for the
$\theta_{1}\leftrightarrow\theta_{2}$ exchange.\footnote{Note that this definition is a bit arbitrary, as we did not fix the
$\mathcal{O}(1)$ term. The coefficient of the $1/(\theta_{1}-\theta_{2})^{2}$
term should rather be a function which depends only on $\theta_{1}+\theta_{2}.$} The source of this double pole singularity is that the original form
factor $F(\vartheta_{2}+i\pi,\vartheta_{1}+i\pi,\theta_{1},\theta_{2},\theta)$
(before taking its finite part in the $\varepsilon$-s, where $\vartheta_{j}=\theta_{j}+i\varepsilon_{j}$)
has another, independent pole structure. Namely we can use the kinematical
axiom between its first and third argument, and simultaneously, between
its second and fourth argument. Thus, it has an expansion also in
terms of $(\vartheta_{1}-\theta_{2})^{-1},(\vartheta_{2}-\theta_{1})^{-1}$,
which looks very similar to (\ref{eq:Fcexpansion}), i.e. the expansion
in terms of $(\vartheta_{1}-\theta_{1})^{-1},(\vartheta_{2}-\theta_{2})^{-1}$.
These two expansions - for the two different pairings of the arguments
- can be used independently, even if all four arguments are close,
which follows from the kinematical singularity axiom. By using the
permutation axiom, and expanding the kinematical pole structure of
$F(\vartheta_{2}+i\pi,\vartheta_{1}+i\pi,\theta_{2},\theta_{1},\theta)$
between its second and third, and its first and fourth arguments,
respectively (based on (\ref{eq:Fcexpansion})), we get the most singular
term
\begin{align}
 & F(\vartheta_{2}+i\pi,\vartheta_{1}+i\pi,\theta_{1},\theta_{2},\theta)/F\\
= & S_{12}F(\vartheta_{2}+i\pi,\vartheta_{1}+i\pi,\theta_{2},\theta_{1},\theta)/F\sim S_{12}\left(\frac{i}{\vartheta_{1}-\theta_{2}}\frac{i}{\vartheta_{2}-\theta_{1}}A_{21}+\ldots\right),\nonumber 
\end{align}
where $A_{21}=s_{2}s_{1}$. If we simply put $\vartheta_{1}=\theta_{1},\vartheta_{2}=\theta_{2}$
(i.e. perform the $\varepsilon$ limit) the above term behaves as
\begin{equation}
-s_{1}s_{2}(\theta_{1}-\theta_{2})^{-2}+\mathcal{O}\left((\theta_{1}-\theta_{2})^{-1}\right),
\end{equation}
and we can read off that $R(\theta_{1},\theta_{2})=-s_{1}s_{2}$.
The minus sign came from the $S$-matrix, since $S_{12}=S(0)+\mathcal{O}(\theta_{1}-\theta_{2})$,
and we dropped the higher order corrections.

\section{Contour deformation for the third order result}

\label{sec:Contour-deformation-for}

In this section we explain the difficulties in evaluating the third
order graphs. In doing so we pick one of the difficult ones, namely
diagram 11 in Figure \ref{NNLOcontrib}, which has the contribution
\begin{align}
 & F^{2}(\vartheta_{3}+i\pi,\vartheta_{2}+i\pi,\theta_{1},\theta_{2},\theta)/F^{2}\times\label{eq:N3val}\\
 & F^{1}(\vartheta_{1}+i\pi,\theta_{3},\vartheta-i\pi)/F^{1}S(\vartheta_{2}-\vartheta_{1})S(\vartheta_{3}-\vartheta_{1})S_{3}e^{-imy(\sinh\vartheta_{1}-\sinh\theta_{3})}.\nonumber 
\end{align}
 The finite part operation for $\varepsilon_{1},\varepsilon_{3}$
is trivial, only the $\varepsilon_{2}$ limit will differentiate a
single $S$-matrix:
\begin{equation}
F^{2}(\theta_{3}+i\pi,\vartheta_{2}+i\pi,\theta_{2},\theta_{1},\theta)/F^{2}S_{12}S(\vartheta_{2}-\theta_{1})\times F^{1}(\theta_{1}+i\pi,\theta_{3},\theta-i\pi)/F^{1}S_{31}S_{3}e^{-iy(p_{1}-p_{3})},
\end{equation}
 where for convenience we also used the permutation axiom for the
form factor of the second operator. We now expand in $\varepsilon_{2}$
as
\begin{align}
 & F^{2}(\theta_{3}+i\pi,\vartheta_{2}+i\pi,\theta_{2},\theta_{1},\theta)/F^{2}S_{12}S(\vartheta_{2}-\theta_{1})\nonumber \\
= & \left(\frac{1}{\varepsilon_{2}}\left(1-S_{21}S_{32}S_{2}\right)F^{2}(\theta_{3}+i\pi,\theta_{1},\theta)/F^{2}+F^{2}(\theta_{2}\vert\theta_{3}+i\pi,\theta_{1},\theta)\right)\left(1-\varepsilon_{2}\varphi_{12}\right)+\ldots,
\end{align}
where we introduced $F^{2}(\theta_{2}\vert\theta_{3}+i\pi,\theta_{1},\theta)$
as the finite part of the above five-particle form factor in $\varepsilon_{2}$.
When we approach $\theta_{3}=\theta_{1},$ then this latter object
still contains the kinematical pole between the first and the fourth
argument of $F^{2}(\theta_{3}+i\pi,\vartheta_{2}+i\pi,\theta_{2},\theta_{1},\theta),$
but nothing from the kinematical singularity between the second and
the third, i.e. no terms proportional to $1/\varepsilon_{2}$ which
would appear in a similar expansion (\ref{eq:Fcexpansion}). That
is, we have
\begin{equation}
F^{2}(\theta_{2}\vert\theta_{3}+i\pi,\theta_{1},\theta)=\frac{i}{\theta_{3}-\theta_{1}}A_{2}+F^{2}(\theta_{2},\theta_{1})+\ldots,\label{eq:semiconnected}
\end{equation}
where $A_{2}=F^{2}(2)s_{1}+s_{2}\varphi_{12}$ following from the
definition in Appendix \ref{sec:Graph_rules}.

Thus, after taking the $\varepsilon\to0$ limit, we have two terms:
\begin{equation}
-\varphi_{12}\left(1-S_{21}S_{32}S_{2}\right)F^{2}(\theta_{3}+i\pi,\theta_{1},\theta)/F^{2}F^{1}(\theta_{1}+i\pi,\theta_{3},\theta-i\pi)/F^{1}S_{31}S_{3}e^{-iy(p_{1}-p_{3})},\label{eq:N3sep}
\end{equation}
which clearly has double and first-order poles at $\theta_{1}=\theta_{3}$,
while being regular in the difference of variables $\theta_{1}-\theta_{2}$
or $\theta_{2}-\theta_{3}$; and another one
\begin{equation}
F^{2}(\theta_{2}\vert\theta_{3}+i\pi,\theta_{1},\theta)F^{1}(\theta_{1}+i\pi,\theta_{3},\theta-i\pi)/F^{1}S_{31}S_{3}e^{-iy(p_{1}-p_{3})},\label{eq:N3conn}
\end{equation}
which is singular when any pair of the three variables $\theta_{1},\theta_{2},\theta_{3}$
coincides.

As for the sixth diagram of the second order, we need to consider
the oscillatory behaviour of the exponential factor and the singularity
structure of the form factors, and regularize the integrals accordingly.
This can be done by shifting all three integration rapidities in the
positive imaginary direction by an infinitesimal amount $\theta_{k}\to\theta_{k}+i\delta_{k},\;k=1,2,3$,
and establishing an ordering where $\delta_{1}>\delta_{2}>\delta_{3}>0$.

The exponential factor $e^{-iy(p_{1}-p_{3})}$ implies, that in the
clustering limit we need to shift the $\theta_{1}$ integration below
the real axis, i.e. $\delta_{1}<0$. In the meantime we need to pick
up possible residues at around $\theta_{1}=\theta_{2}$ and $\theta_{1}=\theta_{3}$.

In the first case there is an exponential factor remaining, which
after the $\theta_{1}\to\theta_{2}$ substitution looks like $e^{-iy(p_{2}-p_{3})}$
, and needs to be treated as before. We now need to shift the $\theta_{2}$
integration contour below the real line, i.e. $\delta_{2}<0$, and
pick up a possible remaining residue at $\theta_{2}=\theta_{3}.$
The exponential factor would disappear after taking this residue,
leaving us with a finite result. The remaining two-integral term -
in which we exchanged the ordering of the $\theta_{2}$ and $\theta_{3}$
integrations - gives zero since we still need to shift $\theta_{2}$
below the real line, and in the end the exponential factor will decay
because of $\delta_{2}<0$ and $\delta_{3}>0$; that is $e^{-imy(\sinh(\theta_{2}-i\vert\delta_{2}\vert)-\sinh(\theta_{3}+i\delta_{3}))}\sim e^{-ym(\sin\vert\delta_{2}\vert+\sin\delta_{3})}.$

Clearly, only the second term (\ref{eq:N3conn}) of the $\varepsilon$
limit could contribute in this scenario, but instead of dealing with
the $\theta_{1}=\theta_{2}$ singularity of the object $F^{2}(\theta_{2}\vert\theta_{3}+i\pi,\theta_{1},\theta)$
we rather return to the initial formula (\ref{eq:N3val}). While forgetting
about the $\vartheta_{2}\to\theta_{2}$ limit, we use the kinematical
axiom between the second and third argument of the form factor: 
\begin{equation}
F^{2}(\vartheta_{3}+i\pi,\vartheta_{2}+i\pi,\theta_{1},\theta_{2},\theta)/F^{2}=\frac{i}{\vartheta_{2}-\theta_{1}}\left(1-S_{12}S(\vartheta_{3}-\theta_{1})S_{1}\right)F^{2}(\vartheta_{3}+i\pi,\theta_{2},\theta)/F^{2}+\mathcal{O}(1),
\end{equation}
then we simply put all $\varepsilon$-s to zero. The reasoning behind
this is that we can omit the $1/\varepsilon_{2}$ singularity as we
know that its explicit contribution (\ref{eq:N3sep}) is not singular
for $\theta_{1}=\theta_{2}$.

Now we take the residue at $\theta_{1}=\theta_{2}$

\begin{align}
-2\pi i\text{Res}_{\theta_{1}=\theta_{2}} & \frac{n_{1}}{2\pi}\frac{n_{2}}{2\pi}\frac{n_{3}}{2\pi}\Bigg\lbrace\frac{-i}{\theta_{1}-\theta_{2}}\left(1-S_{12}S_{31}S_{1}\right)F^{2}(\theta_{3}+i\pi,\theta_{2},\theta)/F^{2}\times\\
 & \quad\quad\quad\quad\quad F^{1}(\theta_{1}+i\pi,\theta_{3},\theta-i\pi)/F^{1}S_{21}S_{31}S_{3}e^{-imy(p_{1}-p_{3})}\Bigg\rbrace\nonumber \\
= & \frac{n_{2}^{2}}{2\pi}\frac{n_{3}}{2\pi}\left(1+S_{32}S_{2}\right)S_{32}S_{3}F^{2}(\theta_{3}+i\pi,\theta_{2},\theta)/F^{2}F^{1}(\theta_{2}+i\pi,\theta_{3},\theta-i\pi)/F^{1}e^{-imy(p_{2}-p_{3})},\nonumber 
\end{align}
and then the second one at $\theta_{2}=\theta_{3}$. There is clearly
a double pole in the product of the form factors, which will differentiate
the multiplicative factors, even the square of the filling fraction
$n_{2}^{2}$. As for the sixth graph in the second order, after partial
integration, one can recognize all these terms being proportional
to $n_{3}^{3}$. The result is then a single integral over $\theta_{3}$,
which is not shown here, since it is rather straightforward to derive.

In the second case of the clustering limit, when we already pulled
the $\theta_{1}$ integration below the $\theta_{2}$ one, the first
residue we need to take is at $\theta_{1}=\theta_{3}$. For (\ref{eq:N3val}),
this can be done easily, the singularities come from the product of
the two three-particle form factors again. Even if the second order
pole differentiates the $n_{1}$ factor, after partial integration
the result will be a double integral where the measure factor is $n_{2}n_{3}^{2}$.

For the $\theta_{1}=\theta_{3}$ residue of (\ref{eq:N3conn}), we
also need to consider the $(\theta_{1}-\theta_{3})^{-1}$ pole shown
explicitly in (\ref{eq:semiconnected}):
\begin{equation}
-2\pi i\text{Res}_{\theta_{1}=\theta_{3}}\frac{n_{1}}{2\pi}\frac{n_{2}}{2\pi}\frac{n_{3}}{2\pi}\left(\frac{i}{\theta_{3}-\theta_{1}}\left(F^{2}(2)s_{1}+s_{2}\varphi_{12}\right)+F^{2}(\theta_{2},\theta_{1})\right)\left(\frac{is_{3}}{\theta_{1}-\theta_{3}}+\bar{F}^{1}(3)\right)S_{31}S_{3}e^{-iy(p_{1}-p_{3})},\label{eq:13residue}
\end{equation}
and the result will be proportional to $n_{2}n_{3}^{2}$ again.

Let us make some remarks about a particular term that appears after
we evaluate the residue (\ref{eq:13residue}): 
\begin{equation}
-\int\frac{d\theta_{2}}{2\pi}\int\frac{d\theta_{3}}{2\pi}n_{2}n_{3}^{2}F^{2}(\theta_{2},\theta_{3})s_{3}S_{3}.
\end{equation}
First of all, if we would like to present our result in the basis
of symmetrized connected form factors (see the discussion after (\ref{eq:symmetrization})
and also in Subsection \ref{subsec:Organisation-of-the}), we need
to separate the anti-symmetric part (\ref{eq:antisymmpart}) of $F^{2}(\theta_{2},\theta_{3})$,
as it gets multiplied with a non-symmetric function. Another peculiarity
of this term is that because of the singularity (\ref{eq:connsing})
of the connected form factor we need to keep the regularization $\theta_{2}\to\theta_{2}+i\delta_{2},\quad\theta_{3}\to\theta_{3}+i\delta_{3}$
where $\delta_{2}>\delta_{3}>0$;

\section{Free fermion calculations}

\label{sec:Free-fermion}

In this appendix, we summarise the calculations of the finite volume
form factors in the massive free fermion theory. Although the model
is free, but the non-local $\sigma$ field changes the boundary condition
and interpolates between the Neveu-Schwarz and Ramond sectors. Its
finite volume form factors are highly non-trivial and were determined
explicitly in \cite{Fonseca:2001dc}. In the following we explain
how our approach reproduces this non-trivial result.

The even infinite volume form factors of the $\sigma$ field are 
\begin{equation}
F(\theta_{1},\dots,\theta_{2n})=i^{n}\prod_{j<k}\tanh((\theta_{j}-\theta_{k})/2)\,,
\end{equation}
which can be written alternatively as 
\begin{equation}
F(\theta_{1},\dots,\theta_{2n})=\sum_{\mathrm{all\ pairings}}\quad\prod_{\mathrm{pairs}}F(\mathrm{pairs)}(-1)^{\#}\,,
\end{equation}
where $\#$ merely counts the crossing in the diagram. We need these
form factors in our approach when $n$ particles are incoming, while
$n$ are outgoing and some of them are almost diagonal.

The simplest almost diagonal contribution is 
\begin{equation}
F(\theta_{1}+i\pi+i\varepsilon_{1},\theta_{1})=\frac{2}{\varepsilon_{1}}+\dots\,,
\end{equation}
where the dots represents terms of $O(\varepsilon_{1})$. The simplest
non-diagonal form factor is 
\begin{equation}
F(\theta_{1}+i\pi+i\varepsilon_{1},\theta_{2})=F(\theta_{1}+i\pi,\theta_{2})+\dots\equiv\bar{F}_{12}+\dots\,,
\end{equation}
which is singular for $\theta_{1}\to\theta_{2}$: 
\begin{equation}
\bar{F}_{12}=\frac{2i}{\theta_{1}-\theta_{2}}+O(\theta_{1}-\theta_{2})\,.
\end{equation}

The next simplest diagonal form factor is 
\begin{equation}
F(\theta_{1}+i\pi+i\varepsilon_{1},\theta_{1},\theta_{2}+i\pi+i\varepsilon_{2},\theta_{2})=\frac{2}{\varepsilon_{1}}\frac{2}{\varepsilon_{2}}-\bar{F}_{12}\bar{F}_{21}+F_{12}F_{21}+\dots\,,
\end{equation}
where $F_{12}=F(\theta_{1},\theta_{2})=-F_{21}$. Clearly the connected
form factor, which is the $O(1)$ piece, is nothing but
\begin{equation}
{\cal F}(\theta_{1},\theta_{2})=F_{12}F_{21}-\bar{F}_{12}\bar{F}_{21}=4f_{12}\,,
\end{equation}
which is symmetric by itself. In the general formula 
\begin{equation}
F^{2}(\{\vartheta\}_{B^{+}}+i\pi,\{\theta)F^{1}(\{\vartheta\}_{B^{-}}+i\pi,\{\theta\}_{A^{+}})\,,
\end{equation}
we need to form all possible pairs between the particles and associate
a contribution $F_{ij}$ if they are both incoming or outgoing and
$\bar{F}_{ij}$ if they are different and multiply with an $S$-matrix
factor $(-1)$ for each crossing, keeping also in mind that 
\begin{equation}
F^{2}(\theta_{1},\dots,\theta_{2n})=(-1)^{n}F(\theta_{1},\dots,\theta_{2n})\quad;\qquad F^{1}(\theta_{1},\dots,\theta_{2n})=F(\theta_{1},\dots,\theta_{2n})\,.
\end{equation}

Let us see now why our formula (\ref{eq:sigmavev},\ref{eq:Fsigmasigma})
gives a factorising and exponentiating result. Clearly the measure
factor and the $K$ factor factorise into one-particle terms. Moreover,
each form factor is a sum of terms factorising into two particle terms,
thus the total contribution is a sum of factorised terms.

In order to show exponentiation we need to show that each diagram
comes with its symmetry factor. At $N$th order in the LM type formula
we have the correct $1/N!$ symmetry factor as the connected form
factor is fully symmetric. By expanding this connected form factor
we connect $k$ outgoing particles to the first operator and $N-k$
outgoing particles to the second operator. Then we also connect $k$
incoming particles to the first and $N-k$ outgoing to the second.
Choosing different incoming distributions lead to different form factors.
They lead to the same form factor contribution only if we permute
them together with the outgoing particles. For outgoing particles
we could choose ${N \choose k}$ ways how $k$ rapidities can be connected
to the first and $N-k$ to the second operator. They all contribute
the same as S-matrix factors cancel and each form factor is completely
symmetric (once incoming and outgoing rapidities are permuted together).
Thus form factors come with their $1/k!$ and $1/(N-k)!$ symmetry
factors. During the resolution of the form factor with $k$ incoming
and outgoing particles we form $k_{1},\dots,k_{l}$ cycles. (In defining
the cycle we just follow the indices of the two-particle form factors
in the product and see when do they close). The first cycle with $k_{1}$
particles can be formed ${k \choose k_{1}}$ different ways. The second
cycle with $k_{2}$ particles can be formed ${k-k_{1} \choose k_{2}}$
different ways, and so on. All together the symmetry factor is 
\begin{equation}
\frac{1}{k!}{k \choose k_{1}}{k-k_{1} \choose k_{2}}\dots{k-k_{1}-\dots-k_{l-1} \choose k_{l}}=\frac{1}{k_{1}!\dots k_{l}!}\,,
\end{equation}
which is indeed the symmetry factor of the graph if each cycle appears
ones. It might happen, however, that the $k_{1}$ cycle appears $l_{1}$
times. This means that we have overcounted the terms and we have to
divide by the symmetry factor $l_{1}$. Similar arguments can also
be made for the higher $k_{i}$-s. But this implies that each composition
of disconnected graphs come with the right symmetry factor, which
guaranties exponentiation. We then compare only the connected graphs
to the exponents. We start with the energy and proceed to the form
factors.

\subsection{Energy difference}

\begin{figure}
\begin{centering}
1.\includegraphics[height=4cm]{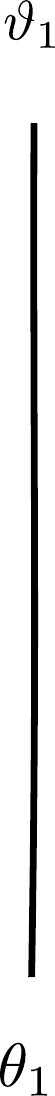}~~~~~~~~~~~~~~2.\includegraphics[height=4cm]{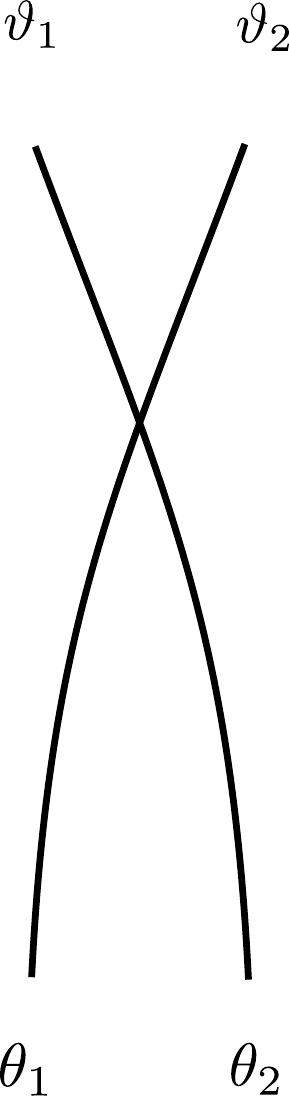}~~~~~~~~~~~~~~3.\includegraphics[height=4cm]{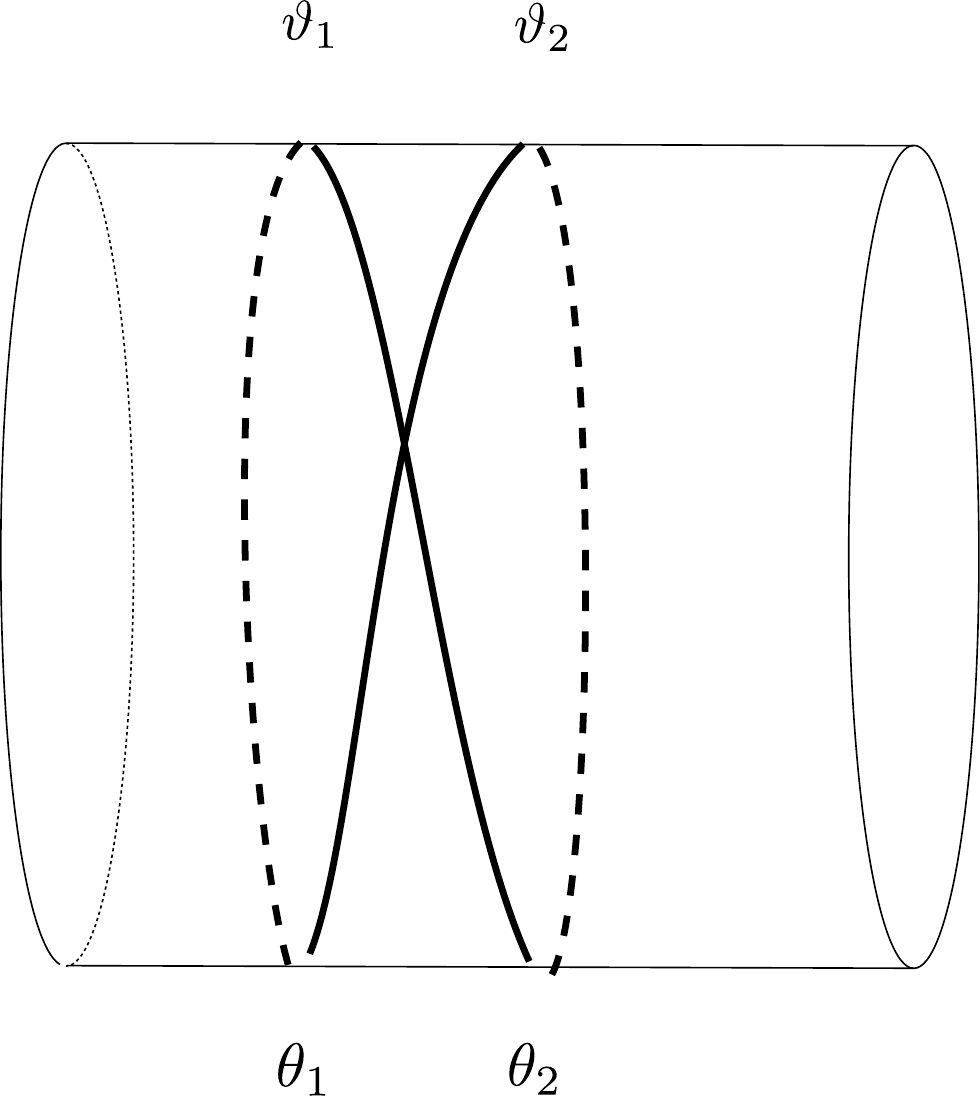}~~~~~~~~~~~~~~4.\includegraphics[height=4cm]{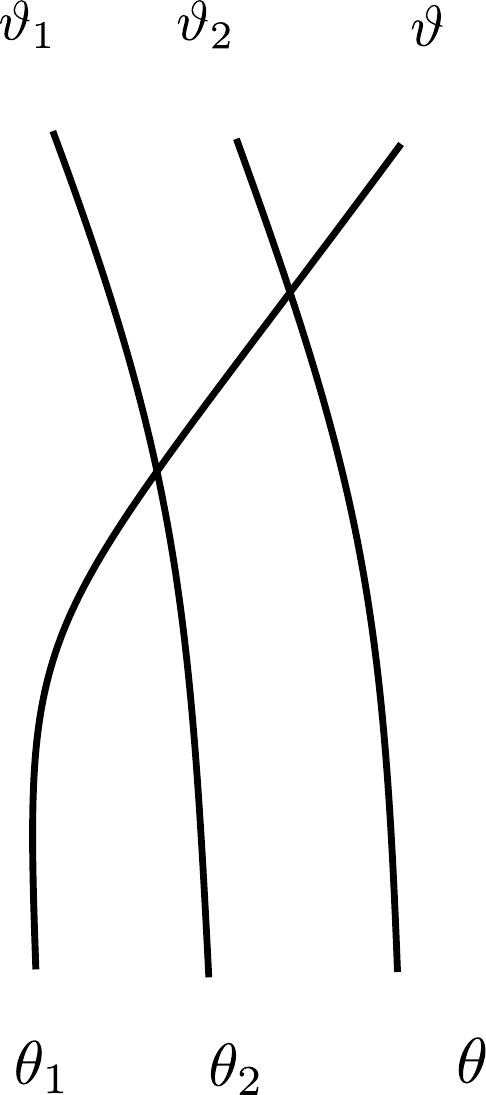}~~~~~~~~~~~~~~5.\includegraphics[height=4cm]{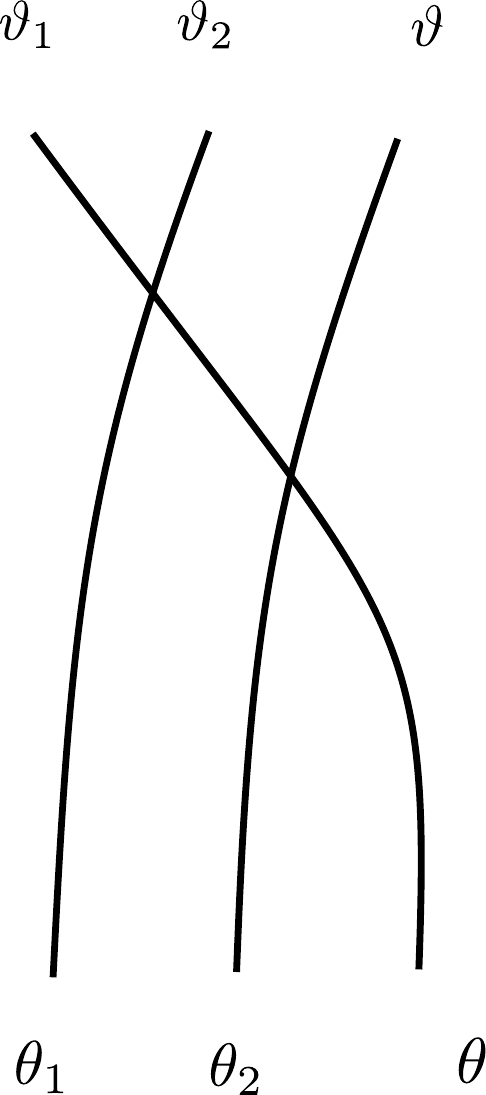}\caption{Low order diagrams contributing to the measure of the energy difference.}
\par\end{centering}
\label{IsingDiag}
\end{figure}

In the case of the energy difference we would like to recover the
\begin{equation}
y\Delta E=-ye_{1}{\cal L}_{1}=-ye_{1}\left(2n_{1}+\frac{(2n_{1})^{2}}{2}+\frac{(2n_{1})^{3}}{3}+\dots\right)\,,
\end{equation}
 expression order by order. We need to show that at $N^{th}$ order
the singly differentiated $K_{y}$ factor comes with a $2^{N}/N$
factor. We proceed inductively in $N$.

At the one-particle level we need the diagram when both particles
are connected to the ${\cal O}_{1}$ operator, which after resolving
the form factor looks like the first diagram in Figure \ref{IsingDiag}
and contributes as 
\begin{equation}
n_{1}F^{1}(\theta_{1}+i\pi+i\varepsilon_{1},\theta_{1})e^{ye_{1}\varepsilon_{1}}=n_{1}\frac{2}{\varepsilon_{1}}(1+ye_{1}\varepsilon_{1})=\dots+2n_{1}ye_{1}+\dots\,.
\end{equation}
Actually when they are connected to ${\cal O}_{2}$ it gives a term
$-\frac{2n_{1}}{\varepsilon_{1}}$, which cancels the singular piece,
while when they are connected to different operators the contribution
will not survive in the $y\to\infty$ limit.

At the two particle level we are testing the $\frac{(2n_{1})^{2}}{2}$
term. We thus need the connected diagrams which differentiate the
exponent. Similarly to the general case our convention is that we
put all $\varepsilon_{i}$ to zero and then shift the contour $\theta_{1}+i\delta_{1}$
such that $\delta_{1}>0$. The term which is growing in this limit
is 
\begin{equation}
F^{2}(\theta_{2}+i\pi,\theta_{1})F^{1}(\theta_{1}+i\pi,\theta_{2})(-1)e^{-iym(\sinh\theta_{1}-\sinh\theta_{2})}=-\bar{F}_{21}\bar{F}_{12}e^{-iy(p_{1}-p_{2})}\,,
\end{equation}
and can be drawn after the resolution of the form factor as the second
diagram in the Figure \ref{IsingDiag}. In this term we shift the
contour below the real line and pick up the residue
\begin{equation}
-i\frac{n_{2}^{2}}{2}\text{res}\frac{2i}{\theta_{2}-\theta_{1}}\frac{2i}{\theta_{1}-\theta_{2}}(-1)(1-iye_{2}(\theta_{1}-\theta_{2})+\dots)\,,
\end{equation}
where we focused only on the surviving odd term, which gives the expected
result
\begin{equation}
\frac{(2n_{2})^{2}}{2}ye_{2}\,.
\end{equation}
Let us think of the contributing diagram on the cylinder when $\vartheta_{1}$
is connected to $\theta_{1}$ and $\vartheta_{2}$ to $\theta_{2}$,
see the third diagram in Figure \ref{IsingDiag}. Observe that we
have one single loop which wraps twice around the cylinder. Clearly
this graph is the only connected graph of this sort.

At the three particle level we would like to reproduce $\frac{(2n_{1})^{3}}{3}$.
This should come from terms when two integrals are eliminated by contour
shifts and residues. We now shift the integrals in an ordered way:
$\theta_{1}+i\delta_{1}$, $\theta_{2}+i\delta_{2}$ such that $\delta_{1}>\delta_{2}>0$.
We shift the $\theta_{1}$ integral through the $\theta_{2}$ and
$\theta_{3}$ integrals and then the $\theta_{2}$ integral through
the $\theta_{3}$ integral. In order to have a triple residue term
$\vartheta_{1}$ should be connected to ${\cal O}_{1}$, while $\theta_{3}$
to ${\cal O}_{2}$. We need a connected diagram, in which after resolving
the form factors wraps three times. Such contribution is displayed
on the fourth diagram in Figure \ref{IsingDiag}. 
\begin{equation}
(-1)\frac{n_{1}n_{2}n_{3}}{3!}\bar{F}_{12}\bar{F}_{23}\bar{F}_{31}e^{-iym(\sinh\theta_{1}-\sinh\theta_{3})}\,.
\end{equation}
By deforming the $\theta_{1}$ contour and picking up $(-i)$ times
the single residue we arrive at 
\begin{equation}
(-1)\frac{2n_{2}^{2}n_{3}}{3!}\bar{F}_{23}\bar{F}_{32}e^{-iym(\sinh\theta_{2}-\sinh\theta_{3})}\,,
\end{equation}
which is $(\frac{2n_{2}}{3!})$ times the diagram we already calculated
at the second order. Actually there is another diagram, the fifth
in Figure \ref{IsingDiag}, with the same contribution, which comes
from 
\begin{equation}
\frac{n_{1}n_{2}n_{3}}{3!}\bar{F}_{13}\bar{F}_{32}\bar{F}_{21}(-1)e^{-iym(\sinh\theta_{1}-\sinh\theta_{3})}\,.
\end{equation}
Together they correctly reproduce the $\frac{(2n_{3})^{3}}{3}$ factor.

At the generic $k$ particle level we need to reproduce the $\frac{(2n_{1})^{k}}{k}$
factor. Clearly after resolving all the form factors we need all diagrams
which wrap around the cylinder $k$ times. These diagrams can be characterised
how we wrap. We always start with $1$ from the top and go through
all other rapidities. Clearly there are $(k-1)!$ terms of this sort.
This cycle can be represented as 
\begin{equation}
1\to i_{2}\to i_{3}\to\dots\to i_{k}\to1\,,\label{eq:loop}
\end{equation}
where each arrow represents a line going from top to down. The first
$1\to2\to\dots\to k\to1$ gives
\begin{equation}
\frac{n_{1}\dots n_{k}}{k!}\bar{F}_{12}\bar{F}_{23}\dots\bar{F}_{k-1k}(-1)e^{-iym(\sinh\theta_{1}-\sinh\theta_{k})}\,.
\end{equation}
This diagram appears when only $\vartheta_{k}$ and $\theta_{1}$
are connected to operator ${\cal O}_{2},$ $F^{2}(\theta_{k}+i\pi,\theta_{1})$
and the rest to ${\cal O}_{1},$ $F^{1}(\theta_{k-1}+i\pi,\dots\theta_{1}+i\pi,\theta_{2},\dots,\theta_{k})$.
The contribution of this diagram can be evaluated by taking residues
recursively: each residue gives a factor ($2n$).

What we really have to show that each wrapping order can appear only
once via resolving the form factors. This means that in the cycle
(\ref{eq:loop}) we have to associate either operator 2 or operator
1 to each line in order to indicate through which operator the line
went. One can recursively show that 
\begin{equation}
i_{j}-\!\!^{2}\!\!\!\!\!\to i_{j+1}\quad{\rm if}\quad i_{j}>i_{j+1}\quad;\quad i_{j}-\!\!^{1}\!\!\!\!\!\to i_{j+1}\quad{\rm otherwise}\,.
\end{equation}
Then we should count the number of 2s and associate a factor $(-1)$
for each. We then evaluate the residues starting from smaller $\theta_{i}$
to higher. To show that they all contribute the same way follows from
the fact how they transform for the permutation $i_{j}\leftrightarrow i_{j+1}$.
Such flip will change the operator of that arrow but in the same time
it changes also the sign of the residue, so all over they cancel.
Similarly, if by this permutation the neighbouring arrows also change
so do their residues. This completes the calculation of the energy
difference.

\subsection{Form factor part}

In the form factor part, we use again factorisation and compare the
exponent ${\cal L}_{1}{\cal L}_{2}f_{12}$ to the connected component.
In particular, we compare the expansion of only one of the ${\cal L}$s
as the expression must be symmetric. At the $k+1$ particle level
it should give
\begin{equation}
\frac{(2n_{1})^{k}}{k}(2n_{2})f_{12}\,,
\end{equation}
which we test order by order. Since we cannot distinguish between
$n_{1}$ and $n_{2}$ in the calculation, for $k>1$ there is an extra
factor $2$.

At the leading two-particle level we have two connected form factor
contributions
\begin{equation}
\frac{n_{1}n_{2}}{2}\left({\cal F}(\theta_{1},\theta_{2})+{\cal F}(\theta_{1},\theta_{2})\right)=4n_{1}n_{2}f_{12}\,,
\end{equation}
which comes from $F^{2}(\vartheta_{2}+i\pi,\vartheta_{1}+i\pi,\theta_{1},\theta_{2})$
and $F^{1}(\vartheta_{2}+i\pi,\vartheta_{1}+i\pi,\theta_{1},\theta_{2})$,
respectively.

At the $k=2$ level we need 
\begin{equation}
\frac{n_{1}n_{2}n_{3}}{3!}()\to2n_{2}^{2}n_{3}4f_{23}\,.
\end{equation}
That is we should take one residue and a connected form factor should
remain. In the connected form factor we always have a loop-like term
$F_{12}F_{21}$ and a crossed term $-\bar{F}_{12}\bar{F}_{21}$. Since
in the resolution we have to take all possible connections the crossed
terms will appear automatically , once we have a loop-like term. So
we focus only on the loop-like term. A loop-like term can originate
for example from a term 
\begin{equation}
\bar{F}_{13}F_{32}F_{12}\,.
\end{equation}
This means that the outgoing $\theta_{2}$ and $\theta_{3}$ are connected,
while the incoming $\theta_{1}$ and $\theta_{2}$ are also connected
and the outgoing $\theta_{1}$ is connected to the incoming $\theta_{3}$.
This diagram can originate only from 
\begin{equation}
F^{2}(\theta_{3}+i\pi,\theta_{2}+i\pi,\theta_{1},\theta_{2})F^{1}(\theta_{2}+i\pi,\theta_{3})e^{-iym(\sinh\theta_{1}-\sinh\theta_{3})}\,,
\end{equation}
thus it comes with an extra $(-1)$ factor from $F^{1}$: taking the
residue of the $\theta_{1}$ integral at $\theta_{3}$ gives the expected
contribution
\begin{equation}
\bar{F}_{13}F_{32}F_{12}(-1)e^{-iym(\sinh\theta_{1}-\sinh\theta_{3})}\to-2F_{12}F_{12}\to2{\cal F}_{12}\,.
\end{equation}
We then should check how many times we can connect two in coming and
two outgoings, such that the remaining incoming-outgoing line connects
different particles, for which we will take the residue. There are
exactly six combinations 
\begin{align}
\bar{F}_{12}F_{32}F_{13}\quad;\quad\bar{F}_{13}F_{32}F_{12}\quad;\quad\bar{F}_{23}F_{31}F_{12} & \,,\nonumber \\
\bar{F}_{31}F_{21}F_{21}\quad;\quad\bar{F}_{21}F_{31}F_{23}\quad;\quad\bar{F}_{32}F_{21}F_{13} & \,.
\end{align}
For each term we found a unique diagram where it came from and by
evaluating the residues they all contributed the same way. Altogether
they reproduced the expected combinatorial factor.

At the $k+1$ particle level we need to have a cycle of size $k-1$,
such that after the contour deformations only one form factor remains
(with two rapidities). Such term can be read starting from the top
$1$ and following its connections. We associate a double arrow if
a rapidity is connected on the same side (both outgoing or incoming,
such that they contribute to the form factor) and single arrow if
they are between different outgoing/incoming rapidities. A typical
cycle looks like
\begin{equation}
1\to i_{2}\to\dots\to i_{j}\Rightarrow i_{j+1}\Rightarrow i_{j+2}\to\dots\to i_{k}\to1\,,
\end{equation}
where $j$ can be any of $1,\dots,i_{k-1}$. Since we need just one
cycle, the two double arrows should come after each other. We then
again need to associate operators to the arrows. Single arrows should
be numbered as before. The double arrows should have the same numbers
(as we resolved a form factor) and can be contracted formally as
\begin{equation}
i_{j}\Rightarrow i_{j+1}\Rightarrow i_{j+2}\quad\longrightarrow\quad i_{j}\Rrightarrow i_{j+2}\,.
\end{equation}
The numbering rule for the triple arrow is the same as for the single
one. The resulting diagram of length $k$ looks similar than the previously
(for the energy) investigated $k$ cycle, with the exception that
the triple arrow now does not encode any singularity, so via contour
deformation it cannot pick up residue. Actually it should remain the
last connection for which residue is not taken since it contributes
to the remaining form factor. First of all, there are $(k-1)!$ cycle
of length $k$. At each cycle there is always one specific connection,
which is the last. That last connection should be the triple arrow,
which can be elevated to two double arrows by inserting all possible
$k+1$ choices as the middle term. This all together gives $(k+1)(k-1)!$
terms. Argumentations as before guaranties that all contribute the
same way and with the $1/(k+1)!$ prefactor they provide the required
$1/k$ factor.

\subsection{Excited state calculations}

In this subsection we check the excited state form factor contribution
$\kappa(\bar{\theta})$ order by order. At the first non-trivial order
we have the first and fourth diagrams on Figure (\ref{1storder}),
which give the same contributions. Let us focus on the first. The
form factor can be resolved as 
\begin{align}
F(\theta_{1}+i\pi+i\varepsilon_{1},\theta_{1},\theta) & =F(\theta_{1}+i\pi+i\varepsilon_{1},\theta_{1})-F(\theta_{1}+i\pi,\theta)+F(\theta_{1},\theta)\nonumber \\
 & =\frac{2}{\varepsilon_{1}}+F_{c}(1)+\dots=\frac{2}{\varepsilon_{1}}-\frac{2i}{\sinh(\theta_{1}-\theta)}+\dots\ .
\end{align}
The connected part after the analytical continuation $\theta\to\bar{\theta}+\frac{i\pi}{2}$
gives
\begin{equation}
\frac{2}{\cosh(\theta-\theta_{1})}
\end{equation}
and together with the measure $n_{1}$ reproduces the first order
result.

At second order we gain contributions from the sixth diagram on Figure
(\ref{NLOcontrib}). The second operator's form factor has a decomposition
\begin{equation}
F^{2}(\theta_{2}+i\pi,\theta_{1},\theta)=\bar{F}_{21}-F(\theta_{2}+i\pi,\theta)+F(\theta_{1},\theta)\ ,
\end{equation}
while the first ones
\begin{equation}
F^{1}(\theta_{1}+i\pi,\theta_{2},\theta-i\pi)=\bar{F}_{12}-F(\theta_{1}+i\pi,\theta-i\pi)+F(\theta_{2},\theta)\ .
\end{equation}
After the contour deformation the residue comes either from $\bar{F}_{21}$
or from $\bar{F}_{12}$ and the result is four times the contribution
of the previous order, which together with the $\frac{n_{2}^{2}}{2}$
measure factor gives the correct result.

At the $k^{th}$ order we need to reproduce $(2n_{1})^{k}/k$. The
calculation is very similar to the calculations for the energy and
for the form factor. We need to pick up the residue of the contour
deformations consecutively $k-1$ times. Keeping in mind that we shift
the integrals as $\theta_{j}\to\theta_{j}+\delta_{j}$ with $\delta_{j}>\delta_{j+1}$
we need to start the deformations with $\theta_{1}$. In order to
have the appropriate number of singular terms we need again loops
which wind around the cylinder. Following the lines from above to
below we represent the loop as 
\[
1\to i_{2}\to\dots\to i_{k}\to{\cal O}_{12}\to1
\]
where by ${\cal O}_{12}$ we mean that the loop should end with the
two operators, out of which one is connected with $1$. There are
exactly $(k-1)!$ such terms, which all contribute the same way. By
taking a residue we always pick up a factor $(2n_{i})$. Using previous
arguments one can show that the labelling of the arrows with the operators
is unique. Actually there are twice as many terms as we could start
the sequence with $1$ and follow the lines from the bottom. Altogether
they give the correct measure factor.

\bibliographystyle{JHEP}
\bibliography{paper}

\end{document}